\documentclass[a4paper, 11pt]{article}
\usepackage{amsmath,amssymb,bbold,mathtools,manfnt,epsfig,enumerate,enumitem,empheq}
\usepackage[mathscr]{euscript}
\usepackage{graphicx,subcaption}
\graphicspath{ {figures/} }
\allowdisplaybreaks[2]
\newcommand*\widefbox[1]{\fbox{\hspace{2em}#1\hspace{2em}}}
\numberwithin{equation}{section}
\usepackage{titlesec,sectsty}
\subsubsectionfont{\normalfont\itshape}

\usepackage[backend=biber,style=numeric-comp,mcite=true,natbib=true,sorting=none,subentry=false,maxbibnames=24]{biblatex} 
\usepackage[hidelinks]{hyperref}

\newcommand{\cA}{\ensuremath{\mathcal{A}}}
\newcommand{\cB}{\mathcal{B}}
\newcommand{\cC}{\mathcal{C}}
\newcommand{\cF}{\mathcal{F}}

\newcommand{\cH}{\ensuremath{\mathcal{H}}}
\newcommand{\cI}{\ensuremath{\mathcal{I}}}
\newcommand{\cL}{\ensuremath{\mathcal{L}}}
\newcommand{\cM}{\ensuremath{\mathcal{M}}}
\newcommand{\cN}{\ensuremath{\mathcal{N}}}
\newcommand{\cO}{\ensuremath{\mathcal{O}}}
\newcommand{\cQ}{\ensuremath{\mathcal{Q}}}
\newcommand{\cR}{\ensuremath{\mathcal{R}}}
\newcommand{\cS}{\ensuremath{\mathcal{S}}}
\newcommand{\cT}{\ensuremath{\mathcal{T}}}
\newcommand{\cU}{\ensuremath{\mathcal{U}}}

\newcommand{\cW}{\ensuremath{\mathcal{W}}}
\newcommand{\cY}{\ensuremath{\mathcal{Y}}}
\newcommand{\cZ}{\ensuremath{\mathcal{Z}}}
\newcommand{\bA}{{\bf A}}
\newcommand{\bB}{{\bf B}}
\newcommand{\bC}{{\bf C}}
\newcommand{\bL}{{\bf L}}
\newcommand{\bM}{{\bf M}}

\newcommand{\bR}{{\bf R}}

\newcommand{\bQ}{{\bf Q}}
\newcommand{\Tr}{{{\operatorname {Tr} }}}

\newcommand{\ket}[1]{\lvert#1\rangle}
\newcommand{\bra}[1]{\langle#1\rvert}
\newcommand{\ip}[2]{\ensuremath{\langle#1\lvert#2\rangle}}
\newcommand{\nonu}{\nonumber \\}
\newcommand{\simbreak}{\nonu&\qquad{}}
\newcommand{\breakeqn}{\right.\nonu&\qquad{}\left.}
\newcommand{\breakeqntwice}{\right.\right.\nonu&\qquad{}\left.\left.}
\newcommand{\breakeqnthrice}{\right.\right.\right.\nonu&\qquad{}\left.\left.\left.}
\newcommand{\bxi}{\xi^\top\!}
\newcommand{\xigh}{\xi^{gh}\!}
\newcommand{\bxigh}{\xi^{gh\top}\!}
\newcommand{\egh}{\eta^{gh}\!}
\newcommand{\kt}{\tilde{\kappa}}
\newcommand{\ktgh}{\tilde{\kappa}^{gh}}
\newcommand{\bktgh}{\tilde{\kappa}^{gh\top}}
\newcommand{\mgh}{\ensuremath{m_0^{gh}}}
\newcommand{\Mgh}{\ensuremath{M_0^{gh}}}
\newcommand{\ninz}{\ensuremath{n\in\mathbb{Z}}}
\DeclareMathOperator{\arctanh}{arctanh}
\newcommand{\sech}{{{\operatorname {sech}}}}
\newcommand{\shkt}{\ensuremath{\sinh \, t \tilde{\kappa}}}

\newcommand{\sktgh}{\ensuremath{\sech \, t \tilde{\kappa}^{gh}}}
\newcommand{\ckt}{\ensuremath{\cosh \left(t \tilde{\kappa}\right)}}
\newcommand{\shktgh}{\ensuremath{\sinh t \tilde{\kappa}^{gh}}}
\newcommand{\tktgh}{\tanh t \tilde{\kappa}^{gh}}
\newcommand{\cktgh}{\cosh t \tilde{\kappa}^{gh}}
\newcommand{\ke}{\kappa_e}
\newcommand{\ko}{\kappa_o}

\newcommand{\ignore}[1]{} 
\newcommand{\p}{\partial}
\newcommand{\s}{\sigma}
\newcommand{\begh}{\eta^{gh\top}}
\newcommand{\lgh}{\lambda^{gh}}
\newcommand{\lghtp}{\lambda^{gh\top}}
\newcommand{\blgh}{\lambda^{gh\top}}
\newcommand{\aqua}{\frac{1}{4}}
\newcommand{\half}{\ensuremath{\frac{1}{2}}}
\newcommand{\mone}{\ensuremath{\mathbb{1}}}
\newcommand{\mzero}{\ensuremath{\mathbb{0}}}
\newcommand{\ortwo}{\ensuremath{\frac{1}{\sqrt{2}}}}
\newcommand{\sign}{\ensuremath{{\rm ~sgn}}}

\newcommand{\sX}{\ensuremath{\mathscr{X}}}

\newcommand{\bls}{\ensuremath{\Big[}}
\newcommand{\brs}{\ensuremath{\Big]}}
\newcommand{\blp}{\ensuremath{\big(}}
\newcommand{\brp}{\ensuremath{\big)}}
\newcommand{\ep}{\ensuremath{e^\prime}}
\newcommand{\op}{\ensuremath{o^\prime}}

\newcommand{\fg}{\ensuremath{\mathfrak{g}}}
\newcommand{\fh}{\ensuremath{\mathfrak{h}}}

\newcommand{\apri}{\alpha^{\prime}}

\def\name{Albin James}
\hypersetup{
  colorlinks = false,
  urlcolor = black,
  pdfauthor = {\name},
  pdfkeywords = {open string field theory, moyal star representation, boundary conformal field theory,
    perturbative off-shell amplitudes},
  pdftitle = {\name: Concerning the ghost contribution to the one-loop integrands in open string field theory},
  pdfsubject = {string field theory paper},
  pdfpagemode = UseNone
}
\begin{document}
	\title{Concerning the ghost contribution to the one-loop integrands in\\ open string field theory
	}
\author{Albin James \\
		{}\\
		{\small{\it University of Southern California,}}\\
		{\small{\it Los Angeles , CA 90089, USA}} \\
		{}\\
		{\small{\it Elite Prep, Arcadia, CA 91007, USA}}\\
		{}\\
		{\small{E-mail: {\ttfamily jamesalbin44 @ gmail.com}}}
        	}
\date{}
\maketitle\thispagestyle{empty}\addtocounter{page}{-1}

\begin{abstract}
\noindent
We examine the ghost contribution to the one-loop integrands in open string field theory using the Moyal representation of the star product. We primarily focus on the open string tadpole integrand, which is an intrinsically off-shell quantity. Due to the closed string tachyon, the full amplitude is badly divergent from the closed string degeneration region $t\to0^+$ of the Schwinger parameter. We obtain expansions for the finite factors from the squeezed state matrix  ${\bf R}(t)$ characterizing the ghost part of the tadpole in Siegel gauge.  The analytic structure of the integrands, as a function of the Schwinger parameter, captures the correct linear order behaviour near both the closed and open string degeneration limits. Using a geometric series for the matrix inverse, we obtain an approximation for the even parity matrix elements. We  employ an expansion based on results from the oscillator basis to construct Pad\'e approximants to further analyse  hints of non-analyticity near this limit. We also briefly discuss the evaluation of ghost integrands for the four string diagrams contributing to the one-loop $2$-point function in open string field theory.
\end{abstract}

\newpage
\begingroup
\hypersetup{hidelinks}
\tableofcontents
\endgroup


\section{Introduction and summary}\label{sec:intro and summary}
String field theories provide an off-shell formulation of string theory that is conceptually simple and very similar in structure to conventional gauge field theories. By construction \mcite{construction,*Gaberdiel:1997ia,*konopka2016construction,Thorn:1988hm,cftbackgrounds,*Taylor:2003gn,*Okawa:2012ica,*Ohmori:2001am,*Rastelli:2000iu,Witten:1985cc}, these furnish a field theory of strings with a spacetime action, and aspire to describe different regions of the parameter space of string theory using a universal set of degrees of freedom, encoded in the string field $\ket{\Phi}$. The best understood covariant string field theory is the bosonic open string field theory (OSFT) with Witten type \cite{Witten:1985cc} cubic vertices. Remarkably, starting from a few  axioms, this OSFT defines an interacting theory for an \textit{infinite} number of fields by virtue of the underlying worldsheet conformal symmetry---which is closely tied to its spacetime gauge invariance. 

In this paper, we revisit the perturbative structure of OSFT \mcite{BVmachinery,*Thorn:1986qj,*Gomis:1994he,samuel,*Bluhm:1989ws,*Samuel:1987uu,*Samuel:1989fea,*Freedman:1987fr,Ellwood:2003xc,Ellwood:2008jh,Konopka:2015tta} at the one-loop level, with only one or two external states. This requires evaluating the one-loop 1-point function (tadpole) and the one-loop 2-point function (string propagator). The latter receives contributions from four diagrams---three planar and one non-planar---due to the rigid nature of the Witten vertex. We have made analytical and numerical progress primarily on the ghost sector contribution to the tadpole integrand, which also appears as a subdiagram in two of the planar one-loop 2-point functions, and is an intrinsically off-shell quantity.

In \cite{Ellwood:2003xc}, Ellwood et al. carry out a careful study of the (open string) tadpole state using boundary conformal field theory (BCFT) and oscillator methods. In the Siegel gauge $b_0\ket{\Phi} = 0$ that we shall be working in, the integrand is expressible as a function of the Schwinger parameter $t$ (the length of the propagator loop in Fig. \ref{fig:tadpole}) and exhibits  essential singularities  at its limiting values, namely $0$ and $\infty$. Physically, these divergent pieces may be understood in terms of degenerating string diagrams from the boundary of moduli space \mcite{Thorn:1988hm,Bluhm:1989ws} and arise from the open string tachyon ($t \to \infty$), the closed string tachyon, and massless closed string states ($t \to 0$) propagating in the loop.

One method to study the tadpole diagram near $t = 0$ is to approximate it by using an appropriate \textit{boundary state} $\ket{\cB}$ in a BCFT analysis. This explicitly includes the closed string oscillators  $c_n, \tilde{c}_n, b_n, \tilde{b}_n$, $a_n$, $\tilde{a}_n$ and can be organized into levels. The chain of conformal maps employed reproduce the correct divergence structure; we refer the interested reader to \cite[\S3]{Ellwood:2003xc} where the leading divergence (for the D25 brane case) was carefully derived to be:
\begin{equation}\label{eq:closedtachyon}
|\mathcal{T}(t)\rangle\sim \frac{e^{+2\pi^2/t}}{t^6} \exp\left[-\half a^\dagger_n C_{nm}a^\dagger_m - c^\dagger_n C_{nm}b^\dagger_m\right]\hat{c}_0|\hat{\Omega}\rangle.
\end{equation}
Here, the ket state on the RHS is the Shapiro-Thorn closed string tachyon state that arose in the work of \cite{Freedman:1987fr} and later analyzed in detail in \cite{Ellwood:2008jh,Thorn:1988hm,Ellwood:2003xc} and $C$ is the twist matrix $(-)^n \, \delta_{nm}$. As discussed in \cite[App A]{Ellwood:2003xc} (see also \mcite{Thorn:1986qj,qocha,*Munster:2011ij,*Kajiura:2003ax,*vallette2012algebra+,*Muenster:2013ptn}), it also contributes to a BRST anomaly $\displaystyle Q_B |\cT\rangle \neq 0$, that is also present in the bosonic OSFTs based on the lower dimensional (unstable) Dp-branes.
For a generic value of the parameter $t$, however, the expressions could only be represented in terms of implicit line integrals (which may however be inverted numerically). See also the earlier treatment in \cite{Bluhm:1989ws} using off-shell conformal theory. Additionally, as described in \cite{Ellwood:2003xc} there is operator mixing induced by conformal transformations, since the boundary state is \emph{not} a conformal primary, and this leads to mixing of divergences from the massless and tachyon sectors.

In the oscillator construction of the 3-vertex, using squeezed state methods for inner products \cite{Kostelecky:2000hz}, the state was shown to be \cite[\S4]{Ellwood:2003xc}
\begin{equation}\label{eq:tadpoleket}
|\mathcal{T}\rangle = \int_0^{\infty}\, dt\, e^t \frac{\det(1-S\widetilde{X})}{(Q\det(1-S\widetilde{V}))^{13}} ~ \exp \left[-\frac{1}{2}a^\dagger \bM a^\dagger-c^\dagger \bR b^\dagger\right]\hat{c}_0 |\hat{\Omega}\rangle,
\end{equation}
where the constituent matrices are expressible in terms of Neumann matrices, as we shall describe later in \S\ref{subsec:osc results}. The determinant and the $e^t$ factors contribute to divergences in the $t \to 0$ and the $t\to \infty$ limits respectively. There are also additional subleading IR divergences from the massless fields. The somewhat complicated nature of the Neumann matrices makes analytic study of the matrix $\bR(t)$ difficult; it also suffers from an order of limits issue while considering expansions around $t=0$ (see \cite[App B]{Ellwood:2003xc} or \S\ref{subsec:osc results}) which leads to factor of $2$ difference for the leading term from BCFT.

The computations we perform are  based on the \textit{Moyal} $\ast$ (star) representation of the vertex \mcite{moyal,*Bars:2001ag,*Bars:2002yj,moyalrep,*Bars:2002nu,*Bars:2003sm,Bars:2003gu,Bars:2002qt,Erler:2003eq,spectroscopy,*Rastelli:2001hh,*Belov:2003df,*Douglas:2002jm} developed by Bars et al. Although Witten's formulation of OSFT is very elegant and only requires a cubic interaction, explicit calculations are made difficult by the somewhat complicated structure of the $3$-string vertex that encodes the gluing condition of strings. By choosing a convenient diagonal basis $\xi\coloneqq (x_{2n},p_{2n})$ for the degrees of freedom (matter + ghosts), the formalism redefines the interactions in terms of the simple ``Moyal product'' \cite{Bars:2001ag,bekaert2005universal} between string fields.
As suggested by Ellwood et al. in \cite{Ellwood:2003xc}, it would therefore be interesting to explore the analytic structure of the tadpole state in the Moyal/diagonal basis, where the interaction term simplifies.

Although the BCFT analysis in \cite{Ellwood:2003xc} reveals a lot of information about the structure of the state $\ket{\cT}$, it is also a useful exercice to understand it purely from the open string perspective as we do in the somewhat algebraic approach here. Another motivation for our work has been to test the validity of the Moyal representation at the one-loop level by
extending the off-shell tree level results \cite{Bars:2003sm,Bars:2003gu,Bars:2002qt} and the computation of Neumann coefficients \cite{Bars:2002nu,Bars:2003gu}.

Since the tachyonic divergences are artefacts of the bosonic theory, we shall limit  our attention in this paper to the finite factors from the squeezed state matrix  $\bR(t)$ characterizing  the Fock space state in the ghost sector. Its matrix elements may be extracted by taking inner products with pure ghost excited states:
\begin{equation}
\bra{\hat{\Omega}}\hat{c}_m \hat{b}_n\lvert \cT(t)\rangle = -\bR_{nm}(t)\times S_0(t),
\end{equation}
where $S_0(t)$ would be a scalar piece, dependent on the Dp brane system. We will be interested in  hints of non-analyticities in $\bR_{nm}(t)$, such as the exponentially suppressed sub-leading terms from \eqref{eq:subleading} that are expected from closed string physics.  Furthermore, in Siegel gauge, it is consistent to restrict to twist even and $SU(1,1)$ singlets \mcite{suoneone,*Zwiebach:2000vc,*Siegel:1985tw,*Hata:2000bj} for the test states, which translates to
\begin{equation}
\bR_{2n,2m-1} = 0,\qquad m\,\bR_{nm} = n \,\bR_{mn}.
\end{equation}

\medskip
The Feynman rules in non-commutative $\xi$ space (\S\ref{subsec:Feynman rules}) may be used for summing over a complete set of states $e^{ i\bxi \eta -\bxigh \egh}$. The evaluation in the Moyal basis then involves transforming certain coefficient matrices having substructure in terms of some simple matrices---which in turn obey a set of rather simple relations (See \cite{Bars:2003gu} or \S\ref{subsec:moyal structures}). This has produced alternate expressions for the integrands that we have used as the starting point for an independent analysis. The calculations are simplified due to a monoid subalgebra \cite{Bars:2003gu}, which is significantly easier to handle than the operator algebra used in the oscillator analysis.  It is noteworthy that since the matrix relations are satisfied even at finite level, we have a \emph{consistent truncation} for numerical checks, although we lack gauge invariance and are limited to machine precision due to the size of the constituent matrices and their substructure. Since gravity is an inconvenience, we shall concern ourselves with only the flat D25 brane background.

We show that the formal expressions involving matrix inverses correctly capture the linear order behaviour near both limits $t\to 0$ and $t\to \infty$ of $\bR(t)$. The qualitative difference near the closed string region between the Moyal and oscillator expressions is that, in case of the oscillators an intermediate matrix becomes singular but in the Moyal case the matrix becomes singular trivially due to the whole matrix vanishing. Interestingly enough, the peculiar nature of the Virasoro operator $L_0$ in the diagonal basis \eqref{eq:peculiar propagator} leads to a pole-zero cancellation and results in
\begin{equation}\label{eq:linear behaviour}
\bR_{nm}(t) = C_{nm} - n \, C_{nm} \, t + \cO(t^2),
\end{equation} 
whose linear term carries information about the conformal mappings used for the incoming external states in the BCFT prescription, and serves as a consistency check on the expansions that follow. Again, the monoid algebra renders the treatment of excited states more tractable and allows one to extend the existing (very detailed) results from the tachyon case \cite{Bluhm:1989ws,Samuel:1989fea}, at least  numerically. Somewhat surprisingly, associativity is also seen to hold to this order \S\ref{subsec:nonassoc}. 

Using a geometric series, we are able to expand the matrix $\bR(t)$ in terms of special functions owing to the simple nature of the constituent matrices. This leads to a discussion of vanishing but non-analytic contributions at $t=0$, such as $t^k \log(t)$.  By going to the continuous $\kappa$-basis, we also verify the correct linear behaviour in $q\coloneqq e^{-t}$ close to zero or $t\to \infty$, that matches with the oscillator construction. In order to probe for hints of non-analyticity in the complex $q$ plane, we use the oscillator expression \eqref{eq:Rnmosc expr} to obtain the coefficients of the general matrix element $\bR_{nm}(q)$ till $q^{18}$ using the \textit{NCAlgebra} \cite{NCAlgebra} package. We move onto construct the associated Pad\'{e} and  Borel-Pad\'{e} approximants and perform various consistency checks.

Quite a lot of work (see \cite{Ellwood:2008jh,Munster:2011ij} and references therein) has been done to understand the one-loop structure of the theory since the work in Refs. \cite{Ellwood:2003xc,Thorn:1986qj,Bluhm:1989ws}.
An analysis was also done using open-closed string field theory of Zwiebach \cite{Zwiebach:1997fe} in \cite{Ellwood:2003xc} where it was shown that it naturally incorporates the shift in the closed string background---just like in gauge field theories. In this regard, we must mention the somewhat recent work of Sachs et al. \cite{Munster:2011ij} where the quantum (in)consistency of OSFT has  been precisely characterized in the language of QOCHA (quantum open-closed homotopy algebras).

\medskip
We must also mention in passing another more recent gauge choice called the Schnabl gauge, where  tree amplitudes (and loop amplitudes to some extent) simplify immensely; both the kinetic term and the interaction terms become more tractable in this conformal frame. This gauge was originally chosen while constructing the non-perturbative tachyon vacuum solution of OSFT  in terms of surface states called wedge states \mcite{schnabl,*Schnabl:2002gg,*Schnabl:2005gv,*Erler:2014eqa}. Additionally, at the one-loop level new interesting geometrical structures arise \cite{Kiermaier:2008jy} which may help with computations in the more physical open superstring field theories where the tachyon would be projected out, but the gauge structure is much more intricate. 

We however continue to choose the Siegel gauge since the computational techniques are more readily available in this frame. Due to the severe divergences from the closed string tachyon (and the absence of winding states, etc. See the discussion by Okawa in \cite[\S5]{Okawa:2012ica}.), the Witten type OSFT is truly inconsistent at the quantum level when one starts considering loop diagrams. Hence, the quantization procedure is necessarily formal but one can hope that the divergences are just an artefact of the bosonic theory and we can still learn helpful lessons from this kind of exercices. We refer the interested reader to the seminal work of Thorn \cite{Thorn:1986qj,Thorn:1988hm} and the construction of quantum effective actions using the Batalin-Vilkovisky (BV) machinery \cite{BVmachinery,konopka2016construction} therein. Since we have been unable to connect our analysis to the one in terms of the effective action,  we do not discuss the role of these subtle boundary contributions to the one-loop tadpole graph, which is calculated ``from scratch'' in the work by Thorn. See also \cite[\S6.3]{Ellwood:2003xc} for a discussion of the issue of gauge invariance at the one-loop level and some comments on possible mismatch with the earlier analysis \cite{Thorn:1988hm} in the Siegel gauge.

The rest of the paper is organized as follows: In \S\ref{sec:perturbative osft}, we first review some essential aspects of perturbative OSFT and the Moyal representation. This is followed by a description of some known results from the oscillator analysis of the tadpole and a summary of our notations for quick reference. In \S\ref{sec:tadpole}, we apply the Feynman rules in Moyal space  to the tadpole and present algebraic expressions for the  integrand, with focus on the ghost sector. We shall be quite explicit throughout the discussions since we are also seeking to clarify a minor mismatch with the oscillator construction, and because most of the operations are elementary block matrix multiplications or Gaussian integrations. Next, we analyse the squeezed state matrix $\bR(t)$ in \S\ref{sec:squeezed} using  a geometric series expansion and present some illustrations of the procedure. We compare various approximation methods near limiting cases numerically. These two sections contain the main results of the paper. Several discussions and intermediate steps may be skipped altogether and the attention be restricted to the final form of the expressions. Due to the  rigid nature of the Witten vertex, we have four diagrams contributing to the 2-point function and their ghost sector is discussed briefly in \S\ref{sec:prop}. Finally, we close by making some  comments in relation to our results and directions for future work in \S\ref{sec:disc apology}.





\section{Algebraic structure of perturbative OSFT}\label{sec:perturbative osft}

In this section, we review some essential aspects of open string field theory that provides context for the subsequent discussions and 
also in order  to set the notations.  For the general structure of the theory, we follow closely the very excellent lectures by Zwiebach and Taylor \cite{Taylor:2003gn}. See also  \cite{Okawa:2012ica,Rastelli:2000iu,Ohmori:2001am,Konopka:2015tta} for  modern developments and \cite{Thorn:1988hm,Kostelecky:1986xg,Gomis:1994he} for classic treatments of the subject. We shall then review the Moyal representation of the $\ast$ product   \cite{Bars:2001ag,Bars:2002yj,Bars:2003gu,bekaert2005universal}  using which most of the calculations in this paper are done. Next, we recall some results \cite{Ellwood:2003xc} from the closely related oscillator formalism, where alternate expressions can be written down for the physical quantities we study, and which we seek to improve upon. We close this section by collecting together some oft used notations and slight modifications from prior conventions.

\subsection{Gauge choice and quantization }\label{subsec:Sea gull}
String field theories are spacetime formulations for interacting strings that are similar in spirit to the quantum field theories. Two  essential requirements demanded from such theories are that a)
the kinetic term should lead to the correct physical states , and
b) the interacting action must  reproduce the S-matrix elements of the Polyakov first quantized string theory by providing a single cover of the associated moduli space. A very useful toy model to study is the open string field theory for bosonic strings and where these statements have been rigorously proven \mcite{singlecover,*Zwiebach:1990az,*Giddings:1986wp}.

\subsubsection*{Basic ingredients of OSFT}\label{subsubsec:osft defns}
\vspace{-2mm}
Open string field theory is a second-quantized formulation of bosonic open string theory that has as its dynamical variable  the classical \textit{string field} $\Phi$, which may be represented as  an element of the state space of a matter-ghost boundary conformal field theory (BCFT):
\begin{equation}
|\Phi\rangle \in \mathcal{H}_{\text{BCFT}} = \mathcal{H}_{matter}\otimes \mathcal{H}_{ghost},
\end{equation}
and contains a component field for every state in the first quantized string Fock space. An elegant covariant formulation of this theory has been given by Witten with the following classical action:
\begin{equation}\label{eq:OSFT action}
S_{cl}[\Phi] =-\half \langle \Phi,  Q_{B} \Phi \rangle_{\rm bpz} - \frac{g_o}{3}\langle \Phi, \Phi\ast \Phi\rangle_{\rm bpz},
\end{equation}
which has the general structure of  a Chern-Simons theory. It employs the BRST quantization procedure which ensures that the underlying worldsheet theory is physically equivalent to the one in covariant quantization.  The string field may also be thought of as being valued in a graded algebra $\cA$ which is chosen as the space of string functionals of the embedding coordinates (matter) and the reparametrization ghost field arising from fixing the worldsheet metric to conformal gauge ($\gamma_{ab}\sim\delta_{ab}$), i.e.
\begin{equation}
\cA = \left\{\Phi[X^\mu(\sigma); c(\sigma)]\right\},
\end{equation}
where $\sigma \in [0,\pi]$ denotes the canonical worldsheet parameter of the open string. We shall be focussing on the ghost sector primarily and hence discuss it in more detail in  \S\ref{subsub:fermionic star} and App \ref{sec:bcbetaosc} below. Additionally, we shall take the underlying boundary conformal field theory (BCFT) to be that of the flat D25 brane theory, although OSFTs may be defined for any matter system with $c = 26$.

The basic ingredients\footnote{See \cite{Gaberdiel:1997ia} for  a precise treatment of these algebraic structures.  In recent formulations that have proven useful, this would define a differential graded algebra (DGA), which encodes the maps \cite{Munster:2011ij,Kajiura:2003ax}. The requirement of associativity may be relaxed to obtain a \textit{homotopy associative algebra} or a cyclic $A_\infty$ structure \cite{vallette2012algebra+,Konopka:2015tta}.} of the above action are the first-quantized BRST operator $Q_{B}$, the BPZ inner product $\langle \, .\, , .\, \rangle_{\rm bpz}$ (or the $\int$ operation), and an associative but non-commutative $\ast$ product between the string fields subject to the following ``Witten axioms'':
\begin{description}\label{eq:osft dga axioms}
	\item [Grading:] The string fields are subject to a $\mathbb{Z}$ grading for the ghost number, $G_\Phi$ and $\mathbb{Z}_2$ for Grassmannality.	The c ghost and the b anti-ghost  are assigned ghost number charges of $+1$ and $-1$ respectively, and are Grassmann odd.  The classical string field $\ket{\Phi} \in \cH_{\rm BCFT}$ at ghost number $+1$ and is also Grassmann odd.
	\item [Differential:] The BRST operator $Q_{B} = \oint \frac{dz}{2\pi i}\, j_{B}(z)$  defines a map
	$Q_{B} : \Lambda^n \mapsto \Lambda^{n+1},$
	i.e. it's a degree one operator under the grading. It is nilpotent:
	$Q_{B}^2 \equiv0$,
	and satisfies the derivation property:
	\[Q_{B}(\Phi_1\ast\Phi_2)= (Q_{B}\Phi_1)\ast\Phi_2+(-)^{G_{\Phi_1}}\Phi_1\ast(Q_{B}\Phi_2).\]
	\item [Associativity:] The binary $\ast$ product is assumed to satisfy:
	$(\Phi_1 \ast \Phi_2) \ast \Phi_3 = \Phi_1\ast(\Phi_2\ast\Phi_3)$.
	\item [BPZ inner product:] This is an invariant, bilinear form of ghost number $-3$ that is graded-symmetric. In terms of the $\int$ operation it induces a map $\int:\cA\to\mathbb{C}$ that respects the following relations: $\int Q_{B} \Phi =0$, $\int \Phi =0$ if $G_\Phi \neq +3$,
	and cyclicity:\[	\int \Phi_1\ast\Phi_2 =(-)^{G_{\Phi_1}G_{\Phi_2}}\int \Phi_2\ast\Phi_1.
	\]
\end{description}
These axioms uniquely determine the action by the requirement of extending the gauge symmetry from the free theory to the interacting case.

This field theory reproduces a single covering of the moduli space of Riemann surfaces generated by the underlying matter-ghost boundary conformal field theory (BCFT). Hence, all on-shell scattering amplitudes are guaranteed to be generated through a Feynman diagrammatic expansion.  It also encodes rich non-perturbative string physics even at the classical level, as has been shown in the study of tachyon condensation \cite{Schnabl:2005gv,Ohmori:2001am} and the computation of gauge invariant observables, called Ellwood invariants \cite{Ellwood:2008jh}, for example. 

\medskip
\noindent
Next, let us turn towards the $\ast$ product which is one of the central aspects of Witten's OSFT.
\vspace{-4mm}
\subsubsection*{The $\ast$ product operation}
\vspace{-2mm}
The interaction between open strings is implemented by using the $\ast$ product which endows the state space $\cH_{\text{BCFT}}$ with the structure of a non-commutative algebra \cite{Witten:1985cc}. For the matter functionals, this can be imagined as by imposing delta function overlap between the two halves of each string: the right half of the first string matches with the left half of the second string, which requires the following connection conditions: 
\begin{equation}
X^{(r)}(\sigma) -X^{(r-1)}(\pi-\sigma) = 0, \qquad P^{(r)}(\sigma)+P^{(r-1)}(\pi -\sigma) = 0,
\end{equation}
for the matter sector and in the ghost sector:
\begin{align}\label{eq:ghost overlap}
c^{\pm(r)}(\sigma) + c^{\pm (r-1)}(\pi-\sigma) = 0, \qquad b^{\pm(r)}(\sigma) - b^{\pm (r-1)}(\pi-\sigma) = 0,
\end{align}
where now the parameter $\sigma$ is restricted to $0 \leq \sigma \leq  \pi/2$ and $r = 1, 2, 3$. See \cite{Kishimoto:2001ac} (and references therein) for a careful treatment of the ghost sector and of the general $N$-string vertex case.

It is worth mentioning that in concrete calculations, the delta function overlap above is implemented by evaluating correlation functions of the BCFT on canonical domains such as the upper half plane (UHP) where the Neumann functions may be constructed explicitly. In particular, the three half-discs corresponding to the three open string worldsheets can be glued together consistently using conformal maps discussed in \cite{Rastelli:2000iu,Okawa:2012ica,Ohmori:2001am} to obtain the $3$-vertex.

A wealth of information has been gained about the structure of the theory using powerful Riemann surface theory employing
elegant conformal mapping techniques. To appreciate how non-trivial the construction of the interacting SFTs is, even for the bosonic open string is, it is necessary and instructive to understand the geometry of the conformal frame dictated by the underlying worldsheet theory. However in this work which focusses on the algebraic approach, it suffices to remark that
since the conformal frame has a somewhat complicated geometry, it introduces  non-trivial conformal factors and branch-cut structure in both the matter and the ghost sectors. This makes explicit study of the string diagrams highly non-trivial in general, especially for loop amplitudes requiring constructions involving higher genus Riemann surfaces \cite{Samuel:1989fea,Bluhm:1989ws}.

\subsubsection*{Siegel gauge}
\vspace{-2mm}

From the resemblance of Witten type OSFT to the Chern-Simons action and p-forms, one can infer that the classical action in \eqref{eq:OSFT action} is invariant under the following gauge transformation, once the Witten axioms are satisfied:
\begin{equation}
\delta_\Lambda \Phi = Q_{B}\Lambda + \Phi \ast \Lambda -\Lambda \ast \Phi,
\end{equation}
where $\Lambda$ is a ghost number zero, Grassmann even string field. Conversely, the cubic action is the unique action allowed by extending the linear gauge symmetry $(\delta_\Lambda \Phi  = Q_{B}\Lambda$) to the non-linear level.

Because of this huge gauge symmetry, we must first fix a gauge before deriving the Feynman rules of this theory. A venerable gauge choice is the \emph{Siegel gauge} where the kinetic term $\langle \Phi, Q_{B}\Phi\rangle$ simplifies drastically. This is obtained by dictating that\footnote{This can  be accomplished by a gauge transformation, at least at the linear level \cite{Ohmori:2001am,Thorn:1988hm}.} the string field satisfies:
\begin{equation}
b_0 \ket{\Phi} = 0,
\end{equation}
where $b_0$ is the anti-ghost zero mode. Then we can rewrite $\Phi$ as $\Phi = b_0 c_0\Phi$ by virtue of the anti-commutation relation $\{b_0,c_0\} =1$. Now, the kinetic term can be rewritten in terms of the total matter + ghost Virasoro zero mode:
\begin{equation}
\hat{L}_0 = \hat{L}_0^X + \hat{L}_0^{gh}
\end{equation}
by making use of the relation $\{Q_{B},b_0\} = L_0$ as
\begin{align}
S_{\rm kin}=\bra{\Phi}\hat{c}_0 (\hat{L}_0-1)\ket{\Phi}
\end{align}
where we  revert to the first quantized operator language for convenience. Now, one may express the propagator in terms of a Schwinger parameter as:
\begin{equation}\label{eq:Schwinger rep}
\apri b_0 (L_0)^{-1}= \apri b_0 \int_0^\infty
\, dt \,e^{-tL_0},
\end{equation}
where we assume that the integral exists. We can interpret the action of the operator $e^{-tL_0}$ as to create a rectangular worldsheet strip of length $t$ and width $\pi$, the canonical range for $\sigma$. The cubic term representing the $\ast$ product now results in a   Riemann surface or \textit{string configuration} constructed out of three such rectangular strips, which is flat everywhere, except for a curvature singularity at the common joining point. The external states in a given interaction can now be placed as vertex operators on the appropriate semi-infinite strips to evaluate the correlators \mcite{lppzr,*LeClair:1988sp,*LeClair:1988sj,*Giddings:1986bp,*Kostelecky:1986xg}.


\subsection{Moyal representation of the star product}\label{subsec:moyal structures}
The operator formalism \mcite{operators,*Gross:1986fk,*Gross:1986ia,*Kishimoto:2001ac} in terms of explicit matter-ghost oscillators, $\hat{\alpha}^\mu_n, \hat{b}_n, \hat{c}_n$ for a given BCFT, provides another concrete realization of the Witten type overlap relations (in addition to the one based on worldsheet path integrals above). The correlation functions on the canonical domains are now expressed in terms of the nine Neumann matrices, which are infinite matrices derived from the \textit{Neumann functions} for the corresponding domain. These come with state space and mode number labels. Since these are quite challenging to handle analytically, the interactions were difficult to analyze in this language for hand-calculations.

In \cite{Bars:2001ag} a basis for the open string degrees of freedom was studied by Bars which diagonalizes the interaction vertex, and makes the connection to non-commutative geometry as originally proposed by Witten rather manifest. The $\ast$ product was implemented as the Moyal product in the phase space of even string modes. This could also explain the spectroscopy of the Neumann matrices studied in \cite{Rastelli:2001hh}. These algebraic transformations correspond to diagonalizing the reparametrization operator $K_1$ (see \S\ref{subsubsec:kappa-basis}) which fixes the special mid-point $\sigma=\pi/2$ or $z = +i$ (in the canonical half-disc coordinates) as may be expected from the geometric picture. This leads to a reduction in the effective number of Neumann matrices.

\vspace{-2mm}
\subsubsection{Weyl ordered polynomials and the Moyal product}\label{subsubsec:Xavier Weyl} 
\vspace{-2mm}
We start with the Heisenberg algebra $\fh_N$ generated by the $N$ pairs of phase space operators $X^i, P_j$ and a central element $C \eqqcolon i \theta$, satisfying the canonical commutation relations:
\begin{equation}
[X^i, P_j] = C \,\delta^i{}_{j},\qquad [C, X^i] = 0 = [C, P_j],
\end{equation}
and is hence a $2N + 1$ dimensional Lie algebra. It is also an associative algebra as may be seen from the Jacobi identity. A very useful construction out of this is its universal enveloping algebra\footnote{We closely follow \cite[\S2]{bekaert2005universal} in this discussion to motivate the Moyal product.}, which is the Weyl algebra $A_N \cong \cU(\fh_N)$. Its elements are the formal polynomials in $X^i$ and $P_j$ modulo the canonical commutation relations.

Let us denote the generators of $\fh_N$ by $T_i$. Then, a natural basis for $A_N$ is the collection of all distinct \textit{Weyl-ordered} formal homogeneous  polynomials
\begin{equation}
T_{i_1}\ldots T_{i_k} + \mbox{~permutations},
\end{equation}
which makes it isomorphic to the symmetric algebra $\odot(\fh_N)$. This naturally leads one to consider an association with variable $t_i$ (that would become the Moyal coordinates $x_{2n}^\mu, p_{2n}^\mu$ later) and an identification with the polynomial algebra $\mathbb{K}[t_i]$ with elements
\begin{equation}
P(\vec{t}) = \sum_{k = 0} \Pi^{i_1\ldots i_k} t_{i_1}\ldots t_{i_k}
\end{equation}
with symmetric coefficients $\Pi^{i_1\ldots i_k}$ valued in the field $\mathbb{K}$ (which will be taken as $\mathbb{C}$ for OSFT).

The non-commutativity of $\cU(\fh_N)$ means that the product of two Weyl ordered polynomials would require further reordering. This induces a deformation of the usual commutative product in $\mathbb{K}[t_i]$ and results in a $\ast$ algebra. The Lie bracket in $\fh_N$ (or the algebra $\fg$ in general) uniquely fixes this product and using the BCH formula, an explicit representation in terms of a \textit{bidifferential} operator may be obtained (see \cite[\S2.3]{bekaert2005universal}). This would then be the Moyal product for the $\ast$ algebra which for the ghost sector of OSFT is given by \eqref{eq:ghost star} or \eqref{eq:ghoststar} for brevity. 

The generalization to OSFT requires an infinite number of modes (to realize the Virasoro algebra that guarantees its consistency) and hence we are essentially considering $\cU(\fh_\infty)$. For physically interesting string configurations, one also needs to enlarge from the space of polynomials to exponential functions (\S\ref{subsubsec:subalgebra}). Hence, convergence properties with these differential operators become more challenging in these limits. See \cite{omori2001singular} for some relevant treatment.

\subsubsection{The discrete Moyal basis in OSFT}\label{subsubsec:discrete moyal}
\vspace{-2mm}
We discuss the discrete Moyal formalism, extensively developed in \cite{Bars:2002bt,Bars:2002nu,Bars:2003gu,Erler:2003eq} by Bars et al. first. Consider the open string field as a  functional of the $X^\mu(\sigma), c(\sigma)$ degrees of freedom. This may be made explicit by going to the oscillator representation in terms of the constituent Fourier modes $x_e, x_o, c_e, c_o$ and the zero modes: $x_0 = \frac{1}{\pi} \int d\sigma \, X^\mu(\sigma)$ for matter and  $c_0$ for the $c$-ghost. In the Siegel gauge that we choose for perturbation theory, we can consistently drop the pieces proportional to $c_0$. Now, the discrete Moyal map is obtained by first taking  a half-Fourier transform with respect to ``half'' of the degrees of freedom to convert the string field $\Phi(x,c)$ defined in coordinate space to Moyal space $A(\bar{x},\xi,\xigh)$. Then the Moyal star product is applied on the string fields valued in the phase space  doublets $\xi = (x_e, p_e)$ of ``even'' string modes. The maps between the even (e) and odd (o) modded subspaces are implemented by the matrices:
\begin{equation}
T: \cH_o \mapsto \cH_e,\qquad R: \cH_e \mapsto \cH_o
\end{equation}
where the modes (for Neumann boundary conditions) are obtained from:
\begin{equation}
X(\sigma) = x_0 +\sqrt{2}\sum_{\ninz_+}x_n \cos n\sigma, \qquad P(\sigma)=\frac{1}{\pi} \left(p + \sqrt{2}\sum_{\ninz_+}p_n \cos n\sigma\right)
\end{equation}
The $\ast$ product then becomes \textit{diagonal} after this change of variables and additionally, the product is local in the midpoint coordinate $\bar{x}$ (and the $\xi_0$ variable corresponding to the $b_0$ dependence). We will later discuss some of the infinite matrices related to $T$ that arise naturally in this transformation to phase space variables. These matrices will be crucial for the evaluation of string diagrams attempted in this paper.

Although the Moyal map employs infinite linear combinations in string mode space and hence is defined formally, it captures several aspects of the physics OSFT including subtle contributions from the midpoint \cite{Bars:2002bt,Erler:2003eq}. It provides a concrete realization of the split-string picture \cite{Gross:2001rk} while giving one prescription for treating the midpoint anomalies by providing a consistent truncation \cite{Bars:2002bt,Bars:2002nu,Bars:2003gu}. For the reduced star product \cite{Kishimoto:2001ac} in Siegel gauge, the ghost Witten vertex is equivalent to the discrete Moyal basis star representation.  It is one of the aims of this paper to test the applicability of this basis at the one-loop level.

The $b$ anti-ghost, which is analogous to the embedding coordinate, satisfies overlapping conditions and the $c$ ghost, which is similar to the momenta, satisfies \emph{anti-overlapping} conditions as expressed earlier in \eqref{eq:ghost overlap}. Hence, we can expect some slight asymmetry between the two sets (see \eqref{eq:xpyq modes}) of odd Moyal coordinates $(x_o, p_o)$ and $(y_o, q_o)$  we  provide in Appendix \ref{sec:bcbetaosc} on the $bc$ system. Since we are mostly interested in the ghost contributions in this paper, we have only illustrated the general idea in the matter sector (which was developed first historically, see \cite{Bars:2002nu}) before focussing on the treatment of the ghosts. Some relevant matter contribution would be presented  in Appendix \ref{sec:determinant}. The continuous $\kappa$ basis would be briefly reviewed in  \S\ref{subsubsec:kappa-basis}. Another basis which uses integral kernels was developed in \cite{Bars:2014vca}; see also the discussion in \cite[\S2]{Erler:2003eq} and \cite{Bars:2002yj} concerning this basis.

\vspace{-2mm}
\subsubsection{The fermionic Moyal product}\label{subsub:fermionic star}
\vspace{-2mm}
To go from the string field $\ket{\Phi}$ defined in Fock space of $(b, c)$ ghosts to Moyal space, one performs a Fourier transform over half the number of degrees of freedom:
\begin{align}\label{eq:Fourier}
&\hat{A}(\xi_0, x_o, y_o, p_o, q_o) \nonu&= \int \, d\bar{c} \, e^{-\xi_0 \bar{c}}A(\bar{c}, x_o, -p_o/\theta^\prime, y_o, -q_o/\theta^\prime)\nonu
&= 2^{-2N} (1+w^\top w)^{-\aqua} \int \, d c_0 \prod_{e>0}^{2N}\left(-i dx_e dy_e\right) e^{-\xi_0 c_0 +\xi_0 w^\top y_e +\frac{2}{\theta^\prime}p_o S^\top x_e +\frac{2}{\theta^\prime}q_o R y_e}\Phi(c_0,x_n,y_n),
\end{align}
where $\xi_0$ is a fermionic object encoding the zero mode dependence, and $\theta^\prime$ is the common non-commutativity parameter in ghost space. This operation 
may also be implemented by taking an inner product with a bra $\bra{\xi,\xigh,\xi_0}$ defining the Moyal basis. The matrix $S$ will be defined below. Restricting to the ghost sector in Siegel gauge, we roughly identify:
\begin{align}
A(\xi^{gh}) &\coloneqq \langle \xi^{gh} | \Phi\rangle\nonu
&\sim\int d x^b_o \, \Phi[c,b],
\end{align}
by isolating the zero mode dependence as $\hat{A}(\xi_0,\xigh) = \xi_0 A(\xigh)$.

We find that it is more convenient to work with objects having \emph{even} labels instead of the  \emph{odd} parity elements that appear naturally in the ghost sector. We emphasize that these are \emph{not} the original even degrees of freedom but special (infinite) linear combinations:
\begin{equation}
\boxed{x^c_e = T_{eo}y_o, \qquad p^c_e = R^\top_{eo}q_o, \qquad x^b_e = \ke^{-1}S_{eo}x_o, \mbox{~and}\qquad p^b_e = \ke S_{eo}p_o},
\end{equation}
where
\begin{subequations}
	\begin{align}\label{eq:T,R,S matrices}
	T_{eo} &= \frac{4}{\pi}\int_{0}^{\frac{\pi}{2}} d\sigma \, \cos e\sigma \cos o\sigma =  \frac{4 o  \, i^{o-e+1}}{\pi(e^2-o^2)}, \mbox{~and its inverse}\\
	R_{oe} &= \frac{4}{\pi}\int_0^{\frac{\pi}{2}}d\sigma \, \cos o\sigma \left(\cos e \sigma-\cos \frac{e\pi}{2}\right) = \frac{4e^2 \,i^{o-e+1}}{\pi o (e^2-o^2)}, \mbox{~and}\\
	S_{eo} &= \frac{4}{\pi}\int_{0}^{\frac{\pi}{2}}d\sigma \, \sin e\sigma \sin o\sigma = \frac{4 i^{o-e+1}e}{\pi(e^2-o^2)},
	\end{align}
\end{subequations}
with mixed parity labels, in the open string limit $N\to \infty$. These  matrices satisfy the relations:
\begin{align}
T\, R = \mone_e, \qquad R\, T = \mone_o, \qquad S\, S^\top = \mone_e,\qquad S^\top \, S = \mone_o,
\end{align}	
along with a few more useful relations that we collect below in \S\ref{subsec:notations}. See \cite[\S2.1.3]{Bars:2003gu} for the finite versions and a careful presentation of many more relations. Here $(\,)^\top$ refers to the matrix transpose which differs from the $(~\bar{}~)$ notation used in \cite{Bars:2003gu}.
The infinite vectors $w, v$ are given by:
\begin{equation}
w_e = \sqrt{2}i^{-e+2}, \qquad v_o = \frac{2\sqrt{2}i^{o-1}}{\pi o}=\ortwo T_{0o}.
\end{equation}

\vspace{2mm}
\noindent
After this preparation, the $\ast$ product among string fields valued in Moyal space is implemented by the bidifferential operator
\begin{equation}\label{eq:ghost star}
(A\ast B)(x^b_e, p^b_e, x^c_e, p^c_e) = A\exp\left[\frac{\theta^\prime}{2}\left(\frac{\overleftarrow{\partial}}{\partial x^b_e}\frac{\overrightarrow{\partial}}{\partial p^b_e}+\frac{\overleftarrow{\partial}}{\partial x^c_e}\frac{\overrightarrow{\partial}}{\partial p^c_e}+\frac{\overleftarrow{\partial}}{\partial p^b_e}\frac{\overrightarrow{\partial}}{\partial x^b_e}+\frac{\overleftarrow{\partial}}{\partial p^c_e}\frac{\overrightarrow{\partial}}{\partial x^c_e}\right)\right] B
\end{equation}
where $\overleftarrow{\partial}$ and $\overrightarrow{\partial}$ are respectively the left right and left fermionic derivatives obeying the standard anti-commutation rules, and $\theta^\prime$ is the non-commutativity parameters for ghosts.

As mentioned earlier, the ghost part of the string fields may be more succinctly obtained by taking the inner product with the oscillator bra $\bra{\xigh}$ defining the Moyal basis:
\begin{equation}\label{eq:xigh bra}
\langle\xi_0, \xi^{gh}\rvert = -2^{-2N}\left(1+w^\top w\right)^{-\aqua}\bra{\Omega}\hat{c}_{-1}e^{-\xi_0 (\hat{c}_0-\sqrt{2}w^\top \hat{c}_e)}e^{-\bxigh \Mgh \xigh -\bxigh \lgh},
\end{equation}
and  we have the vectors
\begin{equation}
\lgh_1 =  \left(\begin{array}{c}\sqrt{2}R^\top \hat{b}_o
\\ -2\sqrt{2}\ke^{-1}\hat{b}_e+2\ke^{-1}w \xi_0
\end{array}  \right), \qquad \lgh_2 =  \left(\begin{array}{c}\sqrt{2}R^\top \ko \hat{c}_o
\\ 2\sqrt{2}i\hat{c}_e
\end{array}  \right).
\end{equation}
Here we have transformed to the \emph{even} basis the expression given in \cite{Bars:2003gu} by utilizing some simple algebraic relations satisfied by the relevant matrices. Next, we define the off-block diagonal matrix $\sigma$, labelled by even mode integers:
\begin{equation}
\sigma \coloneqq -\theta \, \sigma_2 \otimes \mone_e,
\end{equation}
where $\sigma_2$ is the second Pauli matrix and we have chosen the non-commutativity parameter $\theta^\prime = \theta = +1$ for convenience, by a choice of units. Excluding the matter sector, we can now write the Moyal star product between two fields as:
\begin{equation}\label{eq:ghoststar}
\boxed
{\blp A\ast B\brp[\xigh] =  A\exp\left(\half \frac{\overleftarrow{\partial}}{\partial \xigh} \Sigma \frac{\overrightarrow{\partial}}{\partial \xigh}\right)B;\qquad \mbox{where now~~} \Sigma \coloneqq -i \varepsilon \otimes \sigma.
}
\end{equation}
The trace operation\footnote{Although we do not use the zero mode dependence, it is instructive to mention the structure here in the normalization using the odd modes
	\begin{equation}
	\int d\xi_0\, \Tr \left(\hat{A}(\xi_0,\xi)^\dagger \ast \left(\frac{\partial}{\partial\xi_0}-\frac{\theta^\prime}{2}v^\top\frac{\partial}{\partial q_0}\right)\hat{A}(\xi_0,\xi)\right) = \bra{\Phi}\hat{c_0}\ket{\Phi} = 1.
	\end{equation}}
associated with the Fock space inner product 
\begin{equation}
_1\bra{\Phi_1}\otimes \cdots \otimes \, _n\bra{\Phi_n} V_n\rangle\sim \Tr \left(\hat{A}_1\ast \cdots \ast \hat{A}_n\right)
\end{equation}
is then represented as integration over Moyal (phase) space with the appropriate measure:
\begin{equation}
\Tr \coloneqq \frac{\det \sigma^\prime}{\lvert\det(2\pi\sigma)\rvert^{d/2}} \int \, (d\xi) \, (d\xigh)
\end{equation}
where we have restored $\theta^\prime$ for generality, by defining $\sigma^\prime \coloneqq \theta^\prime \sigma_1\otimes\mone_e$.
\subsubsection*{\bf Metric in ghost space}\label{subsubsec:ghostmetric}
\vspace{-2mm}
We can  now combine the ghost (non-zero) modes into the two doublet vectors:
\begin{equation}\label{eq:ghost doublets}
\xi^1 = \left(\begin{array}{c}
x^b_e\\ -p^c_e
\end{array} \right) \mbox{~and}\qquad \xi^2 = \left(\begin{array}{c}
x^c_e\\ p^b_e
\end{array} \right) 
\end{equation}
which we denote  together again by $\xigh$. Under an $SO(4)$ rotation to the new basis, 
\[
\label{new basis}
\xigh = \left[\begin{array}{c}
x^b_e\\
p^b_e 
\\ x^c_e
\\ p^c_e
\end{array}  \right] \to \qquad \xi^1 = \left[\begin{array}{c}x^b_e
\\	-p^c_e
\end{array} \right], \qquad\xi^2=\left[\begin{array}{c}x^c_e
\\+p^b_e 

\end{array} \right]\]
the block matrices transform as:
\begin{align}
\varepsilon \otimes \alpha &\to -i \varepsilon \otimes i \alpha, &\varepsilon \otimes \beta &\to I_2 \otimes -\sigma_3 \beta, 
\nonu
I_2 \otimes \alpha &\to I_2 \otimes \alpha, &I_2 \otimes \beta &\to -i \varepsilon \otimes i \sigma_3 \beta,
\end{align}
where $\alpha$ is block diagonal and $\beta$ is off block diagonal. 

This allows one to use the $Sp(2)$ metric $+i\varepsilon_{ab}$ (with $\varepsilon_{12}=-1=-\varepsilon^{12}$) in the $(b,c)$ ghost phase space suggested in \cite{Bars:2003sm} and makes the $SU(1,1)$ symmetry manifest. The presentation also becomes cleaner due to the similarity of the algebraic expressions with the matter sector. Notice that we have the canonical $\ast$ anti-commutator in the ghost Moyal plane:
\begin{equation}
\{\xi^n_i,\xi^m_j\}_\ast = -i \varepsilon^{nm}\sigma_{ij}
\end{equation}
Hence, it is consistent to impose an $-i \varepsilon \otimes$ tensor product factor while defining dot products. \textit{In all fermionic bilinears and quadratic terms, this metric factor would be understood to be present. }

\vspace{-2mm}
\subsubsection{Monoid subalgebra and regularization}\label{subsubsec:subalgebra}
\vspace{-2mm}
A very interesting feature of the discrete Moyal basis is the consistent regularization developed in \cite{Bars:2002bt} involving  a cutoff prescription in the number of string modes $2N$ defining the phase space doublet. It allows  for a \textit{deformation of the spectrum} from the frequencies valued in the non-negative integers, to  a sufficiently reasonable set\footnote{It is somewhat interesting to compare this to the spectrum of the so called fractal strings \cite{lapidus2008search}.} of frequencies $\kappa_n$. The finite versions of the $N\times N$ matrices $T, R, S$ and $N\times 1$ vectors $w_e, v_o$ from \eqref{eq:T,R,S matrices} are in general dependent on all the frequencies $\kappa_n$, $n = 1,\ldots, 2N$.  These are solved for by requiring that the following relations are satisfied:
\begin{equation}
R = \ko^{-2}T^\top \ke^2,\qquad R = T^\top+v w^\top ,\qquad v = T^\top w, \qquad w = R^\top v
\end{equation}
along with a few more general relations for the ghost sector \cite{Bars:2003gu}. The explicit finite forms \eqref{eq:finiteNmatrices} will be given in App \ref{sec:determinant} when used for numerical checks towards the overall robustness of the regularization. 

This deformation results in a \textit{preservation of associativity} while taking double sums. It essentially removes the null elements of the algebra by hand and hence is topologically different from the string field algebra, even in the open string limit. It would therefore be interesting to study this structure on its own and because it correctly captures certain aspects of perturbative OSFT as shown in \cite{Bars:2003sm} and as we shall see in the following.

The regularization also leads to a (Moyal) star subalgebra constructed out of \emph{finite} number of modes\footnote{In the full open string field theory, all star subalgebras necessarily contain an infinite number of modes for consistency with the Witten axioms. Here we are only considering the deformed theory. We can relegate the subtleties of the closure of sub-algebras in string field theory by working at a finite value of $N$, which is somewhat similar to level truncation, and hence amounts to imposing a UV cut-off. Although this regularization cannot realize the Virasoro algebra and breaks the gauge invariance, it does preserve the \emph{non-linear Gross-Jevicki} matrix identities \cite{Gross:1986ia} (see also \eqref{eq:GrossJevicki} below) satisfied by the infinite Neumann matrices. This is because the fundamental matrices continue to satisfy the same relations as their infinite $N$ counter-parts (whenever they are regular) even after the deformation.}. The elements of this subalgebra are string configurations corresponding to  quadratic exponentials---which are the analogues of the Hermite polynomials in the functional formalism---but now defined in Moyal space:
\begin{equation}
A_{\cN, M,\lambda}(\xi)\coloneqq \cN e^{-\xi^\top M\xi -\xi^\top \lambda}.
\end{equation}
Here, the string field is parametrized by complex (anti-)symmetric matrices $M$, a complex vector $\lambda$ and the normalization factor $\cN$, which is independent of the $\xi$. These form a monoid or a semi-group structure i.e it is closed under the $\ast$ product, is associative, and has the unit element\footnote{corresponding to the identity state $\ket{\tilde{I}}$ under the \textit{reduced} star product studied in \cite{Kishimoto:2001ac}.} (the number $1$). It is in general non-commutative and may not have an inverse element, although the generic elements do have inverses. Thus, being just short of forming a group due to the lack of an inverse, it is a monoid or a semi-group containing many interesting string fields.

In particular, the perturbative vacuum state for the ghost sector (in the Siegel gauge) belongs to this class and is given by the monoid element:
\begin{equation}\label{eq:pert vacuum}
\hat{A}_0^{gh}(\xi^{gh}) = \xi_0\,  \cN_0^{gh}\exp\left[-\bxigh M_0^{gh}\xigh\right],
\end{equation}
where the matrix $\Mgh$ has the block diagonal form in the purely even basis:
\begin{equation}\label{eq:vacuum monoid}
M_0^{gh} = -\left[\begin{array}{cc}
\half R^\top \ko R&  0\\ 0
& 2\ke^{-1}
\end{array} \right],\qquad \lgh = 0,
\end{equation}
and $\cN_0^{gh}$ is a normalization factor.
\medskip

The subalgebra is a helpful structure for evaluating string diagrams, to which we turn next. The rules would be provided later in \eqref{eq:monoid algebra} while illustrating the tadpole computation in \S\ref{subsec:ghost sector}.

\subsubsection{Procedure for evaluating string diagrams}\label{subsec:Feynman rules}
In order to study string diagrams using this formalism, we require the gauge fixed action written in Moyal space:
\begin{equation}
S_{\rm GF} = - \int d^d \bar{x}\,\,\Tr\left(
\frac{1}{2\alpha^\prime}A(\bar{x},\xi)\ast(L_0-1)A(\bar{x},\xi)+\frac{g_o}{3}A(\bar{x},\xi)\ast A(\bar{x},\xi)\ast A(\bar{x},\xi)\right),
\end{equation}
where $A(\bar{x},\xi)$ contains only the non-zero ghost modes and the full string field has the explicit zero-mode dependence $\hat{A}(\bar{x},\xi_0,\xi) =\xi_0 A(\bar{x},\xi)$. We remark that this form of the action is also applicable for the finite $N$ truncations.

The monoid subalgebra \cite{Bars:2002nu} \eqref{eq:monoid algebra} and the propagator rules (given later in \eqref{eq:propagated monoid}) allow for  one straightforward way of writing down the integrands for Feynman graphs in the non-commutative $\xi$ space. Due to the interplay between kinetic term and the interaction term in OSFT, the propagator becomes complicated in $\xi$ space and involves a potential term. The external states $A_i(\xi)$ that correspond to the operator insertions on  the semi-infinite strips are joined together using the $\ast$ product.

The intermediate string fields are propagated using the operators $q_i^{L_0}$ (see \eqref{eq:Schwinger rep}) and the final trace operation (Gaussian integration over $\xi,\xigh$)  implements  the inner product. Here the variables $q_i = e^{-t_i}$ encode the modular parameters of the intermediate strips $t_i$. See \cite{Bars:2003gu,Bars:2002qt} for more details and examples. 

\medskip
In case of diagrams with loops, one also needs to perform a state sum; if we consider the contribution only from the ordinary ghosts, we can insert a (normalized) Fourier basis $e^{+i\bxi \eta+ip\bar{x}}\, e^{-\bxigh \egh}$\footnote{These have to be understood in the form of a distribution due to the singular normalization involved and the vanishing quadratic term in the exponents.}. A $(d \eta)$ integration at the end then implements the state sum. For example, the tadpole diagram that we will be focussing on in \S\ref{subsec:ghost sector} can be obtained by joining two legs of the off-shell 3 vertex to form a loop and inserting a complete set of states.

The string diagrams are thus evaluated at fixed modular parameters $t_i$.  We note that for the purpose of numerical calculations, it's also useful to consider the Feynman rules in the Fourier basis given in \cite{Bars:2003gu,Bars:2002qt}.

\subsection{Results from the oscillator basis}\label{subsec:osc results}

In the oscillator construction \cite{Gross:1986ia}, the Fock space of open (bosonic) string fields is constructed by acting with the creation operators $\alpha^\mu_{-k}, b_{-n}, c_{-m}$ on the vacuum $\ket{\hat{\Omega}}$. The star product is then implemented by using $n$-vertices belonging to the tensor product of the dual spaces $\cH^{(i)\ast}$. In particular, we have the three-vertex $\bra{V_3}$ and the two-vertex $\bra{V_2}$ whose explicit structure encodes the Witten-style overlapping conditions (see \cite{Kishimoto:2001ac,Taylor:2003gn} and references therein). The $3$-string vertex fixes all the interactions that may arise in the theory. For the purpose of this paper, we provide only the relevant ghost part \mcite{vertex,*Samuel:1986wp,*Cremmer:1986if,Gross:1986fk} appearing in the combined vertex:
\begin{align}
\bra{V_3} = ^X\!\!\bra{V_3}\otimes \,^{gh}\bra{V_3},\nonu
^{gh}\bra{V_3} \sim \,\, _{123}\!\bra{\Omega}\exp(-E_3^{gh}),
\end{align}
where $E_3^{gh}$ is a quadratic form coupling the ghosts involving the ghost Neumann matrices $X^{rs}_{nm}$:
\begin{align}
E_3^{gh} = \sum_{r,s =1}^3\sum_{\substack{n=1\\
		m =  0}}^{\infty} c_n^{(r)} X^{rs}_{nm} b_m^{(s)}.
\end{align}
Furthermore only the coefficient matrices for the non-zero modes (in the Siegel gauge) would concern us. These are the ghost Neumann matrices denoted by $X^{rs}_{nm}$, with $r, s\in\{1,2,3\}$ and by their symmetry and cyclicity properties, we can restrict to $X^{11}=X^{(0)}, X^{12} = X^{(+)}$ and $X^{21} = X^{(-)}$. They are algebraic valued and can be obtained efficiently from CFT using contour integral representations \cite{Gross:1986ia,Taylor:2003gn}.


\bigskip
The one-loop tadpole can be represented as a ket (or more properly as a bra) which involves an exponential purely quadratic in the  creation operators. These special states then belong to the class of squeezed states in the Hilbert space. 
\begin{equation}\label{eq:tadpoleket}
|\mathcal{T}\rangle = \int_0^{\infty}\, dt\, e^t \frac{\det(1-S\widetilde{X})}{(Q\det(1-S\widetilde{V}))^{13}} ~ \exp \left(-\frac{1}{2}a^\dagger \bM a^\dagger-c^\dagger \bR b^\dagger\right)\hat{c}_0 |\hat{\Omega}\rangle.
\end{equation}
where the $t$ dependence in $Q$ and the infinite matrices $\bM, \bR,  \widetilde{X}$, and $\widetilde{V}$ are understood. The relevant inner product involving reflector $\bra{V_2}, \ket{V_3}, \hat{L}_0$, etc. is presented  in \eqref{eq:osc tadpole ket}.

We quote the following form for $\bR(t)$ derived in \cite[\S4]{Ellwood:2003xc} using squeezed state methods presented in \cite{Kostelecky:2000hz}:
\begin{equation}\label{eq:Rnmosc expr}
\bR(t) = X^{11}+\left[\begin{array}{cc}
\hat{X}^{12}(0,t)& \hat{X}^{21}(0,t)
\end{array} \right]\frac{1}{\mone-S \widetilde{X}}S\left[\begin{array}{c}\hat{X}^{21}(t,0)
\\ \hat{X}^{12}(t,0)
\end{array} \right]
\end{equation}
The ``hatted'' matrices are simply the  Neumann matrices dressed with the $t$ dependent propagator factors of the following form:
\begin{equation}\label{eq:hatted X}
\hat{X}_{nm}^{i_k j_l}(t_k,t_l)\coloneqq e^{-n t_k/2}X_{nm}^{i_k j_l}e^{-m t_l/2}.
\end{equation}
In terms of these, the infinite matrix $\widetilde{X}$ is given by
\begin{equation}
\widetilde{X}(t) = \left[\begin{array}{cc}\hat{X}^{11}(t,t)
&  \hat{X}^{12}(t,t)\\ 
\hat{X}^{21}(t,t)&\hat{X}^{11}(t,t) 
\end{array} \right],
\end{equation}
and  $\displaystyle S = \mathbb{1}_2\otimes C $, where again $C_{nm}=(-1)^n \delta_{nm}$ is the twist matrix, that arises from the specific overlap conditions imposed by the Witten type vertex  in the matter and ghost sectors.
The above matrices become $2L\times2L$ dimensional in an oscillator level truncation, which roughly corresponds to using $4N\times4N$ dimensional matrices in the discrete Moyal representation for finite $N$.
\vspace{-2mm}
\subsubsection*{\bf Expansion around $t=0$}
\vspace{-2mm}


As observed in \cite{Ellwood:2003xc}, the infinite matrix $\bR(t)$ cannot be reliably expanded around the point  $t = 0$ (or $q\coloneqq e^{-t} = 1$)  that we are interested in. This is because an intermediate matrix to be inverted,  $\mone-S\widetilde{X}(0)$, for the expansion point becomes singular due to a subset of the Gross-Jevicki non-linear relations satisfied by the unhatted matrices $\displaystyle
\mathcal{M}_{0,\pm}\coloneqq - C X^{0,\pm}$ in the ghost sector:
\begin{subequations}\label{eq:GrossJevicki}
	\begin{align}
	\cM_0^{}+\cM_+^{} + \cM_-^{}&=\mathbb{1},\qquad & \cM_+^{}\cM_-^{}=\cM_0^{2}-\cM_0^{},\\
	\cM_0^2+\cM_+^2+\cM_-^2&=\mathbb{1},\qquad & \cM_0\cM_++\cM_+\cM_-+\cM_-\cM_+=0,\\
	\cM_\pm^2-\cM_\pm&=\cM_0\cM_\mp.
	\end{align}
\end{subequations}
These are mutually commuting matrices and in the limit $t\to0$, when we have
\begin{equation}
\mathbb{1}-S\widetilde{X}\rvert_{t=0}=\left[\begin{array}{cc}
\mathbb{1}-\mathcal{M}_-&-\mathcal{M}_0  \\ -\mathcal{M}_0
&\mathbb{1}- \mathcal{M}_+
\end{array} \right], 
\end{equation}
this allows us to express the determinant in terms of the constituent blocks by the usual formula for $2\times 2$ matrices:
\begin{align}
\det(\mathbb{1}-S\widetilde{X})\rvert_{t=0} &= \det(\mathbb{1}+ \mathcal{M}_-\mathcal{M}_+ - \cM_--\cM_+ - \cM_0^2)\nonu
& = \det\left(\cM_0-\cM_0^2+\cM_-\cM_+\right)\nonu
&=\det(\mathbb{0})=0,
\end{align}
which makes the Taylor series ill-defined.
This fact is also carefully pointed out in \cite[App B]{Ellwood:2003xc}. The authors  study these expressions numerically and comment on why a level truncated analysis would differ from the correct numerical behaviour which matches with a BCFT based expansion \eqref{eq:R linear behaviour} as the level is increased. Since  the identities only hold in the infinite $L$ limit, the problem does not arise at finite level,  which effectively acts as a UV cutoff for $t=0$.

Thus, the order of limits $t\to0$ and the level $L\to \infty$ do \emph{not} commute and subsequently the infinite level result gives a factor of $-2n$ for the linear term instead of $-n$ as confirmed by numerical studies at finite level.  As we shall see in \S\ref{subsec:linear order}, the Moyal expressions do not suffer from this order of limits issue (at least at the leading order) and leads to the correct linear coefficients. In the consistent truncation we use, something similar happens with the inverse, but this time the full matrix to be inverted vanishes at $t=0$ even for finite $N$ thus altering the UV behaviour.

\vspace{-2mm}
\subsection{Summary of notations used}\label{subsec:notations}
\vspace{-2mm}
Here we collect some of the notations and conventions that will be used in the rest of the paper.
\begin{description}
	\item[Phase space basis vectors $\xi, \xigh \colon$] The string field  $A(\bar{x},\xi, \xigh)$  is valued in the non-commutative phase space $\xi^\mu_i =(x_2^\mu,\ldots. x^\mu_{2N}, p^\mu_2,\ldots p^\mu_{2N}), \xigh _i= (\xi^1_{2n}, \xi^2_{2n})$ labelled by even integers. The doublet structure $(x,p)$ would be understood in the following which for ghosts is in \eqref{eq:ghost doublets}. The zero mode dependence is factored out in Siegel gauge through $\hat{A} = \xi_0 A(\xigh)$.  We shall  suppress the Lorentz indices unless required.
	
	\noindent
	The integration or the BPZ inner product is mapped to the trace in this phase space. Also, we shall denote $(d\xi), (d\eta)$, etc. for integration over the Moyal space modes\footnote{The $\eta, \xi$ here are (unfortunately) unrelated to the $\eta(z), \xi(z)$ conformal fields defined by FMS \cite{Friedan:1985ge} and in the recent developments in  superstring field theories.}  $d^d\xi_1 \ldots d^d\xi_{2N},\, d^d\eta_1 \ldots d^d\eta_{2N}$, and suppress the measure factors of $\displaystyle  \frac{1}{2\pi i}$ unless necessary.
	\item [Constant matrices:] The spectrum is denoted by a $2N\times2N$ diagonal matrix $\kappa$, which in the parity basis is $\kappa = {\rm Diag}\{\ke, \ko\}$ and the labels $e, o$ refer to the even and odd integer mode numbers. In the open string limit $N\to \infty$, we shall set $\kappa_{2n} = 2n$ and $\kappa_{2n-1} = 2n-1$, corresponding to the perturbative spectrum. A useful matrix in the ghost sector is:
	\begin{equation}\label{eq:ghost spectral}
	\ktgh =\left[\begin{array}{cc}R^{\top}\ko T^\top
	&  0\\ 0
	& \ke
	\end{array} \right]
	\end{equation}
	The linear transformations to go to the discrete diagonal basis requires the use of certain (constant) infinite matrices whose elements are simple functions of the mode labels $e$ and $o$. In the regulated theory, these have their finite dimensional analogues which in general depend on the frequency matrices $\ke$ and $\ko$. The infinite $N$ limit of the matrices is sufficient to see their fall off behaviour at large mode numbers  in infinite sums:
	\begin{equation}
	T_{eo}=\frac{4}{\pi}\frac{o\, i^{o-e+1}}{e^2-o^2},\qquad R_{oe}= \frac{4}{\pi}\frac{e^2\,i^{o-e+1}}{o(e^2-o^2)}, \qquad w_e = i^{2-e},\qquad v_o = \frac{2\sqrt{2}}{\pi}\frac{i^{o-1}}{o}.
	\end{equation}
	these satisfy the relations presented in \cite{Bars:2003gu} of which we mainly use:
	\begin{equation}\label{eq:matrix relations}
	T \, R = \mone_e,\qquad R\, T = \mone_o ,\qquad R = \ko^{-2}\,T^\top\, \ke^2,\qquad  T\,T^\top = \mone_e - \frac{w w^\top}{1+w^\top w}
	\end{equation}
	\item [Monoid elements:] We shall be primarily using the monoid subalgebra \S\ref{subsubsec:subalgebra} for our calculations. These are shifted Gaussian functionals of the form
	\[A(\xi)=\cN e^{-\bxi M\xi -\bxi\lambda}.\]
	For the perturbative ghost vacuum $\hat{A}_0^{gh} = \xi_0 A_0^{gh}$, where $A_0^{gh}$ has parameters $\cN_0^{gh} = 2^{-2N}(1+w^\top w)$ and $M = \Mgh$ \eqref{eq:vacuum monoid}.
	For external states built on the perturbative vacuum, it is sufficient to consider a generating functional with $M^X = M_0^X$ and $M^{gh} = -i\varepsilon \otimes \Mgh$ with a general $\lambda^X,\lgh$ and construct states as polynomials $\wp(\xi,\xigh).$

   We shall assume that the  interchanging functional operations as usually done in QFT can be performed here as well, although it does not seem that straightforward.
	\item [Normalization factor:] The Witten type vertex and the Moyal vertex are related by a (divergent) factor. By using the regularized theory, it leads to a renormalization of the bare coupling $g_0$ to the physical $g_T$ when considering the D25 brane reference BCFT. The two vertices are related as:
	\begin{equation}
	\ip{\Phi_1}{\Phi_2\ast\Phi_3} = \mu_3^{-1}\Tr\left[A_1\ast A_2\ast A_3\right]
	\end{equation}
	where we have chosen the Siegel gauge and $A_i(\xi)\coloneqq \ip{\xi}{\Phi_i}$ and 
	\begin{equation}\label{eq:vertexnormalization}
	\mu_3 = -2^{2N(d-2)}(1+w^\top w)^{-\frac{d-6}{8}}\left(\det(3+tt^\top)\right)^{-d}\left(\det(1+3t t^\top)\right)^2.
	\end{equation}
	\item [Modular parameters:]We use the variable $t$ for the worldsheet lengths so as to match the usual convention for the nome $q=e^{2\pi i \tau}$. This requires that $\tau \mapsto i t/2\pi$.
	\begin{equation}
	q = e^{-t}, \qquad q_1 = e^{-t_1},\qquad q_2 = e^{-t_2}, \mbox{~~etc.}
	\end{equation}
	Some simple matrix functions that would be convenient for writing down integrands can then be defined in terms of $q$ and the mode label $n$ as $f_i(n; q)$ ($i = 1, 2, 3, 4$) to be given in \eqref{eq:f1234 definitions} and some auxiliary functions $h_i(n; q)$ and $g_i(n; q)$ in \eqref{eq:auxfns}.
\end{description}
We shall be quite explicit in the following, since we are seeking to resolve the minor mismatch in the oscillator construction and since many of the steps are simple block matrix multiplications or Gaussian integrations. The reader can skip these intermediate steps and essentially consider the final form of the expressions if desired. We shall also retain the ``gh'' superscript although it is usually clear from the context when the quantities refer to the ghost contribution. When the superscript is not used, it refers to the matter sector which we shall sometimes denote by an ``X'' superscript.  

In the next section, we shall apply the Feynman rules  in Moyal space (described in \S\ref{subsec:Feynman rules}) to write down the formal analytic expression for the tadpole integrand that will serve as the starting point for our analysis.

\section{One-loop tadpole graph}\label{sec:tadpole}
In this section, we write down the  one-loop tadpole integrand in the Moyal representation while focussing on the ghost contribution. 
This can then serve as a starting expression for examining the non-analyticities in the ghost sector, as a function of the modular parameter $t$.
\subsection{Ghost sector expressions in the Moyal basis}\label{subsec:ghost sector}
We wish to obtain an expression for the one-loop contribution to the tadpole graph in bosonic open string theory. Since this is an intrinsically off-shell quantity, we need to work in the framework of a string field theory and we choose the Witten type OSFT reviewed in the previous section. 
\begin{figure}[h!]
	\centering
	\includegraphics[width=0.44\textwidth]{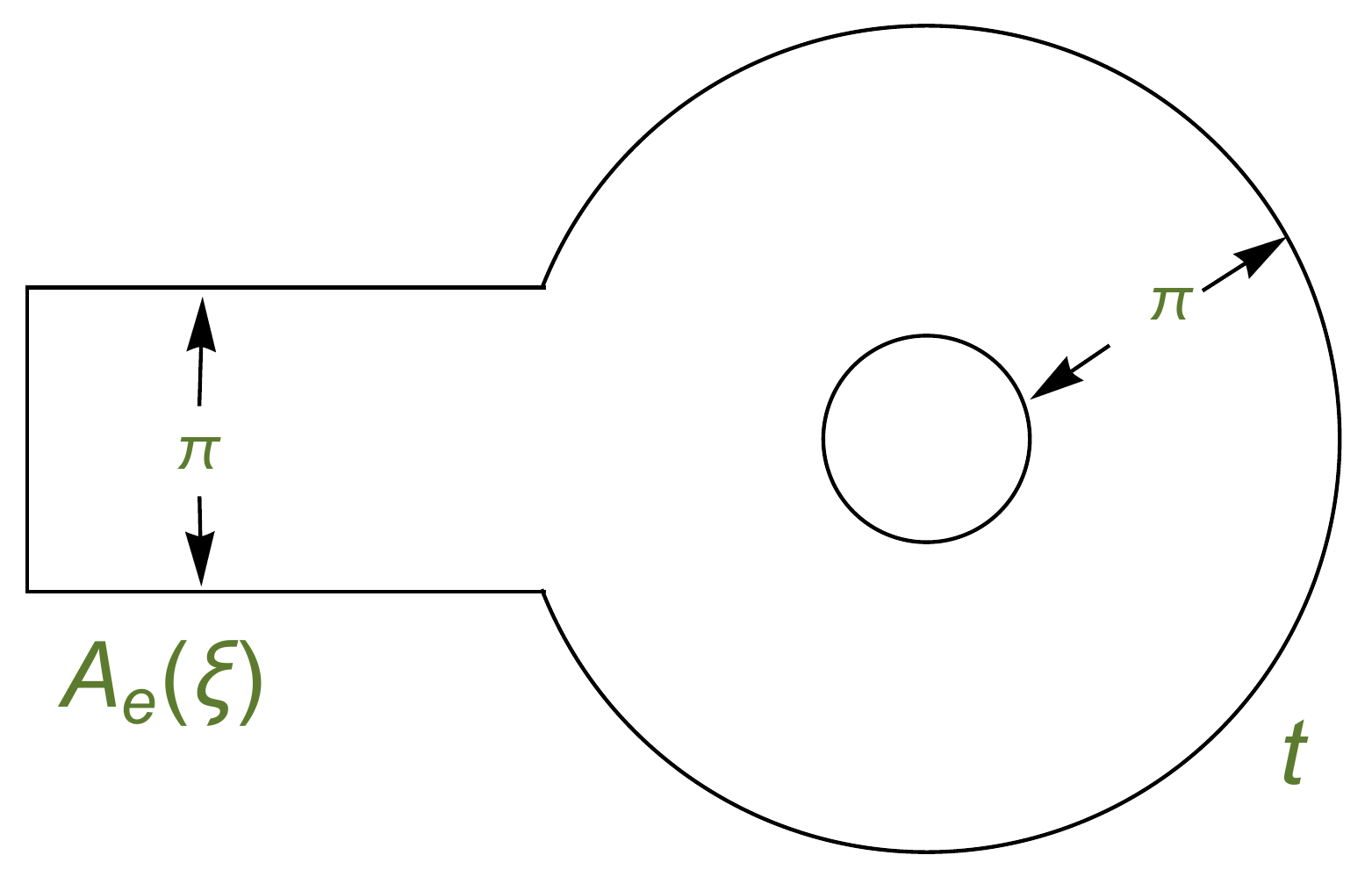}
	\caption{The open string tadpole diagram at a given modular parameter $t$ for an external state $A_e(\xi)$ in Moyal space. The width of each strip is fixed to $\pi$ and the curvature singularities are suppressed.}
	\label{fig:tadpole}
\end{figure}
The string diagram for this process is depicted in Fig.\ref{fig:tadpole} where an open string state at zero momentum ($p_e = 0$) in a D25-brane background appears from vacuum, splits into two open strings and then annihilate each other, just like in QFT. It is parametrized by a single modular parameter associated with the length of the internal propagator. The corresponding integrand at a fixed modular parameter $t$, may be obtained by identifying two legs of an off-shell $3$-point diagram and integrating over a complete set of (normalized) quantum states $e^{+i\bxi \eta+ip\bar{x}}\, e^{-\bxigh \egh}$ as described in \S\ref{subsec:Feynman rules}.


Following the Feynman rules for OSFT perturbation theory in the Feynman-Siegel (FS) gauge, we can formally write down the unintegrated amplitude  corresponding to an external state $A_e(\xi)$ as follows:

\begin{equation}\label{eq:tadpole integrand}
\cI_e(t)=-\frac{g_o}{3}\int d^d \bar{x} \, \frac{d^d p}{(2\pi)^d} \, \frac{(d\eta)}{(2\pi)^{2dN}} \,(d\egh)\,  \Tr\left[A_e\ast\left(e^{-i\xi^\top \eta+\bxigh \egh}e^{-ip\cdot\bar{x}}\right) 
\ast\left(e^{-t (L_0-1)} \left(e^{i\bxi\eta-\bxigh\egh}e^{ip\cdot\bar{x}}\right)\right)\right]
\end{equation}
where $\Tr$ denoted $\xi$ integrations and $\displaystyle L_0 = L_0^X +  L_0^{gh}$ is the total propagator. Here, we have set $l_s = \sqrt{2}$ so that $\alpha^\prime =1$. Additionally, in the discrete formalism we may be allowed to rescale the modes appearing in the matter and ghost degrees of freedom in order to set the non-commutativity parameters $\theta = 1 = \theta^\prime$.

The ghost number saturation condition for the Witten vertex dictates a total of $+3$ ghost number charge at each vertex. Since we restrict to off-shell states of ghost number $1$, this then leads to the projection onto ghost number $(3-1)/2 = +1$ states for all the states propagating in the loop. Hence, the Fourier basis we chose would be sufficient, with the additional insertion of the ghost zero mode $-\xi_0$ which is always understood to be present.

\medskip
\noindent
The expressions for the  ghost sector contributions are naturally simpler compared to the matter sector due to the absence of the ghost zero mode $c_0$ (in the Feynman-Siegel gauge). Additionally, the ghost contribution is in a sense universal. Hence, we restrict to pure ghost external states in this section and consider the matter contribution later in App \ref{sec:determinant}. The fields we consider would now be of the form $\wp(\xigh)A_0^{gh}(\xigh)$, where $\wp(\xigh)$ represents a polynomial in $\xigh$ and $A_0^{gh}$ is the vacuum monoid defined in \eqref{eq:vacuum monoid}:

\begin{equation}
A_0^{gh}=\cN_0 \exp\left[-\bxigh M_0^{gh}\xigh-\bxigh \lgh \right]
\end{equation}
These integrands may therefore be obtained by evaluating from a generating functional $\cW(\lgh, t)$ dependent on an element valued in the monoid subalgebra
\begin{equation}\label{eq:gen string field}
A_1(\xigh) = \cN \exp\left[-\bxigh M_0^{gh}\xigh -\bxigh \lgh\right],
\end{equation}
differentiating with respect to this  parameter $\lgh$ appropriately and then setting it to zero at the end of a calculation:
\begin{equation}
\wp(\xigh)A_0^{gh} = \left(\wp\left(-\frac{\vec{\partial}}{\partial \lgh}\right)A_1\right)\Big\rvert_{\lgh=0},
\end{equation}
as done in usual quantum field theory calculations.
Hence, it would be sufficient to  analyse the class of monoids of the form $A_1(\xigh)$. Furthermore, we restrict $A_e(\xigh)$ to be in the $SU(1,1)$ singlet sector \mcite{suoneone,Bars:2003gu} of twist even pure ghost excitations, since the tadpole state is a twist even singlet.

\subsubsection*{Method of evaluation}\label{subsub:evaluation method}
\vspace{-2mm}
Interchanging the order of integration (between $\eta$ and $\xi$) in \eqref{eq:tadpole integrand} and using associativity of the $\ast$ product allows us to obtain various formally equivalent expressions:
\begin{enumerate}[label=(\alph*)]
	\item $A_1\ast A_2 \to A_{12}[\eta,\xi] \to A_{12}A_3(t) \to \Tr \to \int d\eta$
	\item $A_2\ast A_3(t) \to A_{23}[\eta,\xi,t] \to \int \, d\eta \to A_1 A^\prime_{23}[\xi,t]\to \Tr $
	\item$A_2\ast A_3(t) \to A_{23}[\eta,\xi,t]\to A_1 A_{23}\to \Tr \to \int d\eta$, 
	\item $A_3(t)\ast A_1 \to A_{31}[\eta,\xi,t]\to A_{31}A_2\to \Tr \to \int d\eta$, and
	\item $A_3(t)\ast A_1 \to A_{31}[\eta,\xi,t]\to A_{31}A_2 \to \int d\eta \to \Tr$,
\end{enumerate}
where the last two are possible due to the assumed cyclicity of the trace.

\medskip
\noindent
We choose the first sequence due to its relative simplicity. The second one allows us to identify the Fock space state by integrating $A_{23}^\prime(\xi)$ with $|\xi\rangle$ but it involves a somewhat complicated inverse nested inside another inverse which makes direct evaluations difficult. It does lead to the correct behaviour near $t=0$ as we shall mention in  \S\ref{subsec:nonassoc} while examining associativity. The remaining forms result in awkward expressions that turn out to be  rather unwieldy for our purposes.

If one employs the finite $N$ regularization from \S\ref{subsubsec:subalgebra} and  makes the assumption  that all physical quantities appear as Cauchy sequences in $N$, one can ensure uniform convergence of the integrand as a function of $t$. Perhaps this could justify some of the algebraic manipulations we use, but in general one cannot avoid subtleties associated with order of limits, namely the non-analyticities from closed string physics may be extracted only in the open string limit. We shall return to this point in the later sections.
\vspace{-2mm}
\subsubsection{Overlap amplitude in Moyal space}\label{subsubsec:ghostoverlap}
After this preparation, let us list  the three monoid elements appearing in the amplitude along with their parameters:
\begin{subequations}
	\begin{align}
	\label{A1}
	A_1^{gh}&= \mathcal{N}_0 \exp \left[-\bxigh M_0^{gh}\xigh-\bxigh \lgh \right], & M_1 &= M_0^{gh},& \lambda_1 &= \lambda^{gh}, & \cN_1 &= \cN_0^{gh}, \\
	\label{A2}
	A_2^{gh} &= e^{+\bxigh \egh}, & M_2 &= 0,& \lambda_2 &= -\egh,& \cN_2 &= 1,\\
	\label{eq:A3}
	A_3^{gh} &= e^{-\bxigh \egh},&M_3&=0,&\lambda_3 &= +
	\egh,& \cN_3&=1. 
	\end{align}
\end{subequations}
Here we recall that $M_0^{gh}$ is a symmetric matrix but the metric in ghost space is set to be $+i\varepsilon_{ab}$ (with $\varepsilon_{12}=-1=-\varepsilon_{21}$) and hence the full structure of the matrix for the quadratic term is of the form $-i\varepsilon\otimes M_0^{gh}$. This makes the combination an anti-symmetric matrix, as required for anti-commuting degrees of freedom. Additionally, we have suppressed a metric factor in the linear term $\bxigh \lgh$, whose explicit form is $\bxigh (-i \varepsilon \otimes \mathbb{1}_{2N} )\lgh$. 

\noindent
To commence evaluation, we first take the $\ast$ product of $A_1$ and $A_2$ to obtain:
\begin{equation}
A_{\cN_{12},M_{12},\lambda_{12}}\coloneqq A_{\cN_1,M_1,\lambda_1}\ast A_{\cN_2,M_2,\lambda_2}\label{eq:A12}
\end{equation}
This can be written down by applying the monoid algebra relations given in Ref.\cite{Bars:2003gu} and mentioned briefly in in \S\ref{subsubsec:subalgebra}. Given two monoid elements in Moyal space, $A_1(\xi)$ and $A_2(\xi)$ from the class of shifted Gaussians (quadratic exponentials with a linear term), the string field obtained through the $\ast$ operation is parametrized by:
\begin{subequations}\label{eq:monoid algebra}
	\begin{align}
	m_{12}&=(m_1+m_2 m_1)(1+m_2 m_1)^{-1}+  (m_2-m_1m_2)(1+m_1m_2)^{-1},\\
	\lambda_{12}& =(1-m_1)(1+m_2m_1)^{-1} \lambda_2 + (1+m_2)(1+m_1m_2)^{-1} \lambda_1, \\
	\cN_{12}&=\cN_1\cN_2\det(1+m_2m_1)\exp \bls+\frac{1}{4}\lghtp_\alpha \sigma K_{\alpha \beta}\lambda^{gh}_\beta\brs \mbox{~where},\\
	K_{\alpha\beta}& = \left[\begin{array}{cc}(m_1+m_2^{-1})^{-1}
	& (1+m_2m_1)^{-1} \\ -(1+m_1m_2)^{-1}
	& (m_2+m_1^{-1})^{-1}
	\end{array} \right], \qquad m_i \coloneqq M_i\sigma.
	\end{align}
\end{subequations}
Applying this rule to the two string fields for our case in \eqref{eq:A12}  immediately leads to the parameters:
\begin{align}
A_{12}(\xigh)&\coloneqq\mathcal{N}_{12}\exp(-\bxigh M_{12}^{gh}\xigh-\bxigh \lambda_{12}), \mbox{~where}\nonu
M_{12}&= M_0^{gh},~~ \lambda_{12}= -(1-\mgh)\eta^{gh}+\lambda^{gh},\nonu
\mathcal{N}_{12}&=\mathcal{N}_0\exp \left(\frac{1}{4}\begh\sigma m_0^{gh} \egh -\frac{1}{2} \blgh \sigma \egh \right).
\end{align}
where we have used $K_{11}= 0, K_{12}=1 =-K_{21}$ and $K_{22}= m_0^{gh}$ and once again the ghost space metric is implicit. 

Next, we need  the $t$ evolved monoid element $A_3(\xigh,\egh,t)$, for which we use the action of $L_0^{gh}$ on a general monoid element $\displaystyle \cN e^{-\bxigh M^{gh} \xigh -\bxigh \lgh}$. Unfortunately, the Virasoro operator $L_0^{gh}$ is  no longer diagonal in this basis:
\begin{equation}\label{eq:peculiar propagator}
L_0^{gh}= \Tr \ktgh - \aqua \left(\frac{\partial}{\partial \xigh}\right)^{\top}M_0^{gh-1}\ktgh \left(\frac{\partial}{\partial\xigh}\right)+\bxigh \ktgh M_0^{gh}\xigh
\end{equation}
unlike the oscillator case: The simplicity in the interaction term has made the kinetic term complicated. Hence $L_0^{X+gh}$ has a non-trivial action on the string fields, which can however be written down in closed form. This leads to the following transformation rules \cite{Bars:2003gu} in terms of hyperbolic functions of the ``spectral matrix'' $\ktgh$ \eqref{eq:ghost spectral}:
\begin{subequations}\label{eq:propagated monoid}
	\begin{align}
	A(t)&\coloneqq e^{-t L_0^{gh}}A_{\cN^{gh},M^{gh},\lgh}(\xigh) =\mathcal{N}(t) \exp \left(-\bxigh M^{gh}(t)\xigh-\bxigh \lambda^{gh}(t)\right), \mbox{~ where}\\
	M^{gh}(t)&= \left[\shktgh +\left(\shktgh+M_0^{gh}M^{gh-1}\cktgh\right)^{-1}\right]\sktgh M_0^{gh}, \\
	\lgh(t)&= \left[\cktgh + M^{gh}M_0^{gh-1}\shktgh\right]^{-1}\lgh,\\
	\cN^{gh}(t)&=\cN^{gh}\exp\bls+\aqua \lambda^{gh\top}(M^{gh}+\coth t\ktgh)^{-1}\lgh\brs\nonu
	&\qquad{}\times \det\left[\frac{1}{2}(1+M^{gh}M_0^{gh-1})+\half(1-M^{gh}M_0^{gh-1})e^{-2t\ktgh}\right],
	\end{align}
\end{subequations}
and a very similar expression in the matter sector, except for the extra dependence on the zero mode momentum $p^\mu$ and the vector $w$. Notice that the correct boundary conditions for $t=0$ and $t=\infty$ are taken into account in the above rules.

\medskip
\noindent
Now, applying this transformation on \eqref{eq:A3}, for which the matrix of parameters $\Mgh$ vanishes, we readily obtain the string field:
\begin{align}
A_3(t)&= \mathcal{N}_3(t) \exp \left(-\bxigh M_3(t)\xigh-\bxigh \lambda_3(t)\right), \mbox{~ with parameters}\nonu
M_3(t)&= \tktgh M_0^{gh},~~ \lambda_3(t)= +\sech(t \ktgh)~ \eta^{gh},\nonu
\mathcal{N}_3(t)&=  2^{-2N} \prod_{n=1}^{2N}(1+e^{-2tn}) \exp \left(\aqua \begh M_0^{gh-1}\tktgh \egh\right).
\end{align}
We can now use the property that the remaining $\ast$ product between $A_{12}$ and $A_3(t)$ may be dropped as total derivative pieces contribute only to boundary terms under a trace ($\xi$ integration with appropriate measure factors inserted).
We therefore define  a new string field configuration $A_{12}A_3(t)\eqqcolon A_{123}^{gh}(t)$, under the ordinary (local) product in function space, with parameters:
\begin{align}
M_{123}(t)&=M_{12}+M_3(t),\nonu
\lambda_{123}(t)&=\lambda_{12}+\lambda_3(t),\nonu
\cN_{123}(t)&=\cN_{12}\cdot \cN_3(t).
\end{align}
Hence, the trace in \eqref{eq:tadpole integrand} simply results in
\begin{align}\label{eq:ghost trace}
\Tr [A_{123}(t)] &= \mathcal{N}_{123} \det(2M_{123}(t)) \exp\left[+\frac{1}{4}\lambda^\top_{123}M_{123}^{-1}\lambda_{123}\right]\nonu
&\eqqcolon \cC_\eta\exp\left[-\eta^{gh\top}\cQ_\eta \egh +\lghtp\cL_\eta^{\top}\eta^{gh}\right].
\end{align}
In order to perform the remaining Gaussian integration over $\egh$, we have  separated the quadratic, linear and zero degree terms in $\eta^{gh}$ by collecting  the contributions from $\cN_{123}$ and the argument of the exponential in the first line of \eqref{eq:ghost trace} above.  In terms of the matrices that are used in the Moyal representation \S\ref{subsec:notations}, these are  given by:
\begin{subequations}\label{eq:eta coeffs}
	\begin{align}
	\cQ_\eta(t)&=-\frac{1}{4}\Big[\sigma m_0^{gh} +\sigma m_0^{gh-1}\tktgh \nonu
	&\qquad{}+(m_0^{gh\top}+\sech t \bktgh-1)\sigma m_0^{gh-1}(1+\tktgh)^{-1}(m_0^{gh}+\sktgh-1)\Big],\\
	\cL^{\top}_\eta(t) &= -\frac{1}{2} \sigma\left[1 +m_0^{gh-1}(1+\tktgh)^{-1}(1-m_0^{gh}-\sktgh))\right],\\
	\cC_\eta(t) &=\cN_0 \det(2M_0) \exp\left[+\aqua \lghtp M_{123}^{-1}\lgh\right]\label{eta coeffs}\nonu
	&= \det(M_0)^{\half}\exp\left[+\aqua \lghtp M_0^{-1}(1+\tktgh)^{-1}\lgh\right].
	\end{align}
\end{subequations}
where we have used the subscript $\eta$ to specify the variable in the quadratic form, a convention we shall be following from now onwards\footnote{Here we point out that the $+ve$ sign in the exponential factor in the first line of \eqref{eq:ghost trace} is different from the usual $-ve$ sign for Grassmannian Gaussian integral, since the antisymmetric metric factor $\varepsilon$ adjoining $\lambda_{123}$ produces an extra -ve sign upon taking a transpose. Explicitly, we have the following signs:
	\begin{equation}
	(-i \varepsilon)^\top (-i \varepsilon)^{-1}(-i \varepsilon) = -(-i \varepsilon).
	\end{equation}
	Since we insist on using an $-i \varepsilon$ metric in ghost space, there is the extra $-ve$ sign which makes the exponential part in the Gaussian integral identical to the matter sector.}.

\bigskip

At this point it is convenient to introduce the (Euclidean) nome $q  = e^{+2\pi i \tau} = e^{-t}$ and the functions:
\begin{align}\label{eq:f1234 definitions}
f_1(n; q) &= (1-q^n)^2,\qquad &&f_2(n; q) = 1+q^{2n},\nonu
f_3(n; q) &= 1-q^{2n},\qquad&&f_4(n; q) = (1-q^n)(3-q^n)=\frac{f_1^2+2f_3}{f_2},
\end{align}
%
in order to convert the hyperbolic functions to exponentials for typographical simplicity. We can then rewrite the coefficient matrices obtained above in terms of the matrix functions $f_i(\ktgh; q)$. These have block diagonal structure but contain non-diagonal matrices in the upper block. Additionally, they do \textit{not} commute with  matrices such as $m_0^{gh}$ and $M_0^{gh}$. However, using matrix relations such as 
\begin{equation}
\kt^{gh\top}= (M_0^{gh})^{-1}\ktgh M_0^{gh}, ~~m_0^{gh\top} = -\s M_0^{gh}, \mbox{~and}\qquad\sigma \left[ \begin{array}{cc}\alpha_1
&  0\\ 
0&\alpha_2 
\end{array} \right] \sigma =  \left[ \begin{array}{cc}\alpha_2
&  0\\ 
0&\alpha_1 
\end{array} \right]
\end{equation}
for block diagonal matrices, one can simplify the above  expressions for $\egh$ coefficients as
\begin{align}\label{eq:Q gh}
\cQ_\eta(q) 
&= -\frac{1}{4}\left(\sigma M_0^{gh}\sigma+M_0^{gh-1}\frac{f_3(q)}{f_2(q)}\right)-\frac{1}{8}\left(M_0^{gh-1}\frac{f_1^2(q)}{f_2(q)}-\sigma f_2(q) M_0^{gh}\sigma+\sigma f_1(q) -f_1(q)^\top\s\right)\nonu
&=-\frac{1}{8}\blp \sigma f_3(q) M_0^{gh}\sigma + M_0^{-1}f_4(q) + \sigma f_1(q) - f_1(q)^\top\sigma \brp,\\
\cL_\eta(q)& = \aqua \sigma f_3(q)-\aqua f_1(q)^\top M_0^{gh-1},
\end{align}
where we have dropped one argument of $f_i(q;\ktgh)$ as shall be done in other places as well for typographical simplicity.
\medskip\\
Let us also mention that these functions simply appear through intermediate expressions\footnote{Yet another useful relation is \[
	f_1^2+f_3^2 = 2 f_1 f_2.
	\]}of the form
\begin{subequations}
	\begin{align}
	(1+\tktgh)^{-1}=e^{-t \ktgh}\cktgh &= \frac{1}{2}f_2,  \\
	1-\sktgh  &=\frac{f_1}{f_2},\\
	e^{-t \ktgh}\blp \cktgh +\sktgh -2\brp & = \frac{f_1^2}{2f_2}.
	\end{align}
\end{subequations}
Additionally, we can obtain the half-angle relations by noting that $f_3(n; q) = f_2(n; \sqrt{q}) \, f_3(n; \sqrt{q})$.

\noindent
Finally, we perform the  integration over $\egh$ \eqref{eq:ghost trace}  to obtain a  purely quadratic functional dependence on $\lgh$  in the exponential of the form $\displaystyle +\aqua \blgh\cF \lgh$, where the matrix $\cF$ in the ghost sector can be written as
\begin{equation}\label{eq:Fgh matrix}
\mathcal{F}(t)= M_{123}^{-1}(t)+\cL_\eta^{\top} \mathcal{Q}_\eta^{-1}\cL_\eta.
\end{equation}
Here, the first term $M_{123}^{-1}$ arises from the $\egh$ independent overall factor $\cC_\eta(t)$ in \eqref{eta coeffs} defined above.

\medskip
Collecting all the factors together, the ghost contribution to the generating functional $W[\lambda,\lgh, t]$ is given by:
\begin{equation}\label{eq:ghost amplitude}
\cW[\lgh,t]=(1+w^\top w)^{\aqua} \det(2\cQ_\eta^{gh})\exp\left[+\aqua \lambda^{gh\top}\cF\lambda^{gh}\right].
\end{equation}
We shall  include the matter sector contribution from App \ref{sec:determinant}, which is obtained through a very similar computation---with the only difference being the integration over the zero mode momenta $p^\mu$ along the Neumann directions, and the use of a different set of constant matrices for defining the monoid elements. The matter contribution to the generating functional  serves to provide a  consistency check for our analytical expressions. Only the determinant factors need  be included in numerical checks when considering overlap with the perturbative vacuum state $\lvert\hat{\Omega}\rangle = \hat{c}_1\lvert\Omega\rangle$. And for purely ghost excitations, we use this scalar piece for the matter sector---it contributes to the measure factor and does not affect the structure of the $\bR_{nm}(q)$ factors in \eqref{eq:tadpoleket}, that we are primarily interested in.

\noindent
Finally, the total matter$+$ghost generating functional has the structure:
\begin{equation}\label{eq:matter ghost gen functional}
\cW[\lambda^X,\lgh,t]=\left(1+ww^\top\right)^{\frac{d+2}{8}}e^t \frac{\det(2\cQ_\eta)}{\lvert\det(2\cQ_\psi)\rvert^{d/2}}\exp\left[\aqua \left(\lambda^{X\top}\cF_X\lambda^X+\lambda^{gh\top}\cF_{gh}\lgh\right)\right]
\end{equation}
where $X$ denotes the matter part from the embedding coordinates $X^\mu(z)$, have combined the conjugate variables $\eta^X$ and $p$ into a single ``vector''
\[\psi \coloneqq \left(\begin{array}{c}
\eta^X\\ p
\end{array} \right)\]
and denoted the matter coefficient matrix $\cQ$ with the subscript $\psi$.

\subsubsection{Block matrices}\label{subsubsec:block matrices}
Next, we can consider the block structure of the matrices $\cQ_\eta, \cL_\eta$, and $\cF_{gh}$. To this end, we recall that the matrices  $M_0^{gh}$ and $\ktgh$ (given in \S\ref{subsec:notations}) take the block diagonal form:
\begin{equation}\label{eq:M0 kappa tilde block defn}
M_0^{gh} =- \half \left[\begin{array}{cc}
R^{\top}\kappa_o R&  0\\ 0
& 4\ke^{-1}
\end{array} \right], \qquad \ktgh =\left[\begin{array}{cc}R^{\top}\ko T^\top
&  0\\ 0
& \ke
\end{array} \right].
\end{equation}
We remark that the $M_0^{gh}$ above is given by $i \times M_0^{gh}$ as compared to the one given in \cite{Bars:2003gu}  whereas the matrix $\ktgh$ remains the same. Then the $2N\times 2N$ coefficient  matrices have the explicit constituent block structure:
\begin{align}\label{eq:ghost blocks}
\cQ_\eta(q) &=+ \aqua\left[\begin{array}{cc}
\ke^{-1}f_3(\ke)+T\ko^{-1}f_4(\ko) T^\top & -\frac{i}{2}\left[f_1(\ke)-Tf_1(\ko)R\right]\\
-\frac{i}{2}\left[f_1(\ke)-R^\top f_1(\ko) T^\top\right] & \frac{1}{4}\left[\ke f_4(\ke) + R^\top\ko f_3(\ko) R\right]
\end{array} \right]
, \mbox{~and}\nonu
\cL_\eta(q)&= \frac{1}{2}\left[ \begin{array}{cc}
T\ko^{-1}f_1(\ko)T^\top&  \frac{i}{2}f_3(\ke)\\ 
-\frac{i}{2}R^\top f_3(\ko) T^\top & \frac{1}{4}\ke f_1(\ke)
\end{array}  \right],
\end{align}
where again the blocks are labelled by half-phase space degrees of freedom $(x^{c,b}_e,p^{c,b}_e)$.

Let us observe that the infinite sums over the odd integers $\ko$  in all the four blocks of $\cQ_\eta$ diverge badly for $t<0$ since the functions $f_{1}, f_3,$ and  $f_4$ are unbounded as $\ko$ increases. Hence, these matrix elements are not analytic in a neighbourhood of $0$. Only the $t\to 0^+$ limit is well-defined for which the matrix $\cQ_\eta$ vanishes due to the zeroes of the functions $f_1,f_3$ and $f_4$ at that point (as we shall discuss below). Strictly speaking, this prevents the expansion we seek involving $\cQ_\eta^{-1}$. However, the matrix $\cL(t)$ also vanishes at $t=0$ due to the zeroes in $f_1$ and $f_3$. Hence, the combination $\displaystyle
\cL(t)^\top \cQ(t)^{-1}\cL(t)
$ in $\cF(t)$---which does involve infinite sums---can be taken to vanish at $t=0$ for the purpose of this work. This behaviour signals that the expansion we obtain may be asymptotic and not a convergent expansion, owing to this non ${\rm C}^\infty$ nature.

Additionally, we notice that in the open string limit $N\to \infty$, the order of the pole from the combined determinant factors, $\cQ_\eta$ and $\cQ_\psi$ in \eqref{eq:matter ghost gen functional}, becomes infinite as well. This is consistent with our expectations of an essential singularity at $t=0$ associated with the Shapiro-Thorn closed string tachyon state in  \eqref{eq:closedtachyon}.

In general, due to the relatively simple structure of the $T$ matrix, we can expect combinations of the generalized hypergeometric functions, $_JF_{J-1}$, to arise from the infinite sums in $\cQ_\eta$. The non-analyticity in $\cQ_\eta$ matrix elements would then be a $\log$ branch cut.
The non-diagonal terms, with $n\neq m$ are of the form:
\begin{small}
	\begin{subequations}
		\begin{align}\label{eq:QetaDiag}
		\cQ^{xx}_{2n,2m}&=\frac{(-1)^{m+n}}{4\pi ^2 (m^2-n^2)}\left\{ q \left(q \left(\Phi \left(q^4,1,\frac{1}{2}-m\right)-\Phi \left(q^4,1,\frac{1}{2}-n\right)+\Phi \left(q^4,1,m+\frac{1}{2}\right)\breakeqnthrice-\Phi \left(q^4,1,n+\frac{1}{2}\right)\right)-4 \Phi \left(q^2,1,\frac{1}{2}-m\right)+4 \Phi \left(q^2,1,\frac{1}{2}-n\right)-4 \Phi \left(q^2,1,m+\frac{1}{2}\right)\breakeqntwice+4 \Phi \left(q^2,1,n+\frac{1}{2}\right)\right)-3 \psi ^{(0)}\left(m+\frac{1}{2}\right)-3 \psi ^{(0)}\left(\frac{1}{2}-m\right)\breakeqn+3 \psi ^{(0)}\left(n+\frac{1}{2}\right)+3 \psi ^{(0)}\left(\frac{1}{2}-n\right)\right\},\\
		\cQ^{xp}_{2n,2m}&=\frac{-i(-1)^{n+m} m q}{4\pi^2n(m^2-n^2)}\left\{ -n q \Phi \left(q^4,1,\frac{1}{2}-m\right)+m q \Phi \left(q^4,1,\frac{1}{2}-n\right)+2 n \Phi \left(q^2,1,\frac{1}{2}-m\right)\breakeqn-2 m \Phi \left(q^2,1,\frac{1}{2}-n\right)+n q \Phi \left(q^4,1,m+\frac{1}{2}\right)-m q \Phi \left(q^4,1,n+\frac{1}{2}\right)\breakeqn-2 n \Phi \left(q^2,1,m+\frac{1}{2}\right)+2 m \Phi \left(q^2,1,n+\frac{1}{2}\right)\right\},\\
		\cQ^{pp}_{2n,2m}&=\frac{(-1)^{n+m}}{4\pi^2(m^2-n^2)}\left\{m^2 \left(H_{-n-\frac{1}{2}}+H_{n-\frac{1}{2}}\right)-n^2 \left(H_{-m-\frac{1}{2}}+ H_{m-\frac{1}{2}}\right)\breakeqn+ q^2 m^2 \left(\Phi \left(q^4,1,\frac{1}{2}-n\right)+\Phi \left(q^4,1,n+\frac{1}{2}\right)-q^2 n^2 \left(\Phi \left(q^4,1,\frac{1}{2}-m\right)+\Phi \left(q^4,1,m+\frac{1}{2}\right)\right)\right)\breakeqn+4 \left(n^2-m^2\right) \tanh ^{-1}\left(q^2\right)+4\log (2) (m^2-n^2) \right.\Big\}
		\end{align}
	\end{subequations}
\end{small}
while the diagonal matrix elements are given by:
\begin{small}
	\begin{subequations}
		\begin{align}\label{eq:QetaNonDiag}
		\cQ^{xx}_{2n,2n}&=\frac{1}{8 \pi ^2 n}\left\{q \left(q \Phi \left(q^4,2,\frac{1}{2}-n\right)-4 \Phi
		\left(q^2,2,\frac{1}{2}-n\right)-q \Phi \left(q^4,2,n+\frac{1}{2}\right)+4 \Phi
		\left(q^2,2,n+\frac{1}{2}\right)\right)\breakeqn-\pi ^2 \left(q^{4 n}-1\right)+3 \psi
		^{(1)}\left(\frac{1}{2}-n\right)-3 \psi ^{(1)}\left(n+\frac{1}{2}\right)\right\},\\
		\cQ^{xp}_{2n,2n}&=\frac{i}{8 \pi ^2 n}\left\{ q \left(-q \Phi \left(q^4,1,\frac{1}{2}-n\right)+n q \Phi
		\left(q^4,2,\frac{1}{2}-n\right)+2 \Phi \left(q^2,1,\frac{1}{2}-n\right)\breakeqntwice-2 n \Phi
		\left(q^2,2,\frac{1}{2}-n\right)+q \Phi \left(q^4,1,n+\frac{1}{2}\right)+n q \Phi
		\left(q^4,2,n+\frac{1}{2}\right)-2 \Phi \left(q^2,1,n+\frac{1}{2}\right)\breakeqntwice-2 n \Phi
		\left(q^2,2,n+\frac{1}{2}\right)\right)-\pi ^2 n q^{2 n} \left(q^{2
			n}-2\right)\right\},\\
		\cQ^{pp}_{2n,2n}&=\frac{1}{8 \pi ^2}\left\{2 H_{-n-\frac{1}{2}}+2 H_{n-\frac{1}{2}}+2 q^2 \Phi
		\left(q^4,1,\frac{1}{2}-n\right)-n q^2 \Phi \left(q^4,2,\frac{1}{2}-n\right)\breakeqn+2 q^2 \Phi
		\left(q^4,1,n+\frac{1}{2}\right)+n q^2 \Phi \left(q^4,2,n+\frac{1}{2}\right)+\pi ^2 n
		\left(-4 q^{2 n}+q^{4 n}+3\right)\breakeqn+n \psi ^{(1)}\left(\frac{1}{2}-n\right)-n \psi
		^{(1)}\left(n+\frac{1}{2}\right)-8 \tanh ^{-1}\left(q^2\right)+8\log (2)\right\}.
		\end{align}
	\end{subequations}
\end{small}
In this case, the $_JF_{J-1}$ functions get further expressed in terms of Lerch transcendents $\Phi(z,s,a)$, a generalization of the zeta and the polylog functions, defined classically \cite{bateman1953higher} by the infinite series representation:
\begin{equation}\label{eq:lerch}
\Phi(z,s,a) = \sum_{n=0}^{\infty}\frac{z^n}{(n+a)^s}.
\end{equation}
In all of the above, the $\Phi$ functions with the argument $\Re(a)<0$ is chosen to be  the analytic continuation Hurwitz-Lerch Transcendents, which has by definition, an identical analytic expression\footnote{This would differ from the representation in terms of the original Lerch functions, which take the form
	\begin{equation}
	\Phi^\ast(z,s,a) = \sum_{n=0}^{\infty}\frac{z^n}{[(n+a)^2]^{s/2}}.
	\end{equation}
	for $\Re(a)<0$, where we omit any term with $n+a = 0$.}.

We can now examine some series expansions to notice  that these are functions having a leading logarithmic branch cut $t^4 \log t$ for the blocks $\cQ^{xx}$ and $\cQ^{pp}$, and $t^3\log(t)$ for the blocks $\cQ^{xp} = \cQ^{px\top}$:
\begin{small}
	\begin{align}
	\cQ^{pp}_{88}&=16 t-\frac{6612992 t^2}{11025 \pi ^2}+\frac{512 t^3}{3}+t^4 \left(\frac{4096 \log (t)}{3
		\pi ^2}-\frac{365923328}{33075 \pi ^2}+1024\right)+O\left(t^5\right),\nonu
	\cQ_{24}^{xp}&=-\frac{8 i t^3 (18 \log (t)+19+6 \log (2))}{9 \pi ^2}-\frac{2 i t^5 (360 \log (t)+151+24
		\log (2))}{9 \pi ^2}+O\left(t^6\right).
	\end{align}
\end{small}
Since  the three functions $f_{1,3,4}(t)$ have first order zeroes at $t=0$ (or $q = 1$), and because the single  sum over the odd frequencies $\kappa_o$ still retain the \emph{same order} of zero for both finite and infinite $N$, we can factor out this zero.
Hence, in the open-string limit, corresponding to $N\to \infty$, we can simply\footnote{We do however expect to miss some of the very interesting non-analyticities of the form $\displaystyle e^{-2 j \pi^2/t} = e^{+\frac{2j\pi^2}{\ln q}}$ in our analysis.} divide out by $t$ in order to expand the inverse. The physically correct order of operations would be to consider the expansion only in the open string limit. However, one can also attempt an expansion for the deformed theory defined at finite $N$, and see where it leads us; since both have similar formal structure. The oscillator counterpart of this issue with order of limits, namely level $L\to \infty$ followed by $t\to0$ and its reverse, is discussed in \cite{Ellwood:2003xc}  (and was reviewed earlier in  \S\ref{subsec:osc results}) where it was found that the result does differ from BCFT\footnote{Although the oscillator and Moyal representations are formally isomorphic, there are subtle differences due to the special nature of the Witten type vertex. See \cite{Erler:2003eq,Bars:2002bt} for a careful discussion of these matters and for a detailed analysis of midpoint issues.} by a factor of $2$ already at the leading correction in $t$.

Let us therefore factor out the parameter $t = -\ln q$ and introduce the two matrices:
\begin{equation}\label{eq:cZ, cY defined}
\cZ(q) \coloneqq -\frac{4\cQ_\eta(q)}{\ln q},~~ \cY(q) \coloneqq -\frac{2\mathcal{L}_\eta(q)}{\ln q}
\end{equation}
in order to rewrite the matrix $\cF(q)$ in \eqref{eq:Fgh matrix} as below:
\begin{equation}\label{eq:F matrix}
\boxed{\cF(q)= \frac{1}{2}(M_{0}^{gh})^{-1}f_2(q) -\ln q  \, \mathcal{Y}(q)^\top \cZ(q)^{-1}\mathcal{Y}(q).}
\end{equation}
This form will turn out to be convenient when we study the behaviour of the matrix $\bR(t)$ in the limit $t\to 0^+$ directly in the modular parameter $t$ later in \S\ref{subsec:evenparity}. The matrix $\cF$ has component blocks which would be labelled as $\cF^{xx}, \cF^{xp} = \cF^{px\top}$ and $\cF^{pp}$ in terms of the phase space doublet indices $(x,p)$ as usual.

The matrix $\cZ(q)$ as it appears above is bounded at $t=0$ and hence would still be amenable to an expansion. However, the resulting expression for $\cF$ need  not be analytic because the matrix inverse $\cZ(t)^{-1}$ allows for infinite sums that alter the pole-zero structure. Furthermore, there are double infinite sums involved when this is sandwiched between $\cY^\top$ and $\cY$. Perhaps these non-analyticities may be relatable to the closed string states arising in this degeneration limit geometrically. In order to simplify the analysis, we shall restrict to the case when the $\cY$s contribute only diagonal matrices---corresponding to even parity elements---thus eliminating some of the multiple summations. We hope to look at the other cases in more detail when occasion offers itself.

\vspace{-2mm}
\subsubsection{Remarks on determinant factors}\label{subsec:detremarks}
\vspace{-2mm}
In this work, we are primarily interested in the analytic behaviour of the squeezed state matrix $\bR(t)$ or equivalently $\cF(t)$ in the limit $\displaystyle t\to0^+$, but as a check on the correctness of our expressions, we shall study the determinant factor numerically in App \ref{sec:determinant} using similar methods as in \cite{Ellwood:2003xc}. The determinant part corresponds to the overlap with the \emph{perturbative} vacuum state, i.e $\ket{A_e}=\ket{\hat{\Omega}}$: the open string tachyon at zero momentum, and has interesting divergence structure of its own. However, as encountered in \cite{Ellwood:2003xc}, it is awkward to study this factor analytically due to the essential singularity at  $t=0$. 

The full matrix element contributing to the ghost sector does not lend itself to an expansion because in general each of the matrices  whose determinant would be required would appear as a power series starting at degree $0$ (constant term). As the minimal degree does not decrease or increase along a row or a column, this form of the determinant proves unwieldy for a systematic expansion. We therefore do not perform a series based analysis of the determinant using the diagonal basis in this work and instead focus on the finite factor from the $\bR$ matrix:
\begin{equation}
\bra{\hat{\Omega}}\hat{c}_n \hat{b}_m\lvert \cT(t)\rangle \sim \bR_{nm}(t)\times \det\left(\cdots; t\right)
\end{equation}
Additionally, as part of  a series of papers on \emph{off-shell} conformal field theory (see \cite{Bluhm:1989ws,Samuel:1987uu} and references therein), the $N$-tachyon scattering case has been studied in great detail by Samuel et al. and addresses these questions much more directly using advanced Riemann surface theory upto the one-loop level. In this approach, the measure factors corresponding to the matter $+$ ghost determinants are evaluated in terms of line integrals involving rational combinations of elliptic functions and their derivatives. It may be possible to extend some of their results to the overlap with a general Fock space state other than the tachyon case considered there.

\subsection{Squeezed state matrix elements}\label{subsec:squeezed matrices}
\vspace{-2mm}
In the Siegel gauge, the ghost contribution to the tadpole \emph{state} can be expressed in terms of Fock space kets and Moyal fields as:
\begin{equation}
|\cT(t)\rangle = \int d\xigh \, \cT(\xigh,t)|\xigh\rangle.
\end{equation}
Comparing to \eqref{eq:tadpole integrand}, we have the Moyal string field
\begin{equation}
\mathcal{T}(\xigh,t) \sim \int  (d\eta^{gh})\, \left[e^{+\bxi \eta^{gh}} \ast (e^{-t L_0^{gh}} e^{-\bxi \eta^{gh}})\right],
\end{equation}
where again we have left the overall sign unfixed.  We transform the expression for $\bra{\xigh}$ in the odd basis given in \cite{Bars:2003gu} to the basis labelled by \emph{even} integers, that we use, and write this as  a bra: 
\begin{equation}\label{eq:xigh bra}
\langle\xi^{gh}\rvert = -2^{-2N}\left(1+w^\top w\right)^{-\aqua}\bra{\Omega}\hat{c}_{-1}e^{-\xi_0 (\hat{c}_0-\sqrt{2}w^\top \hat{c}_e)}e^{-\bxigh \Mgh \xigh -\bxigh \lgh},
\end{equation}
where we have the vectors
\begin{equation}
\lgh_1 =  \left(\begin{array}{c}\sqrt{2}R^\top \hat{b}_o
\\ -2\sqrt{2}\ke^{-1}\hat{b}_e+2\ke^{-1}w \xi_0
\end{array}  \right), \qquad \lgh_2 =  \left(\begin{array}{c}\sqrt{2}R^\top \ko \hat{c}_o
\\ 2\sqrt{2}i\hat{c}_e
\end{array}  \right),
\end{equation}
and $\Mgh$ is the matrix defining the perturbative ghost vacuum $A_0^{gh}$: 
\[ \Mgh = -\half \mbox{Diag}\{ R^\top \ko R, 4\ke^{-1}\}.\]
We remind the reader of the metric convention we have been using---where the $-i\varepsilon$ factor is implicit---and hence $\bxigh \Mgh \xigh = -2i \xi^{1\top}\Mgh \xi^2$ as well as $\bxigh \lgh = -i (\xi^{1\top}\lambda^2 - \xi^{2\top}\lambda_1)$.

To probe the structure of the state $\ket{\cT(t)}$, one usually finds its overlap with various Fock space basis states $\bra{\varphi}$. Hence, we must consider the corresponding overlap amplitudes in Moyal space and then transform back to Fock space.

In order to convert the amplitude written in Moyal space, \eqref{eq:tadpole integrand} to the one in terms of Fock space states, we need to construct the appropriate perturbative string fields $A_e(\xi)$. To this end, we give the corresponding expressions\footnote{Upto a $t$ independent normalization factor to which we return in \S\ref{subsubsec:Factorization}.} in the oscillator formalism:
\begin{equation}\label{eq:osc amplitude}
\langle A_e|\mathcal{T}(t)\rangle \sim \int d\xi \, \langle A_e|\xigh\rangle\langle \xigh | \cT(t)\rangle,
\end{equation}
where we have denoted the external state by $\ket{A_e}$ and introduced a complete set of states $\bra{\xigh}$---the  appropriately normalized bra defining the Moyal basis in ghost space to be given below in \eqref{eq:xigh bra}.

We recall that we can  restrict to the $SU(1,1)$ symmetric \cite{suoneone} combination of pure ghost external states, since the tadpole state is a singlet under this symmetry. In particular, the matrix $\bR_{nm}$ defining the quadratic form in the exponential of the squeezed state  satisfies:
\begin{equation}
m\,\bR_{nm}  = n\, \bR_{mn},
\end{equation}
i.e $\bR \,\kappa$ is a symmetric matrix. This does not demand the full $SU(1,1)$ but can be achieved by restricting to the discrete $\mathbb{Z}_{4}$ subgroup.

\subsubsection*{\bf The ghost sector matrix $\bR(t)$}\label{subsubsec:betaosc}
The $\hat{\beta}_{e,o}^{c,b}, \hat{\beta}_{e,o}^X$ oscillators (described in App \ref{sec:bcbetaosc}) can now be directly used to construct the perturbative string fields $A_e(\xi, \xigh)$ that correspond to the matrix elements  $\bR_{nm}(t)$ (and $\bM_{nm}(t)$ in the matter sector) when written in terms of Fock space states. The pure ghost fields would be of the form
\[\wp(\xigh)A_0^{gh}(\xigh)\]
where $\wp(\xigh)$ is an appropriately normalized polynomial, which would be the analogue of Hermite polynomials acting on Gaussians in a representation in terms of position space functionals $\Phi[X^\mu(\sigma),c(\sigma)]$.

In terms of these, the relevant matrix elements get mapped to the following Moyal polynomials with ghost bilinear pieces:
\begin{small}
	\begin{subequations}
		\label{osc matrix map moyal}
		\begin{align}
		\bR_{ee^\prime}&\leftarrow   -\delta_{ee^\prime}+\frac{8i}{\kappa_{e^\prime}}p^c_e p^b_{\ep},\\
		\bR_{oo^\prime}&\leftarrow \delta_{oo^\prime} + 2i  (\ko R x^bx^{c\top}R^\top)_{o\op},\\
		i	\bR_{eo}&\leftarrow - 4 i(p^c x^{c\top}R^\top)_{eo},\\
		i	\bR_{oe}&\leftarrow + 4 i (\ko R x^b p^{b\top})_{oe}.
		\end{align}
	\end{subequations}
\end{small}
acting on the perturbative vacuum field (as an ordinary product). We know  that the mixed parity cases $\bR_{eo}, \bR_{oe}$ terms\footnote{In the above, we have inserted extra factors of $i$ in the mixed parity cases to make the string fields real.}  vanish identically which reflects the \textit{twist} symmetry of the tadpole state $|\cT\rangle$. This can also be seen numerically as we have verified. We remark in passing that we can also obtain the matrix elements for the matter part by using the oscillators given in \cite{Bars:2002nu}  as:

\begin{subequations}
	\begin{align}
	\bM_{ee^\prime}&\leftarrow -\blp\kappa_e \delta_{ee^\prime}-\kappa_e \kappa_{e^\prime}x_e x_{e^\prime}\brp \\
	i \bM_{eo}= i \bM_{oe}&\leftarrow  \blp 4 \kappa_e x_e (p^\top T)_o\brp \\
	\bM_{oo\prime}&\leftarrow \blp\kappa_{o}\delta_{oo^\prime}-16 (p^\top T)_o(p^\top T)_{o^{\prime}}\brp 
	\end{align}
\end{subequations}
which may be useful for future applications.

Now that we know the required form of the polynomials, we can proceed to construct them starting from the generating string field $A_1(\xigh,\lgh)$ given in \eqref{eq:gen string field} using

\begin{equation}
\wp(\xigh)A_0 = \left(\wp\left(-\frac{\vec{\partial}}{\partial \lgh}\right)A_1\right)\Big\rvert_{\lgh=0}
\end{equation}
while taking into account the implicit $-i \varepsilon \otimes\mathbb{1}_{2N}$ metric factors everywhere, including the linear term. Explicitly, we make the replacements:
\begin{align}
\wp\left(x^c,p^c,x^b,p^b\right) \mapsto \wp\left( +i \frac{\partial}{\partial \lambda_{x^b}}, -i \frac{\partial}{\partial \lambda_{p^b}}, -i \frac{\partial}{\partial \lambda_{x^c}}, +i \frac{\partial}{\partial \lambda_{p^c}}\right).
\end{align}
Once we have the matrix $\cF(t)$ defining the quadratic form in $\lgh$ in the exponential of the generating functional $\cW(\lgh,t)$ for the integrand \eqref{eq:ghost amplitude}, we can plug it in the above map which produces the Fock space amplitudes from the ones in Moyal space. Then we can rewrite the matrix element $\bR_{nm}(t)$ corresponding to $\displaystyle \langle \hat{\Omega}|\hat{c}_{m}\hat{b}_{n}|\mathcal{T}(t)\rangle$ (or equivalently the perturbative monoid element $p^b_{2n}p^c_{2m}A_0^{gh}$ for the purely even parity case, etc.) as
follows:
\begin{subequations}\label{eq:Rnm in terms of Fnm}
	\begin{empheq}[box=\widefbox]{align}
	\bR_{2n,2m}&=-\left( \delta_{nm}+\frac{4}{2m}\cF^{pp}_{2n,2m}\right)\\
	\bR_{2n-1,2m-1}&= \delta_{nm}+(2n-1)(R\cF^{xx}R^\top)_{2n-1,2m-1},\\
	\bR_{2n,2m-1}&= -2i(R\cF^{xp})^\top_{2n,2m-1},\\
	\bR_{2n-1,2m} &= -2i(\ko R \cF^{xp})_{2n-1,2m}.
	\end{empheq}
\end{subequations}
where the upper indices on $\cF$ refer to the $N\times N$ blocks in the $2N\times2N$ matrix $\cF(t)$ belonging to the phase space representation used, namely ``momenta''  $p^cp^b$, ``position'' $x^cx^b$ and the mixed cases. The negative sign in the first equation (and implicit in the following) is due to the particular way the ghost zero mode $\xi_0$ is incorporated in the Moyal basis. This gives  a normalization constant $(-\mu_3^{-1})$ (\S\ref{subsec:notations}) that absorbs the extra negative sign.

From the expression for $\cF$, \eqref{eq:F matrix} , we notice that the matrix elements in the purely \textit{momentum} sector, $\cF^{pp}$ are particularly simple since the $N \times N$ block matrices in $\cY$ that contribute to the product are all \emph{diagonal} matrices. Hence, the infinite summations are sidestepped. By using the above map, we find that these correspond to the purely even parity elements of the $\bR$ matrix. In \S\ref{subsec:evenparity}, we shall study the behaviour of these class of matrix elements more closely by taking advantage of the simple forms for the $T$, $R$ matrices(in the infinite $N$ limit).

Because of the twist symmetry of the Witten type vertex $\langle V_3|$ and the reflector $\langle \widetilde{V}_2|$, we have vanishing of the mixed parity elements $\bR_{eo} = 0 = \bR_{oe}$. This requires that block  $\cF^{xp} = 0$, which then translates to a linear constraint relating the three blocks\footnote{Since $\cQ_\eta$ is symmetric and  $\displaystyle (\Mgh)^{-1}f_2(q)$ is already block diagonal, it suffices to consider only three independent blocks.}  in $\cQ_\eta^{-1}$, 
\begin{equation}\label{eq:twist constraint}
\cL_\eta^{\top x\alpha}(\cQ_\eta^{-1})^{\alpha\beta}\cL_\eta^{\beta p} = \mathbb{0}.
\end{equation}
We can now express the relation \eqref{eq:Rnm in terms of Fnm} as:
\begin{align}\label{eq:R explicit}
\bR &= -C+\sigma \left[\begin{array}{cc}
\ko R&  \mathbb{0}\\ 
\mathbb{0}&\mathbb{1} 
\end{array} \right]\cF \left[\begin{array}{cc}
R^\top& \mathbb{0} \\ \mathbb{0}
& -4\ke^{-1}
\end{array} \right]\sigma\nonu
&=-C + C+\left[\begin{array}{cc}
q^{2\ke}&  \mathbb{0}\\ \mathbb{0}
& -q^{2\ko}
\end{array} \right]+\sigma \left[\begin{array}{cc}
\ko R&  \mathbb{0}\\ 
\mathbb{0}&\mathbb{1} 
\end{array} \right]\cL_\eta^{\top}\cQ_\eta^{-1}\cL_\eta \left[\begin{array}{cc}
R^\top& \mathbb{0} \\ \mathbb{0}
& -4\ke^{-1}
\end{array} \right]\sigma\nonu
&=\left[\begin{array}{cc}
q^{2\ke}&  \mathbb{0}\\ \mathbb{0}
& -q^{2\ko}
\end{array} \right] + \aqua \sigma \left[\begin{array}{cc}
f_1(\ko)T^\top&  -\frac{i}{2}\ko f_3(\ko)R\\ \frac{i}{2}f_3(\ke)
& \aqua \ke f_1(\ke)\end{array}\right]\cQ_\eta^{-1}\left[\begin{array}{cc}
T\ko^{-1}f_1(\ko)& -2i \ke^{-1}f_3(\ke) \\ -\frac{i}{2}R^\top f_3(\ko)
& -f_1(\ke)
\end{array} \right].
\end{align}
Here, we have inserted the $\sigma$ matrices simply to interchange the two blocks on the diagonal in order to match our conventions for the parity basis. We have written the above to show that the the $\cL_\eta$ matrices do not result in two more infinite sums---but only one extra infinite sum---which gets simplified by using the $T R = \mone_e$, $R T = \mone_o$ relations after the matrix inverse is expanded as a formal operator series as we do in \S\ref{subsec:evenparity} for the purely even parity case.

\subsection{Matrix elements to linear order}\label{subsec:linear order}
\vspace{-2mm}
One of the interesting results from our analysis is that our starting expressions correctly reproduce the linear order behaviour of the matrices $\bR_{nm}(t)$ and $\bM_{nm}(t)$ that appear in the definition of the one-loop tadpole state in \eqref{eq:tadpoleket} as expected from BCFT. The oscillator and the Moyal formalism are formally isomorphic but this is one of the instances where the subtleties in the definition of the propagator and level truncation result in different forms. It is difficult to say where exactly the isomorphism breaks down but it may be attributable to the level truncation which breaks the gauge symmetry of OSFT and the peculiar nature of the Virasoro zero mode operator $\hat{L}_0$ in the Moyal basis\cite[\S7]{Erler:2003eq}. It is interesting\footnote{Due to non-associativity, a factor of $2$ issue arises also in the computation of the closed string tachyon mass \cite{Hashimoto:2001sm} through the Ellwood-Hashimoto-Itzhaki-Zwiebach invariant.} that the difference for the linear correction term from the two methods is only a factor of $2$.

\subsubsection*{\bf Verification of the linear behaviour}
\underline{{\bf Zeroth Order}}\\
For $t=0$, the matrix $\cF$ becomes simply
\begin{align}
\cF\rvert_{t=0}&= \half M_0^{gh-1}f_2(\ktgh; q)\rvert_{q=1} \nonu&= -\left[\begin{array}{cc}
T\ko^{-1}T^\top&  \mathbb{0}\\ \mathbb{0}
& \aqua \ke
\end{array} \right]\times \left[\begin{array}{cc}
R^\top(1+1)T^\top& \mathbb{0} \\ \mathbb{0}
& \mathbb{2}
\end{array} \right]\nonu
&=-\left[\begin{array}{cc}
2T\ko^{-1}T^\top&  \mathbb{0}\\ \mathbb{0}
& \half \ke
\end{array} \right],
\end{align}
by quickly noting that $f_2(n; 1) = 2, T\,R=\mone_e$ and $R\,T = \mone_o$.
This when substituted into \eqref{eq:Rnm in terms of Fnm} gives the $t$ independent piece to be
\begin{equation}
\bR_{nm}\rvert_{t=0}= (-1)^n\delta_{nm}=C_{nm}
\end{equation}
and by a similar short calculation, we can show that
\begin{equation}
\bM_{nm}\rvert_{t=0}=(-1)^{nm}\delta_{nm}=C_{nm}
\end{equation}
for the matter sector. Here we recall that $C$ is the twist matrix which is crucial in defining the reflector vertex $\langle \tilde{V}_2|$ and arises from BPZ conjugation and the Witten style overlapping conditions. These precisely correspond to the closed string tachyon state \eqref{eq:closedtachyon} which dominates due to the divergence structure arising from the determinant factor near $t=0$.

Here, we have assumed that there are no extra poles from the infinite summations in $\cY^\top(t)\cZ(t)^{-1}\cY(t)$ that cancels the single power of $t$ multiplying it. This will certainly be true for $\bR_{2n,2m}$ associated with the diagonal blocks in $\cY(t)$ but can also be seen to hold for $\bR_{2n-1,2m-1}$ by examining the block structure in  \eqref{eq:R explicit}.  But more importantly, we can take this as the correct prescription since it matches with the BCFT prediction for the structure of $|\cT(t)\rangle$!

\medskip

\noindent
\underline{{\bf First Order}}\\
\medskip
Interchanging the order of summation over $\ko$ (odd integers) and the non-negative integers defining the exponentials $e^{-t}$ of $f_i(n; t)$, appearing in the various blocks in $\cQ_\eta$ and $\cL_\eta$ given in \eqref{eq:ghost blocks}, we can expand them to the lowest order in the parameter $t$:
\begin{align}
\cQ_\eta & = \frac{t}{4}  \left[\begin{array}{cc}
2(\mathbb{1}+TT^\top)&  \mathbb{0}\\ \mathbb{0}
& \aqua 4\ke^2
\end{array} \right]+\cO(t^2)\nonu
&=t\left[ \begin{array}{cc}
\mathbb{1}-\half\frac{ww^\top}{1+w^\top w}&  \mathbb{0}\\ \mathbb{0}
& \aqua \ke^2
\end{array} \right]+\cO(t^2),\\
\cL_\eta&= \frac{t}{2} \left[
\begin{array}{cc}
\mathbb{0}&  i \ke\\ 
-i R^{\top} \ko T^\top & \mathbb{0}
\end{array} \right] + \cO(t^2),
\end{align}
where we have used the relations:
\begin{equation}
TT^\top = \mone -\frac{ww^\top}{1+w^\top w},\qquad R = \ko^{-2}T^\top \ke^2,
\end{equation}
and the off-block diagonal elements in $\cQ_\eta$ do not contribute since $f_1(n; t)$ starts at $\cO(t^2)$. The quantity $w^\top w$ diverges linearly as $\cO(N)$ and expressions involving it should be treated with care to avoid inconsistencies. Hence, we shall keep the $\cO(1/N)$ term and argue when it may be dropped. The matrix
\begin{equation}
\mathbb{V}\coloneqq \mathbb{1}-\half \frac{ww^\top}{1+w^\top w}
\end{equation}
appearing in the first block of $\cQ_\eta$ above can be readily inverted using a Taylor series in $1/\bar{w}w<1$ when $N\geq1$. We make the following ansatz involving a function $\mu(z)$:
\begin{equation}
\mathbb{V}^{-1}_{2n,2m}= \delta_{nm}+\mu(w^\top w) ~w_{2n}(w^\top)_{2m}
\end{equation}
and require $\mathbb{V}\mathbb{V}^{-1}=\mathbb{1}=\mathbb{V}^{-1}\mathbb{V}$ to find 
\begin{equation}\label{eq:W matrix}
\mathbb{W} \coloneqq \mathbb{V}^{-1} = \mathbb{1}+\frac{ww^\top}{2+w^\top w},
\end{equation}
which may then be verified by a direct substitution. Then we find that
\begin{equation}
\cQ_\eta^{-1}=\frac{1}{t}
\left[\begin{array}{cc}
\mathbb{W}& \mathbb{0}\\ \mathbb{0}
& 4 \; \ke^{-2}
\end{array} \right]+\mbox{~finite}+\mbox{~subleading},
\end{equation}
showing that:
\begin{align}
\cL_\eta^\top \cQ_\eta^{-1}\cL_\eta &= - \frac{t}{4}\left[\begin{array}{cc}
4T\ko R \ke^{-2}R^\top \ko T^\top&  \mzero\\ \mzero
& \ke \mathbb{W}\ke
\end{array} \right] +\cO(t^2)\nonu
&= -t \left[\begin{array}{cc}
TT^\top& \mzero \\ \mzero
& \aqua \ke \mathbb{W}\ke\end{array} \right] +\cO(t^2)\nonu
&=-t\left[\begin{array}{cc}
\mone-\frac{ww^\top}{1+w^\top w}&  \mzero\\ \mzero
& \aqua \left(\ke^2 +\frac{\ke w w^\top \ke}{2+w^\top w}\right)
\end{array} \right] +\cO(t^2).
\end{align}
Now we can consider the open string limit for the second block since there are no divergent terms in this expansion, while we retain the $TT^\top$ form for the first block. Isolating the linear term from $\frac{1}{2}(\Mgh)^{-1}f_2(q)$, we obtain
\[+t \left[\begin{array}{cc}
2 \,T\,T^\top&  \mzero\\ \mzero
& \half \ke^2
\end{array} \right]\]
This when substituted into \eqref{eq:F matrix} leads to:
\begin{equation}
\cF= -\left[\begin{array}{cc}
2T\ko^{-1}T^\top - t \, TT^\top&  \mzero\\ \mzero
& \half \ke -\frac{t}{4}\,\ke^2
\end{array} \right] + \cO(t^2).
\end{equation}
Consequently, we can readily write down the squeezed state matrix $\bR_{nm}$ to this order using \eqref{eq:Rnm in terms of Fnm}:
\begin{align}
\bR_{2n,2m}& = -\delta_{nm}-\frac{4}{2m}\times-\half 2n \delta_{nm}-\frac{4t}{2m}\times \frac{1}{4}4n^2\delta_{nm}\nonu
&=\delta_{nm}-2n t \,  \delta_{nm}+\cO(t^2),\\
\bR_{2n-1,2m-1}& = \delta_{nm}+(2n-1)\left[R\left(-2T\ko^{-1}T^\top +tTT^\top\right)R^\top\right]_{2n-1,2m-1}\nonu
&=-\delta_{nm}+(2n-1)t \, \delta_{nm}+\cO(t^2).
\end{align}
The mixed parity cases $\bR_{2n,2m-1}$ vanishes identically as we have argued before. This enables us to express the general matrix element as:
\begin{equation}\label{eq:R linear behaviour}
\bR_{nm}^{(\rm{Moy})}=C_{nm}- n\, C_{nm}\,t+\cO(t^2)
\end{equation}
As shown in \cite{Ellwood:2003xc},  the linear correction in $t$ is completely generated from the conformal transformation of the external Fock space state, and is  determined from a BCFT analysis of the conformal map done near $t=0$. 

It precisely coincides with the above form, whereas a Taylor expansion based on the oscillator expressions gives a linear coefficient off by a factor of $2$:
\begin{equation}
\bR_{nm}^{(\rm{osc})} = C_{nm} - \underline{2} n\, C_{nm}\,t + \cO(t^2).
\end{equation}
As explained carefully in \cite{Ellwood:2003xc}, the two limits involving the level (size of the matrices) and the modular parameter, $L\to \infty$ and $t\to 0$ do not commute in the oscillator case but holds in the Moyal case at least to the order that we have analysed As inferred earlier in this thesis,
the difference in the propagator structure could account for this subtle breakdown of the isomorphism. Hence, the peculiar structure of the propagator in Moyal space merits further investigation. See also the interesting discussion in \cite[\S7]{Erler:2003eq}.

One may also verify this behaviour numerically by repeating the analysis done in \cite{Ellwood:2003xc} for finding a numerical fit near $t=0$. Here we have used the finite $N$ versions of the matrices \eqref{eq:finiteNmatrices} which ensure that the star algebra relations are satisfied. For $N = 84$ (requiring inversion of $168\times 168$ matrices) and $t$ varying from $10^{-4}$ to $16\times 10^{-4}$ in steps of $10^{-4}$, we obtain the linear fit given in Table \ref{tab:linearfit}.
\begin{table}[h!]
	\centering
	\[	\begin{array}{c| c}
	\hline
	\bR_{11}^{(lin)} & -(0.99999974-0.99912772\times t) \\
	\bR_{22}^{(lin)} & +(0.99999858-0.99010797\times 2t)  \\
	\bR_{33}^{(lin)} &-(0.99999768-0.99743118\times 3t)  \\
	\bR_{44}^{(lin)} & +(0.99999516-0.98843886\times 4t)  \\
	\hline
	\end{array}\]
	\caption{Linear behaviour of the matrix elements $\bR_{nm}$ near $t=0$ based on numerical evaluation of $168\times168$ size matrices. The fit reinforces the agreement between the Moyal and the BCFT predictions for the structure of $\bR(t)$.}
	\label{tab:linearfit}
\end{table}
We emphasize that the higher order terms starting at $t^2$ are the ones that really encode any effects of the  Shapiro-Thorn massless closed-string states. Unfortunately, our algebraic approach only allows to successively approximate these coefficients (as we do in \S\ref{subsec:evenparity}) but not exactly. It does clarify the discrepancy noticed in the oscillator case and is in that sense an improvement. However, we remark that in the geometric approach based on a BCFT analysis, it is difficult to isolate their effects as well due to operator mixing under a  conformal transformation of non-primary operators. 

\subsection{Associativity at linear order}\label{subsec:nonassoc}

The various orders for evaluating the overlap mentioned in \S\ref{subsec:ghost sector} could differ if associativity is not strictly satisfied. The alternate order $\displaystyle A_2\ast A_3 \to A_{23}(\eta,\xi) \to \int \, d\eta \to A_1 A^\prime_{23}(\xi)\to \Tr $ corresponds to the manner in which the amplitude would be evaluated in the oscillator method (see \S\ref{subsec:osc results}) where the tadpole state is evaluatedt as:
\begin{equation}\label{eq:osc tadpole ket}
|\cT\rangle = -g_T \, K^3 \int_0^\infty dt \,\, _{1,2}\langle\widetilde{V}_2|b_0^{(2)}e^{-\half t(L_0^{(1)}+L_0^{(2)})}|V_3\rangle_{1,2,3}.
\end{equation}
and the amplitude is obtained by taking the inner product with an external state $\bra{A_e}$. Here the superscripts refer to the string Hilbert spaces in the first quantized formalism.

The corresponding matrix $\cF(t)$ defining the quadratic form in $\lgh$ towards the generating functional, $\cW(\lgh,t)$ in this particular order of evaluation is then
\begin{align}
\cF(t) &= (M_{123}^{\prime})^{-1}, \mbox{~with}\nonu
M_{123}^{\prime} & =  \frac{2}{f_2(q)}M_0^{gh} + \left(\frac{f_1}{f_2}+\frac{f_3}{f_2} m_0^{gh}\right)\left(M_0^{-1}\frac{f_3}{f_2}+\sigma \frac{f_3}{f_2}M_0^{gh}\sigma - 2\sigma \frac{q^{\ktgh}}{f_2(q)}+2\left(\frac{q^{\ktgh}}{f_2}\right)^\top\sigma\right)^{-1}\simbreak \left(\frac{f_1}{f_2}+\frac{f_3}{f_2} m_0^{gh}\right)^\top
\end{align}

\medskip
\noindent
A quick inspection of the above structure reveals that similar to the earlier evaluation order, the matrix to be inverted in $M_{123}^\prime$ \emph{vanishes} at $t=0$. Collecting the linear order terms after some simple algebra results in an identical expression for the linear correction term, namely:
\begin{equation}
\bR_{nm} = C_{nm}-nC_{nm}t +\cO(t^2),
\end{equation}
and hence we conclude that the order of limits problem does not arise in this order of evaluation either. However, further expansions are made awkward by the somewhat complicated form of the above expression, which requires two matrix inverse operations nested one inside the other.

Hence, from the above exercice we can infer that constructing the tadpole state out of the Fourier basis and then combining their contribution to the overlap amplitude (by the $\egh$ integration as we have done earlier) would be preferred over considering the overlap with the state itself, which may be somewhat counter-intuitive. Of course, just the linear order behaviour does not fix a prescription uniquely or prove the correctness of these expressions. Nonetheless, this is an encouraging result showing the subtleties in the map between Moyal space and Fock space. 

It would have been more interesting if associativity was indeed violated in this calculation, which could display the similarity to the oscillator inner product directly. Hence, we have not been able to clarify the order of limits issue completely.
\subsubsection*{\bf Fourier Basis}
\vspace{-2mm}
We must remark that the issue encountered in the oscillator basis also arises if one attempts to expand the amplitude in Fourier space defined by the conjugate variable $\egh$. The Feynman rules in Fourier space were studied and given in detail in Refs. \cite{Bars:2002qt,Bars:2003gu} by Bars et al. The propagator and vertex take of the form:
\begin{align}
&\Delta(\egh,\eta^{\prime gh},t) \sim \exp\left[\begh F^{gh}\egh+\eta^{\prime gh\top}F^{gh}(t)\eta^{\prime gh\top}-2\eta^{\prime gh\top}G^{gh}(t)\egh\right], \mbox{~and}\\
&\Tr\left(e^{-\bxigh\eta^{gh}_1}\ast\cdots\ast e^{-\bxigh \eta^{gh}_n}\right)\sim
\exp\left[-\half\sum_{i<j}\begh_i\sigma \eta^{gh}_j\right]\delta\left(\egh_1+\cdots + \egh_n\right),
\end{align}
respectively, where we are now using the $2N\times 2N$ basis with implicit $-i \varepsilon$ metric and 
\begin{equation}
F^{gh}(t)=-\aqua M_0^{gh-1}\frac{f_2(q)}{f_3(q)},\qquad G^{gh}(t)=-\half M_0^{gh-1} \frac{q^{\ktgh}}{f_3(q)}
\end{equation}
One may again write down an amplitude formally and as $t\to 0$, the matrix to be inverted becomes singular simply due to the linear dependence of the blocks
\begin{equation}
\cQ(t) \sim \frac{1}{t}\left[
\begin{array}{cc}
1&  1\\ 1
& 1
\end{array} \right] \otimes M_0^{gh-1}
\end{equation}
where the first block has vanishing determinant. Hence this form cannot be used as the starting point for a systematic series expansion around $t=0$. However, we remark that for numerical purposes, the Fourier basis provides quicker analytic expressions since the $\ast$ products are already taken care of. The disadvantage is numerical instability due to using much bigger sized matrices as compared to the $\xi$ basis.

\noindent
Thus in summary, we have demonstrated in this section that the expected behaviour from BCFT is correctly reproduced by the Moyal expressions in $\xi$ space. For showing the validity of the relations, we have used the map \eqref{eq:betamap} from th oscillators as operators in Fock space to differential operators in Moyal space.  One really interesting aspect is the non-analyticity of these matrix elements already seen at the quadratic stage: the higher order terms come with factors of $\log(t)$ as we shall show later in \S\ref{subsec:evenparity}. Hence, even the expansion in BCFT can at best be asymptotic and thus allows for explicitly including the closed-string states in the form of exponentially suppressed subleading tails in the form of $e^{-2\pi^2 n/t}$.


\section{Expansions for squeezed state matrix elements}\label{sec:squeezed}
In this section, we wish to study the behaviour of the matrix element factor $\bR_{nm}$---defining the squeezed state in the (ghost) exponential factor of the integrand as appearing  in \eqref{eq:tadpoleket}---near $t = 0^+$  using expansions in various basis functions. Naturally, one can find an absolutely convergent expansion in the nome $q\coloneqq e^{-t}$ for $|q|<1$, corresponding to open string degrees of freedom. Because of  the essential singularity at $t = 0 $ coming from the massive closed string states, the expansion in other basis functions such as $\{t^s,\ln t\}$ or equivalently $\{(-\ln q)^s ,\ln(-\ln q)\}$ would not be a convergent expansion, but could at best be an asymptotic expansion.  This is consistent with our understanding of the \emph{quantum inconsistency} of bosonic OSFT (or any open bosonic string theory) at the loop level.

In the following, we simply explore the utility of the Moyal formulation to directly learn about the structure of the integrand as a function of the parameter $t$. The expressions we obtained in  \S\ref{subsec:ghost sector} do  have the correct qualitative features near $t=0$ and we found that it reproduces the correct zeroth and linear order coefficients, which is somewhat non-trivial. Furthermore, we can develop a series expansion involving special functions to successively approximate the true analytic form for $\bR_{nm}(t)$ by our method.

\subsection{Even parity matrix elements near $t \to 0^+$}\label{subsec:evenparity}
In order to perform an expansion in $t$, let us introduce the following auxiliary functions derived from the functions $f_1, f_2, f_3$ and  $f_4$ employed earlier \eqref{eq:f1234 definitions}
\begin{align}\label{eq:auxfns}
h_i(n; t) &\coloneqq \frac{f_i(n; t)}{t}, &g_i(n; t)&\coloneqq h_i(n;t )-h_i(n; 0)= \frac{f_i(n; t)}{t}-f^{(1)}_i (0), \mbox{~~giving explicitly }\nonu
h_1(n; t)& = \frac{(1-e^{-nt})^2}{t}, &h_2(n; t)&= \frac{1+e^{-2nt}}{t},\nonu
h_3(n; t)& = \frac{1-e^{-2nt}}{t},&h_4(n; t)&= \frac{3-4e^{-nt}+e^{-2nt}}{t}, \mbox{~~and}\nonu
g_1(n;t)& = \frac{(1-e^{-t n})^2}{t} , &g_2(n; t)&=  \frac{1+e^{-2nt}}{t}+2n,\nonu 
g_3(n; t) &= \frac{1-e^{-2nt}}{t} -2n, &g_4(n; t)&= \frac{3-4e^{-nt}+e^{-2nt}}{t} -2n.
\end{align}
These can be thought of as certain basis functions with a well-defined asymptotic behaviour. Also, we notice that $\lvert g_i(n; t)\rvert < 2n$ for $i = 1, 3, 4$; a boundedness property which we will use later.

We remark at this juncture that the functions $\tanh t$ and $\sech~t$  which appear in the original form of the block matrices \eqref{eq:eta coeffs} may be Taylor expanded in terms of the Bernoulli numbers $B_{2n}$ and the Euler numbers $E_{2n}$ around the point $t = 0$. However, reorganizing  the multiple sums and products followed by applying any identities involving them quickly becomes challenging. Therefore, we continue to use the much more straightforward (and uniform) representation in terms of exponential functions in our analysis.

\bigskip
Moving on, we illustrate this expansion scheme for the case of even parity matrix elements, $\bR_{2n,2m}$, for convenience. Since its expression involves the inverse of a matrix function, which is difficult to obtain analytically (at least in the discrete diagonal basis), we employ a formal  series to represent the inverse\footnote{This is justified because the domains of analyticity of the two maps overlap.}. After this step, one can find expansions around the point $t=0$ although the sub-matrices do lead to more terms without any apparent patterns for resummations. To this end, we split the the matrix to be inverted $\cZ(t)$, which appeared in \eqref{eq:cZ, cY defined} as follows:
\begin{align}
\cZ(t)&= \cZ_0+\delta\cZ(t), \mbox{~~ so that we may write}\nonu
\cZ(t)^{-1}&= (1+\cZ_0^{-1}\delta \cZ(t))^{-1}\cZ_0^{-1}\nonu
&\coloneqq (1+\bM(t))^{-1}\cZ_0^{-1}\nonu
&= \sum_{s=0}^{\infty}(-1)^s {\bf M}^s(t) \times \cZ_0^{-1}.
\end{align}
where we have defined a matrix function $\bM(t)\coloneqq \cZ_0^{-1}\delta\cZ(t)$. Here we recall that $\cZ_0$ must be defined as the limit $\displaystyle \lim_{t\to0^+}\frac{\cQ_\eta(t)}{t}$ but note that the matrix $\cQ_\eta(t)$ in \eqref{eq:ghost blocks} is not analytic at $t=0$ due to the insufficient fall-off behaviours of the $T_{2n,2m-1}$ matrix elements as $n, m$ increases
\begin{equation}
T_{2n,2m-1}^\infty = (-)^{n+m}\frac{4 (2m-1)}{\pi (4n^2-(2m-1)^2)}
\end{equation}
which behave like $\displaystyle \frac{1}{2m-1}$ in sums for large $m$. 

\medskip
We also remind the reader  of the block matrix forms from \eqref{eq:W matrix}
\begin{equation}
\cZ_0=\left[\begin{array}{cc}
4\mathbb{V}&  \mathbb{0}\\ \mathbb{0}
& \ke^2
\end{array} \right],  \mathbb{V}\coloneqq  \mathbb{1}-\frac{1}{2}\frac{ww^\top }{1+w^\top w}, \mbox{~and}	\qquad \mathbb{W} \coloneqq \mathbb{V}^{-1} = \mone+\frac{ww^\top}{2+w^\top w},
\end{equation}
giving
\begin{equation}
\cZ_0^{-1}=\left[\begin{array}{cc}
\aqua \mathbb{W}&  0\\ 0
& \ke^{-2}
\end{array} \right].
\end{equation}
The matrix $\delta \cZ(t)$ is then expressed in terms of functions $g_1, g_3$ and $g_4$ in a form very similar to $\cZ(t)$:
\begin{align}
\delta\cZ(t) &= \left[ \begin{array}{cc}
\ke^{-1}g_3(\ke)+T\ko^{-1}g_4(\ko)T^\top &  -\frac{i}{2}\left(g_1(\ke)-Tg_1(\ko)R\right)\\
-\frac{i}{2}\left(g_1(\ke)-R^\top g_1(\ko) T^\top\right) & \frac{1}{4}\left(\ke g_4(\ke) + R^\top\ko g_3(\ko) R\right)
\end{array} \right]\nonu
\Rightarrow \bM(t) &= \left[ \begin{array}{cc}
\frac{1}{4}\mathbb{W}\left(\ke^{-1}g_3(\ke)+T\ko^{-1}g_4(\ko)T^\top  \right)&  -\frac{i}{8}\mathbb{W}\left(g_1(\ke)-T\ko^{-2}g_1(\ko)T^\top \right)\\
-\frac{i}{2}\left(\ke^{-2}g_1(\ke)-T\ko^{-2}g_1(\ko) T^\top\right)& \frac{1}{4}(\ke^{-1} g_4(\ke) + T\ko^{-3} g_3(\ko) T^\top\ke^2)
\end{array} \right],
\end{align}
where we remind the reader that \emph{each matrix element} is in general an infinite sum --- owing  to the matrix products --- and we have used the relation $R = \ko^{-2}T^\top \ke^2$ for rewriting the structure using only the $T$ and $\kappa$ matrices. However, in taking powers of the matrix $\bM$ symbolically it is more helpful to keep the matrix $R$ since then we can readily apply relations such as $R\, T = \mone_o$ and $T \,R =\mone_e$ in order to  reduce the number of terms.

Note that although the individual blocks in $\bM(t)$  have at least a  first order zero at $t = 0$, the products of these blocks still \emph{retain only a first order zero} due to the higher order poles arising from the infinite sums. This is because we are interchanging the order of summation in double sums which are not absolutely convergent. Therefore all the higher matrix powers $(\bM(t))^s$ continue to contribute to the $t^2$ term in $\bR_{nm}(t)$ in our expansion scheme and consequently these coefficients cannot be obtained exactly by the above series. This drawback is again due to the infinite dimensional nature of the problem.

Because of the logarithmic branch points, the terms for various $s$ are not analytic, although they vanish at $t = 0$ as remarked above. However, we expect that the contributions fall off with increasing values of $s$ (as seen from the tractable $s=0, 1, 2$ cases) and must converge since the full function only has a removable singularity as $t \to 0^+$. The matrix $\cY(q)$ presented in \eqref{eq:cZ, cY defined} is now written in terms of the functions $h_1(n; q)$ and $h_3(n; q)$ as follows:
\begin{equation}
\cY(q)=\left[\begin{array}{cc}
T\ko^{-1}h_1(\ko)T^\top&\frac{i}{2}h_3(\ke)  \\ 
-\frac{i}{2}R^\top h_3(\ko)T^\top&\aqua \ke h_1(\ke) 
\end{array} \right].
\end{equation}

\medskip
\noindent
We shall now try to investigate the effect of working with a finite size truncation for the matrices vs directly using the infinite $N$ versions of the expressions. Because the functions involved in the infinite sums satisfy the \emph{boundedness} property: $\lvert g_i(n; t)\rvert < 2n$ for $i = 1, 3, 4$, we find that on examining the structure of the   matrix powers $\bM^s$, the contribution from the extra term $\displaystyle \frac{ww^\top}{2+w^\top w}$ in $\mathbb{W}$ remains subleading and always goes as $N^{-p}$ for some $p \geq 1$. Since we only work with the partial sums for defining the  series representation of the inverse, i.e $s= 0, 1, \ldots, S$, say, these terms do not add up to give extra finite $N$ corrections.  Consequently, we can  drop these $\cO(1/N^p)$ extra terms from our calculations and effectively set $\mathbb{W}=\mone$ to do the relevant infinite sums over odd/even integers. In other words, we have made a choice of order of limits that allows us to use the infinite $N$ expressions consistently.

Now, in order to study these partial sums using a series representation in $t$, we shall now comment on their analytic structure. As the functions arising from the infinite sums over the \emph{odd/even} parity indices are uniformly convergent only for $\Re(t)>0$, term by term differentiation is not justified. Such mathematical niceties would have existed if we kept $N <\infty$ but then one misses the nice non-analytic behaviour expected in the quantum theory which signals the inconsistency attributed to closed string states. In the following, we therefore work directly in the open string limit.  In addition, as we are expecting only an asymptotic expansion due to physical reasons, it may be possible to justify sending $N\to \infty$ at this stage of the calculation for practical reasons.

In concrete terms, the above procedure would result  in an expansion of the form:
\begin{equation}
\bR_{nm}(t) = \sum_{r=0}^\infty (\Lambda_r + \log (t)\tilde{\Lambda}_r)t^r 
\end{equation}
where the coefficients $\Lambda_r$ and $\tilde{\Lambda}_r$ receive contributions from the partial sums over $s$ and the $\log(t)$ piece will be shown to result from the non-analytic behaviour of the special functions that arise.

There are more non-analytic terms than the simple $\log(t)$ dependence that can be admitted (see \eqref{eq:subleading}) since we do not have absolute convergence and hence the individual coefficients $\Lambda_r$ and $\tilde{\Lambda}_r$ may not all exist. We are at this point only looking for hints of non-analytic behaviour and cannot rigorously account for any missing subleading terms.

\medskip 
After these digressions, let us return to the series expansion at hand. In the following we illustrate the general procedure and also display some coefficients that contribute to the final matrix elements. We shall denote the expansion for $\bR(t)$ in terms of the matrix products by a sequence of functions
\begin{equation}
\bR(t)=\sum_{s=0}^\infty \cR^{(s)}(t),
\end{equation}
where we have chosen $\cR^{(0)}$ to match the linear order expansion we derived in \S\ref{subsec:linear order}. We emphasize that this sequence of functions  constructed out of hypergeometric functions does \emph{not} furnish an asymptotic basis as can be seen from the basic criteria for the gauge functions
\begin{equation}
\phi_{n+1}(z) = o(\phi_n(z)), \qquad\left(\mbox{as~~} \frac{1}{z}\to +\infty\right)
\end{equation}
not being satisfied by these. Except the first two terms, the rest all contribute starting at $\cO(t^2)$ and consequently these provide only an asymptotic approximation to the true function.

The simplest block to look at is the \emph{purely even} block $\bR_{2n,2m}$ given by \eqref{eq:Rnm in terms of Fnm} which we provide here again:
\begin{align}
\bR_{2n,2m}=&-\left( \delta_{2n,2m}+\frac{4}{2m}\cF^{pp}_{2n,2m}\right)
\end{align}
This is because it involves $\cZ^{-1}$ sandwiched between $\cY^{xp}$ and $\cY^{pp}$ which are diagonal matrices and hence is easy to keep track of in a power series expansion. In block matrix form, the matrix $\cF^{pp}$ from \eqref{eq:F matrix} corresponding to the even parity elements is given by:
\begin{align}
\cF^{pp}&=-\aqua \ke f_2(\ke)+t \;(\cY^{\top}\cZ^{-1}\cY)^{pp},\mbox{~~where we can expand}\nonu
(\cY^{\top}\cZ^{-1}\cY)^{pp}
&=\cY^{\top px}(\cZ^{-1})^{xx}\cY^{xp}+\cY^{\top px}(\cZ^{-1})^{xp}\cY^{pp}+\cY^{\top pp}(\cZ^{-1})^{px}\cY^{xp}+\cY^{\top pp}(\cZ^{-1})^{pp}\cY^{pp} \nonu
&= -\aqua h_3(\ke)(\cZ^{-1})^{xx}h_3(\ke)+\frac{i}{8}h_3(\ke)(\cZ^{-1})^{xp}\ke h_1(\ke)\nonu
&\qquad{}+\frac{i}{8}\ke h_1(\ke)(\cZ^{-1})^{px}h_3(\ke)+\frac{1}{16}\ke h_1(\ke)(\cZ^{-1})^{pp}\ke h_1(\ke)
\end{align}
The matrix powers $\bM, \bM^2,\ldots$ required for implementing this procedure requires some block matrix multiplications. These can be performed  using the \emph{NCAlgebra} package \cite{NCAlgebra}   and recursively applying the relations satisfied by the $T,R$ matrices such as 
\[T \,R = \mone_e, \qquad R\,T = \mone_o,\qquad T^\top T = \mone_o - vv^\top, \mbox{~etc.}\]
using the ``NCReplace'' series of commands\footnote{The $sth$ power would give $2\times2$ blocks where each block is a sum of $2^{s-1}$ terms. Each such term is a product of $s$ elements from the matrix $\bM$ which in turn have sub-structure.}.

We are not at this point able to explicitly resum the series and demonstrate that this converges  but it is still instructive to look at the functional behaviour of each of these contributing terms separately.

\subsection{Illustrations for geometric series}\label{subsec:geometric}
We have therefore obtained a few lower order terms by this method when the expressions reduce to a sum of terms with infinite sum over a single index (the odd integers). At higher values of $s$, there are many terms which still involve only a single infinite sum but the few remaining terms involving double and triple sums (over both even and odd integers) lead to computational problems. 

In the following, the integer in the superscript corresponds to the power $s$ in the series expansion for the inverse.
\vspace{-2mm}
\subsubsection*{$s=0$ term}\label{subsubsec:seqzero}
\vspace{-2mm}
The first term in the expansion corresponding to $s=0$ is given by:
\begin{align}
\cR_{2n,2m}^{(0)}&= -\delta_{nm}-\frac{4}{2m}\left[-\aqua 2n f_2(2n; t)\delta_{nm}+\frac{t}{16}\left(h_1(2n; t)^2-h_3(2n; t)^2\right)\delta_{nm}\right]\nonu
&=\left(\frac{e^{-4 n t} (n t-1)}{n t}+\frac{e^{-6 n t}}{2 n t}+\frac{e^{-2 n t}}{2 n t}\right)\delta_{nm}\nonu
&=\left(1-2 n t+6 n^3 t^3-\frac{40 n^4 t^4}{3}+\cO\left(t^5\right)\right)\delta_{nm}
\end{align}
which contributes to the leading behaviour in the even sector, namely, $C_{nm}-n C_{nm}t+\cO(t^2)$. The coefficients increase rapidly initially but then decrease as expected due to the factorial suppression.
\vspace{-2mm}
\subsubsection*{$s=1$ term}\label{subsubsec:seqone}
\vspace{-2mm}
For $s=1$, the infinite sums arising from the matrix products over the \emph{odd} integers can be performed using \emph{Mathematica}. 
\begin{align}
\cR_{2n,2m}^{(1)}&=-\frac{4}{2m}\times -t\left[\cY^\top \bM\cZ_0^{-1}\cY\right]^{pp}
\end{align}
Since the matrices $T_{2n,2m-1}$ and  $R_{2n-1,2m}$ have  a relatively simple structure expressible in terms of integers, we expect these to be in general in terms of hypergeometric functions $_JF_{J-1}$ with arguments of the form $q^k;~ k\in \mathbb{Z}_+$.\footnote{For the $s=1$ case, which is very similar to the original matrix $\cZ(q)$(\eqref{eq:QetaDiag}, \eqref{eq:QetaNonDiag}), these functions reduce to the Lerch transcendent representations and the appropriate analytic continuations---see \eqref{eq:lerch}.}

\medskip
\noindent
For the diagonal matrix elements, we obtain:
\begin{small}
	\begin{align}
	\cR^{(1)}_{2n,2n}&=\frac{\left(q^{2 n}-1\right)^2}{64 \pi^2 n^3 \log^2(q)}
	\left\{-4 q^{2 n+2} \Phi \left(q^4,1,\frac{1}{2}-n\right)+2 n
	q^{2 n+2} \Phi \left(q^4,2,\frac{1}{2}-n\right)\breakeqn
	+n q^{4 n+2} \Phi
	\left(q^4,2,\frac{1}{2}-n\right)+4 q \left(q^{4 n}-1\right) \Phi
	\left(q^2,1,\frac{1}{2}-n\right)\breakeqn
	-4 n q \left(q^{4 n}-1\right) \Phi
	\left(q^2,2,\frac{1}{2}-n\right)+4 q^2 \Phi \left(q^4,1,\frac{1}{2}-n\right)\breakeqn-3 n q^2
	\Phi \left(q^4,2,\frac{1}{2}-n\right)+4 q \Phi \left(q^2,1,n+\frac{1}{2}\right)+4 n q
	\Phi \left(q^2,2,n+\frac{1}{2}\right)\breakeqn
	-4 q^{2 n+2} \Phi
	\left(q^4,1,n+\frac{1}{2}\right)-2 n q^{2 n+2} \Phi \left(q^4,2,n+\frac{1}{2}\right)+4
	q^{4 n+2} \Phi \left(q^4,1,n+\frac{1}{2}\right)\breakeqn
	+3 n q^{4 n+2} \Phi
	\left(q^4,2,n+\frac{1}{2}\right)-4 q^{4 n+1} \Phi \left(q^2,1,n+\frac{1}{2}\right)-4 n
	q^{4 n+1} \Phi \left(q^2,2,n+\frac{1}{2}\right)\breakeqn
	-n q^2 \Phi
	\left(q^4,2,n+\frac{1}{2}\right)-16 \pi ^2 n^2 q^{2 n} \log (q)-16 \pi ^2 n q^{2 n}-8
	\gamma  q^{2 n}+14 \pi ^2 n q^{4 n}+4 \gamma  q^{4 n}\breakeqn
	+8 \log (2) \left(q^{2
		n}-1\right)^2+2 \left(q^{2 n}-1\right)^2 \psi ^{(0)}\left(\frac{1}{2}-n\right) (2 n \log
	(q)+1)\breakeqn
	-2 \left(q^{2 n}-1\right)^2 \psi ^{(0)}\left(n+\frac{1}{2}\right) (2 n \log
	(q)-1)\breakeqn
	-n \psi ^{(1)}\left(\frac{1}{2}-n\right) \left(-5 q^{4 n}-2 q^{6 n}+q^{8 n}+4 q^{2
		n} (4 n \log (q)+1)+2\right)\breakeqn
	+n \psi ^{(1)}\left(n+\frac{1}{2}\right) \left(4 q^{2 n}-5
	q^{4 n}-2 q^{6 n}+q^{8 n}+8 n \left(q^{4 n}+1\right) \log (q)+2\right)\breakeqn
	+16 q^{2 n} \tanh
	^{-1}\left(q^2\right)-8 q^{4 n} \tanh ^{-1}\left(q^2\right)+2 \pi ^2 n-8 \tanh
	^{-1}\left(q^2\right)+4 \gamma \right.\bigg\}.
	\end{align}
\end{small}
The above expression would simplify for particular integer values $n$.
A very useful series representation for understanding these special functions is given by Erd\'{e}lyi \cite{bateman1953higher}, which is valid for $|\log(z)|<2\pi$, $s=2,3, \ldots$ and  $a\neq 0, -1, -2, \ldots$
\begin{equation}\label{eq:Erdelyi}
\Phi[z,s,a] = z^{-a}\left\{\sum_{\substack{k=0\\
		k\neq s-1}}^{\infty} \zeta(s-k,a)\frac{\log^k(z)}{k!}+[\psi(s)-\psi(a)-\log(-\log(z))]\frac{\log^{s-1}(z)}{(s-1)!}\right\}
\end{equation}
where $\zeta(s,a)=\Phi(1, s, a)$ is the Hurwitz zeta function. Substituting this into the \emph{Mathematica} output would give us the $\log (t)$ dependence we wanted (as $z = e^{-\#t}$ in our case). The resulting expression can be truncated at a finite $k$ to obtain an expansion in $t$ and $\log t$ as for instance:
\begin{align}
\cR_{22}^{(1)}&=\log t \left[
\frac{16 t^4 }{\pi ^2}-\frac{64 t^5 }{\pi ^2}+\cO\left(t^6\right)\right]\nonu
&\!\!\!\!+\left[2 t^2+\frac{32}{3} \left(\frac{1}{\pi ^2}-1\right) t^3-\frac{4 t^4 \left(34-21 \pi ^2+4\log
	(2)\right)}{3 \pi ^2}+\frac{4 t^5 \left(1408-589 \pi ^2+240 \log (2)\right)}{45 \pi
	^2}+\cO\left(t^6\right)\right].
\end{align}
For $\Re(a) > 0$, the two functions defined by Lerch Transcendents and Hurwitz Lerch transcendent coincide and one can simply replace the former with the latter. This is useful since \emph{Mathematica}
is able to expand Hurwitz Lerch functions with arguments $e^{\#t}$ arguments near $t=0$. This is another way to obtain the series expansions, although it is sightly less computationally efficient.

\medskip
\noindent
Similarly, the non-diagonal elements ($n\neq m$) are expressed as:
\begin{small}
	\begin{align}
	\cR_{2n,2m}^{(1)}&=\frac{(-)^{n+m}}{32 \pi ^2 m (n^2-m^2)  \log ^2(q)}\left\{\frac{\left(q^{2 m}-1\right)^2 \left(q^{2 n}-1\right)^2}{nm}\left[-m^2 q^2 \Phi
	\left(q^4,1,\frac{1}{2}-n\right)\breakeqntwice
	+n^2 q^2 \Phi \left(q^4,1,\frac{1}{2}-m\right)-m^2 q^2
	\Phi \left(q^4,1,n+\frac{1}{2}\right)
	+n^2 q^2 \Phi \left(q^4,1,m+\frac{1}{2}\right)\breakeqntwice
	-4
	m^2 n \psi ^{(0)}\left(\frac{1}{2}-n\right) \log (q)+4 m^2 n \psi
	^{(0)}\left(n+\frac{1}{2}\right) \log (q)-m^2 \psi ^{(0)}\left(\frac{1}{2}-n\right)\breakeqntwice
	-m^2
	\psi ^{(0)}\left(n+\frac{1}{2}\right)+4 m^2 \tanh ^{-1}\left(q^2\right)+2 m^2 \psi
	^{(0)}\left(\frac{1}{2}\right)+4 m n^2 \psi ^{(0)}\left(\frac{1}{2}-m\right) \log (q)\breakeqntwice
	-4
	m n^2 \psi ^{(0)}\left(m+\frac{1}{2}\right) \log (q)+n^2 \psi
	^{(0)}\left(\frac{1}{2}-m\right)+n^2 \psi ^{(0)}\left(m+\frac{1}{2}\right)-4 n^2 \tanh
	^{-1}\left(q^2\right)\breakeqntwice-2 n^2 \psi ^{(0)}\left(\frac{1}{2}\right)\right]+\frac{\left(q^{2 m}-1\right) \left(q^{2 n}-1\right)}{nm} \left(n \left(q^{2 m}+1\right)
	\left(q^{2 n}-1\right)+m \left(q^{2 m}-1\right) \left(q^{2 n}+1\right)\right)\breakeqn
	\times \left[2 n
	q \Phi \left(q^2,1,\frac{1}{2}-m\right)-2 m q \Phi \left(q^2,1,\frac{1}{2}-n\right)-n
	q^2 \Phi \left(q^4,1,\frac{1}{2}-m\right)\breakeqntwice
	+m q^2 \Phi \left(q^4,1,\frac{1}{2}-n\right)-2
	n q \Phi \left(q^2,1,m+\frac{1}{2}\right)+2 m q \Phi \left(q^2,1,n+\frac{1}{2}\right)\breakeqntwice
	+n
	q^2 \Phi \left(q^4,1,m+\frac{1}{2}\right)-m q^2 \Phi \left(q^4,1,n+\frac{1}{2}\right)+n
	\psi ^{(0)}\left(\frac{1}{2}-m\right)-n \psi ^{(0)}\left(m+\frac{1}{2}\right)\breakeqntwice
	-m \psi
	^{(0)}\left(\frac{1}{2}-n\right)+m \psi ^{(0)}\left(n+\frac{1}{2}\right)\right]+ \left(q^{4 m}-1\right) \left(q^{4 n}-1\right) \left[-4 q \Phi
	\left(q^2,1,\frac{1}{2}-m\right)\breakeqntwice
	+q^2 \Phi \left(q^4,1,\frac{1}{2}-m\right)+4 q \Phi
	\left(q^2,1,\frac{1}{2}-n\right)-q^2 \Phi \left(q^4,1,\frac{1}{2}-n\right)-4 q \Phi
	\left(q^2,1,m+\frac{1}{2}\right)\breakeqntwice
	+q^2 \Phi \left(q^4,1,m+\frac{1}{2}\right)+4 q \Phi
	\left(q^2,1,n+\frac{1}{2}\right)-q^2 \Phi \left(q^4,1,n+\frac{1}{2}\right)\breakeqntwice
	-4 m \psi
	^{(0)}\left(\frac{1}{2}-m\right) \log (q)+4 m \psi ^{(0)}\left(m+\frac{1}{2}\right) \log
	(q)-3 \psi ^{(0)}\left(\frac{1}{2}-m\right)-3 \psi ^{(0)}\left(m+\frac{1}{2}\right)\breakeqntwice
	+4 n
	\psi ^{(0)}\left(\frac{1}{2}-n\right) \log (q)-4 n \psi ^{(0)}\left(n+\frac{1}{2}\right)
	\log (q)+3 \psi ^{(0)}\left(\frac{1}{2}-n\right)+3 \psi
	^{(0)}\left(n+\frac{1}{2}\right)\right]\right\}
	\end{align}
\end{small}
By construction, these non-diagonal elements all satisfy the $SU(1,1)$ condition \mcite{suoneone}
\begin{equation}
\bR_{nm}~m = \bR_{mn}~n
\end{equation}
\textit{order by order} in $s$. For specific values of $n, m$, the above expressions do simplify, for instance: 
\begin{small}
	\begin{align}
	\cR_{24}^{(1)}&=\!\frac{\cR_{42}^{(1)}}{2}\!=\frac{1}{288 \pi ^2 q^4 \log ^2(q)}\left\{(q-1)^4 (q+1)^2 \left(q^2+1\right) \left(3 (q+1)^2 \left(q^4+4 q^2+1\right)
	\left(q^2+1\right)^3 \tanh ^{-1}q^2\breakeqntwice+(q-1)^2 q \left(3 q^8+9 q^7+34 q^6+65
	q^5+78 q^4+65 q^3+34 q^2+9 q+3\right)\right)\right.\nonu
	&\qquad{}\left.-3 \left(q^4-1\right)^4 \left(q^4+4
	q^2+1\right) \tanh ^{-1}q\right\}.
	\end{align}
\end{small}
However, we find that after $\Phi(z,1,a)$ simplifies, the $\log t$ terms from those terms are absent in a series expansion for the non-diagonal elements. For example, we find interestingly enough that:
\begin{align}
\cR_{24}^{(1)}&=-\frac{8 t^2 \log (2)}{\pi ^2}+\frac{48 t^3 \log (2)}{\pi ^2}+\frac{t^4 (31-512 \log
	(2))}{3 \pi ^2}+\cO\left(t^5\right)\nonu
\cR_{46}^{(1)}&=-\frac{16 t^2 \log (2)}{\pi ^2}+\frac{160 t^3 \log (2)}{\pi ^2}-\frac{2 t^4 (1408 \log
	(2)-79)}{3 \pi ^2}+\cO\left(t^5\right)\nonu
\cR_{26}^{(1)}&=\frac{8 t^2 \log (2)}{\pi ^2}-\frac{64 t^3 \log (2)}{\pi ^2}+\frac{t^4 (928 \log (2)-61)}{3
	\pi ^2}+\cO\left(t^5\right)\nonu
\cR_{28}^{(1)}&=-\frac{8 t^2 \log (2)}{\pi ^2}+\frac{80 t^3 \log (2)}{\pi ^2}+\frac{t^4 (103-1472 \log
	(2))}{3 \pi ^2}+\cO\left(t^5\right), \mbox{~etc.}
\end{align}
The $\log$ branch cuts from $\arctanh$ terms have cancelled after the sum over the block matrix indices $(x, p)$. The individual infinite sums from $\bM$ all diverge badly for $t<0$ but they combine appropriately for the non-diagonal case to give a \emph{log-free expansion} at this order in $s$.
\vspace{-2mm}
\subsubsection*{$s=2$ term}\label{subsubsec:seqtwo}
\begin{figure}[h!]
	\begin{subfigure}[b]{0.44\textwidth}
		\centering
		\includegraphics[width=\textwidth]{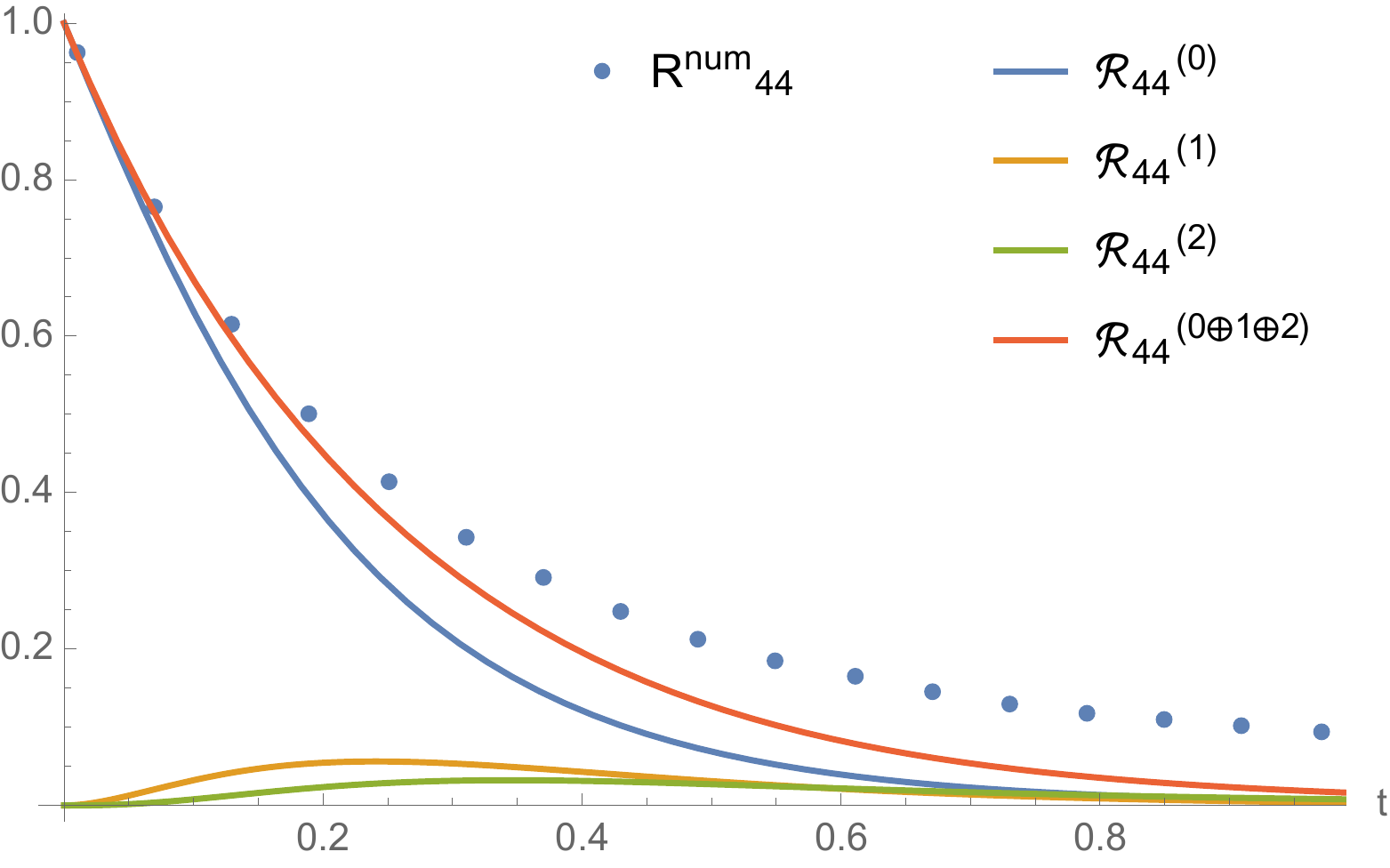}
		\caption{\label{fig:r44s012}}
	\end{subfigure}
	\begin{subfigure}[b]{0.44\textwidth}
		\centering
		\includegraphics[width=\textwidth]{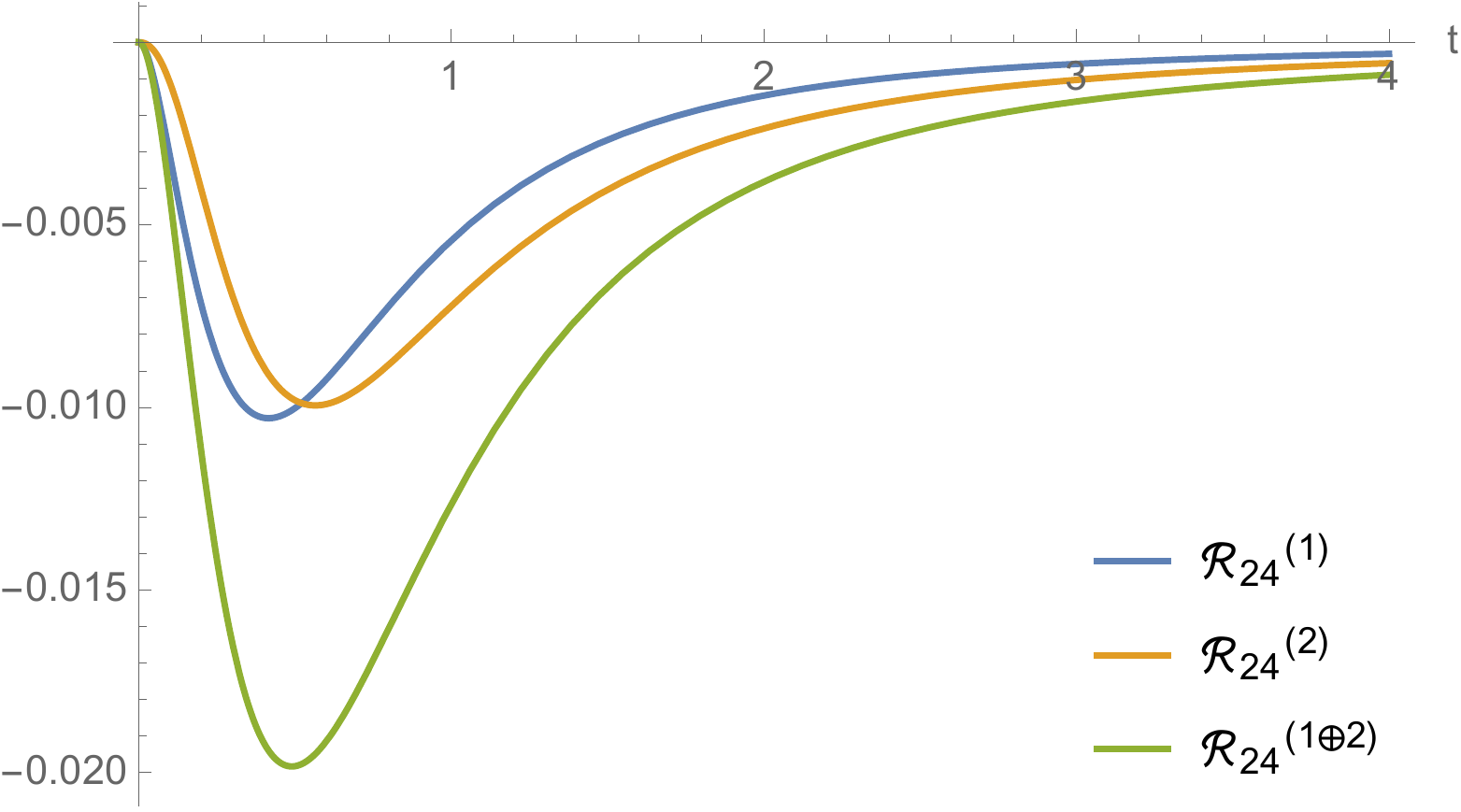}
		\caption{\label{fig:r24s012}}
	\end{subfigure}
	\caption{The individual contributions from the various matrix powers $s = 0, 1, 2$ and their sum is plotted for two matrix elements (\subref{fig:r44s012}) $\bR_{44}(t)$ and (\subref{fig:r24s012}) $\bR_{24}(t)$. The numerical estimate for $N = 64$ is also plotted for the $\bR_{44}$ case and is seen to closely follow the analytic sum. For $\bR_{24}$, the fit is not quite good since it starts only at the quadratic order and more terms would be required to account for the small but comparable contributions.}
	\label{fig:s012}
\end{figure}

\vspace{-2mm}
We are able to construct the matrix elements $\cR_{2n,2m}^{(2)}$ for a general $n, m$ although they are a longer combination of special functions, namely products of Hurwitz Lerch transcendents and Lerch transcendents which are not particularly illuminating. Hence, we  only provide the series expansions for certain matrix elements to show the general numerical structure:
\begin{subequations}
	\begin{align}
	\cR^{(2)}_{22}&=\frac{2 t^2 (44 \log (2)-27 \log (3))}{3 \pi ^2}+t^3 \left(2-\frac{16 \log ^2(2)}{\pi
		^4}-\frac{304 \log (2)}{3 \pi ^2}+\frac{72 \log (3)}{\pi ^2}\right)\nonu&\qquad{}+t^4
	\left(-\frac{32}{3}+\frac{71}{15 \pi ^2}+\frac{64 \log ^2(2)}{\pi ^4}+\frac{9416 \log
		(2)}{45 \pi ^2}-\frac{894 \log (3)}{5 \pi ^2}\right)+\cO\left(t^5\right),\\
	\cR^{(2)}_{24}&=\frac{2 t^2 (27 \log (3)-44 \log (2))}{3 \pi ^2}+t^3 \left(\frac{16 \log ^2(2)}{\pi
		^4}-\frac{108 \log (3)}{\pi ^2}+\frac{76 \log (4)}{\pi ^2}\right)\nonu&\qquad{}+t^4 \left(-\frac{34}{3
		\pi ^2}-\frac{96 \log ^2(2)}{\pi ^4}+\frac{411 \log (3)}{\pi ^2}-\frac{2162 \log (4)}{9
		\pi ^2}\right)+\cO\left(t^5\right).
	\end{align}
\end{subequations}
Next, we can combine the contributions from these three terms and analyse how well they approximate the behaviour by comparing to a numerical evaluation of the same as we do in  Fig. \ref{fig:s012}.
%
%
%
\medskip
However, from the open-closed correspondence we expect the above expansion in terms of the $t^r$ and $\log(t)$ basis to be incomplete. The crucial point is that one cannot dictate that the summation over $s$ and the Taylor series expansions over $r$ above must commute. Hence, the summation over $s$ can lead to the subleading terms\footnote{This physical input from the CFT picture can be taken into account explicitly by the formalism of \emph{Hardy fields} employed in real asymptotics, which allows to amalgamate many ``exponential scales''. See chapter V, App. $1$ of \cite{bourbaki2004elements} and chapters $3, 5$ of \cite{shackell2013symbolic} for details on the theory.} from closed string states of the form:



%



\begin{equation}\label{eq:subleading}
\bR_{nm}(t)=\sum_{k =0}^{\infty}\sum_{j=0}^{\infty}\sum_{s_1,s_2=0}^\infty c_{k|j|s_1s_2}(\log_{s_1}(t^{-1}))^{s_2} t^j e^{-\frac{2\pi^2 k}{t}}
\end{equation}
where $c_{k|j|s_1s_2}$ are some specific (real) coefficients and we have suppressed the mode labels $n, m$ for simplicity.	

By adding the contributions from the higher matrix powers in $s$, one may obtain a subset of  the above  coefficients to  higher accuracy, the ones corresponding to the $k = 0$ level in this expansion. The exponentially small parts from $k \geq 1$ are the ones of most interest to us and which encode information about the \emph{on-shell} closed string states and it would be interesting to recover some information about those states.



In summary, we have provided a formal procedure for successively approximating the coefficients in an expansion near $t = 0$.  We do not claim to the efficiency or numerical control resulting from this method. We must also acknowledge that this procedure does not extend in practice beyond the lowest orders due to some of the double sums (involving generalized hypergeometric functions) that arise from the block matrix multiplications.  Because of the non-analytic behaviour---which has its physical origins in the worldsheet picture---it is intrinsically difficult to identify the divergences or the subleading terms in the algebraic method we have used. Nonetheless, we hope that it has augmented the knowledge levels on the algebraic structure of OSFT at the quantum level.


In the next subsection, we shall study the behaviour of these matrix elements near the other limit of the modular parameter, that is, $t \to +\infty$ by using an expansion in the variable $q = e^{-t}$ near $q = 0^+$. Since the oscillator based expressions are much more suited for this kind of an expansion, we only check till the linear order term for consistency. We will find that our expressions correctly reproduce the numbers that can be generated from the oscillator based expansion.

\subsection{Expansions near $t=\infty$ in the continuous $\kappa$-basis}\label{subsec:moyal q expansion}
In order to obtain the expansion for $\bR_{nm}$ in the large $t$ (or small $q$) limit, as a power series in $q = e^{-t}$, one goes to the continuous $\kappa$-basis \cite{Belov:2003df,Rastelli:2001hh,Bars:2003sm,Erler:2003eq} that we discuss later in \S\ref{subsubsec:kappa-basis}. It corresponds to another basis for the same $j=0$ representation of $SL(2,\mathbb{R})$ associated with the discrete basis defined in terms of mode number labels, that we have been using. We choose it to convert  some of the infinite sums to  integrals for the purpose of numerical evaluation of the series coefficients. In certain cases, one can correctly guess the exact algebraic numbers by using the \emph{RootApproximant } command in \emph{Mathematica} if they stabilize as the ``WorkingPrecision'' is increased.  

\medskip
\vspace{-2mm}
\subsubsection{Even parity elements till linear order in $q$}\label{subsubsec:even parity kappa}
\vspace{-2mm}
To commence the evaluation, let us define a matrix $\cS$ in terms of purely the even frequencies $\ke$ as 
$\cS \coloneqq {\rm diag}\{\ke^{1/2}, \ke^{-1/2}\}$.
Then we can express the matrices $\cQ_\eta$ and $\cL_\eta$ that contribute to $\bR(q)$ as:
\begin{small}
	\begin{align}
	\cQ_\eta &= \frac{1}{4}\cS^{-1}\left[\begin{array}{cc}
	f_3 +tf_4t^{\top} &  -\frac{i}{2}(f_1-tf_1r)\\ -\frac{i}{2}(f_1-r^{\top}f_1t^{\top})
	& \frac{1}{4}(f_4+r^{\top}f_3 r )
	\end{array} \right] \cS^{-1}\eqqcolon \aqua\cS^{-1}\bQ^\kappa\cS^{-1},\nonu
	\cL_\eta &= \frac{1}{2}\cS^{-1}\left[\begin{array}{cc}
	tf_1t^{\top} &  \frac{i}{2}f_3\\ -\frac{i}{2}r^{\top}f_3t^{\top}
	& \frac{1}{4}f_1
	\end{array} \right]\cS^{-1}\eqqcolon \half \cS^{-1}\bL^\kappa \cS^{-1},
	\end{align}
\end{small}
where we have introduced the block matrices $\bQ^\kappa$ and $\bL^\kappa$ after absorbing the numerical factors.
Here, we have employed the matrix $\displaystyle t\coloneqq \sqrt{\ke}\, T \frac{1}{\sqrt{\ko}}$ which is the operator $\displaystyle \tanh \frac{\pi\hat{Q}_1}{2}$ which appears in the continuous Moyal basis\footnote{As we do not use the parameter $t = -\log q$ in this section, we hope the repeated use of the symbol wouldn't give rise to any ambiguities.}  and $r$ is the formal inverse of the matrix $t$, i.e. $\displaystyle r \coloneqq \sqrt{\ko}R\frac{1}{\sqrt{\ke}}$. Then we can rewrite the matrix which appears in the matrix $\cF(q)$ \eqref{eq:F matrix} as:
\begin{small}
	\begin{align}
	\cL_\eta^\top\cQ_\eta^{-1}\cL_\eta & = \cS^{-1}\bL^{\kappa\top} (\bQ^{\kappa})^{-1}\bL^\kappa \cS^{-1}\nonu
	& =  \cS^{-1}\left[\begin{array}{cc}
	tf_1t^{\top} & -\frac{i}{2}t f_3r\\  \frac{i}{2}f_3
	& \frac{1}{4}f_1
	\end{array} \right]\left[\begin{array}{cc}
	f_3 +tf_4t^{\top} &  -\frac{i}{2}(f_1-tf_1r)\\ -\frac{i}{2}(f_1-r^{\top}f_1t^{\top})
	& \frac{1}{4}(f_4+r^{\top}f_3 r )
	\end{array} \right]^{-1}\left[\begin{array}{cc}
	tf_1t^{\top} &  \frac{i}{2}f_3\\ -\frac{i}{2}r^{\top}f_3t^{\top}
	& \frac{1}{4}f_1
	\end{array} \right]\cS^{-1}
	\end{align} 
\end{small}
in block matrix form.

To obtain the inverse, once again we perform a geometric series expansion by separating the degree zero term through
$
\bQ^\kappa(q) \eqqcolon \bQ^\kappa_0+\delta \bQ^\kappa(q).
$
Since as $q\to 0^+$, we have $f_1(q)\to 1^-, f_3(q)\to 1^-$ and $f_4(q)\to 3^-$,
%
\begin{equation}
(\bQ_0^{\kappa})^{-1}= \left[\begin{array}{cc}
\Lambda&  0\\ 0
& 4\Omega
\end{array} \right],
\end{equation}
where we define the infinite matrices:
\begin{equation}
\label{Lambda matrix}
\Lambda \coloneqq \frac{1}{1+3tt^{\top}}, \qquad \Omega \coloneqq \frac{1}{3+r^{\top} r} = \frac{1}{3}(1-\Lambda).
\end{equation}
Next, we  insert the following binomial inverse series 
in order to obtain $\cF(q)$ which in terms of these new matrices become:
\begin{equation}\label{F matrix q Expansion Moyal}
\cF(q) = \half M_0^{gh-1}f_2(q)+\sum_{s=0}^{\infty}(-1)^s\cS^{-1}\bL^{\kappa\top}  \bls(\bQ_0^\kappa)^{-1}\delta \bQ^\kappa\brs^s\,\cdot (\bQ_0^\kappa)^{-1}\bL^\kappa \cS^{-1}.
\end{equation}
We shall denote the expansion as:
\begin{align}
\cF(q)&=\sum_{s=0}^\infty(-1)^s\cF^{\kappa|s}(q), \mbox{~where we set}\\
\cF^{\kappa|0}(q)&=\frac{1}{2}M_0^{gh-1}f_2(q)+\cS^{-1}\bL^{\kappa\top}(q)(\bQ_0^\kappa)^{-1}\bL^\kappa(q)\cS^{-1}\label{eq:Fkappazero}.
\end{align}
Here as well, the simplest element to consider is the purely even block $\cF_{2n,2m}^{pp}$, which contains the explicit term: 
\begin{equation}
(\bL^{\kappa\top} (\bQ^{\kappa})^{-1}\bL^\kappa)^{pp}=\bL^{\kappa\top}_{px}(\bQ^\kappa)^{-1}_{xx}\bL^\kappa_{xp}+\bL^{\kappa\top}_{px}(\bQ^\kappa)^{-1}_{xp}\bL^\kappa_{pp}+\bL^{\kappa\top}_{pp}(\bQ^\kappa)^{-1}_{px}\bL^\kappa_{xp}+\bL^{\kappa\top}_{pp}(\bQ^\kappa)^{-1}_{pp}\bL^\kappa_{pp},
\end{equation}
where we have used the doublet indices $x,p$ as subscripts for typographical convenience.

In the following, we simply restrict to the general structure of the lowest order $s=0$ term since we are primarily interested in certain consistency checks in the $q\to 0^+$ limit. The oscillator expansion is much better suited for expansions near this limit and hence we  return to that method in \S\ref{subsec:osc expansions}. We shall later collect the exact coefficients for $q^0$ and $q^1$ and verify that they match with the exact results from the oscillator expressions. 

\noindent
Upon inserting the constituent block matrices, the momentum block $\cF^{\kappa pp|0}$ in \eqref{eq:Fkappazero} has the structure:
\begin{align}
\cF^{\kappa pp|0}_{2n,2m}(q) &=-\aqua\cdot 2nf_2(2n)\delta_{nm}+2\sqrt{nm}\cdot \aqua \left(f_1(2n)f_1(2m)\Omega_{2n,2m}-f_3(2n)f_3(2m)\Lambda_{2n,2m}\right)\nonu
&=-\frac{n}{2}\delta_{nm}+\frac{\sqrt{n m}}{2}(\Omega-\Lambda)_{2n,2m}\nonu
&\qquad{}+\frac{\sqrt{nm}}{2}\left[-2\Omega_{2n,2m}q^{2n}-2\Omega_{2n,2m}q^{2m} +\left(\Lambda_{2n,2m}+\Omega_{2n,2m}-\sqrt{\frac{n}{m}}\right)q^{4n}\breakeqn
+(\Omega+\Lambda
)_{2n,2m}q^{4m}+4\Omega_{2n,2m}q^{2n+2m}\breakeqn
-2\Omega_{2n,2m}q^{2n+4m}-2\Omega_{2n,2m}q^{4n+2m}+(\Omega-\Lambda)_{2n,2m}q^{4n+4m}\right.\bigg],
\end{align}
without any summations over repeated indices. The next term in the expansion becomes more tedious but starts contributing at $q^1$ (due to the infinite summations over the \emph{odd} index).

To this end, it is worthwhile to note that the constant part of the matrix $\bL^\kappa$ is  of the form:
\begin{equation}
\bL^\kappa(0)=\left[\begin{array}{cc}
tt^\top& \frac{i}{2} \\ 
-\frac{i}{2}& \aqua 
\end{array} \right]
\end{equation}
This allows us to collect the coefficient of $q^0$ from
\begin{equation}
(\bL^{\kappa\top}(\bQ^\kappa)^{-1}\bL^\kappa)_{pp}^{(0)}=-\aqua \ke-\aqua \sqrt{\ke}\frac{1-tt^{\top}}{1+3tt^{\top}}\sqrt{\ke}
\end{equation}
which when substituted into the expression for $\bR_{2n,2m}$ in terms of $\cF_{2n,2m}$ \eqref{eq:Rnm in terms of Fnm}:
\begin{align}
\bR^{(0)}_{2n,2m}&= -\bls\mathbb 1 - \mathbb 1-\sqrt{\ke}\frac{1-tt^{\top}}{1+3tt^{\top}}\frac{1}{\sqrt{\ke}}\brs_{2n,2m}\nonu
&=\sqrt{\frac{n}{m}}\left[\frac{1-tt^{\top}}{1+3tt^{\top}}\right]_{2n,2m},
\end{align}
which is precisely the even parity elements of the Neumann matrix $X^{11}_{2n,2m}$ in \eqref{eq:Rnmosc expr} as was derived in the oscillator formalism.

Similarly, the coefficient of $q^1$ is given by the matrix:
\begin{equation}
\bR^{(1)}_{2n,2m} = -2\sqrt{nm}\left[(\Lambda t)_{2n,1}(\Lambda t)_{2m,1}+\half \left((\Lambda t)_{2n,1}(\Omega r^{\top})_{2m,1}+(n\leftrightarrow m)\right)\right]\times \frac{-4}{2m}
\end{equation}
by considering the $s=1$ power and noticing that the $q^{4n}, q^{2n}$ terms  do not contribute at this order for any $n$. Now, one can show that $\Omega r^\top = \Lambda t$; hence the above reduces to:
\begin{equation}\label{eq:moyal q linear coeff}
\bR^{(1)}_{2n,2m}= 8\sqrt{\frac{n}{m}}(\Lambda t)_{2n,1}(\Lambda t)_{2m,1}.
\end{equation}
\vspace{-2mm}
\subsubsection{Numerical evaluation in the continuous $\kappa$-basis}\label{subsubsec:kappa-basis}
\vspace{-2mm}
The matrix elements of the rational functions involving the $t$ matrix such as
\[\Lambda t =\frac{1}{1+3tt^{\top}}t = t\frac{1}{1+3t^{\top}t}\]
can be obtained by numerical integration by going to  the continuous Moyal basis\footnote{The  notational conflict in using $\kappa$ for the continuous basis and for the spectral matrix would be restricted to this subsubsection.}, known as the $\kappa$-basis. The $\kappa$ basis diagonalizes the operator $K_1 = (L_{1} + L_{-1})$ of $SL(2, \mathbb{R})$ \cite{Belov:2003df,Rastelli:2001hh,Bars:2003sm,Erler:2003eq}: 
\begin{equation}
K_1 |\kappa\rangle = \kappa|\kappa\rangle
\end{equation} 
which commutes with the vertex, and is useful for performing analytic and numerical calculations. 

The $t$ matrix is diagonalized in the infinite $N$ limit to give the eigenvalues: $t_{\kappa} = \tanh (\pi \kappa/4)$. Then we have the integral representation for the matrix elements as
\begin{align}
t_{2n,2m-1} &= \sqrt{2n}\,\,T_{2n-1,2m-1}\,\frac{1}{\sqrt{2m-1}}\nonu& = \int_{-\infty}^{\infty} d\kappa \, v_{2n}(\kappa)\tanh (\pi \kappa/4) v_{2m-1}(\kappa)
\end{align}
where we start\footnote{ The following properties are taken from App A of \cite{Erler:2003eq}.} with defining the overlap functions
\begin{equation}
v_n(\kappa) =\ip{\kappa}{n}=\frac{y_n(\kappa)}{\sqrt{n}\sqrt{\frac{2}{\kappa}\sinh \frac{\pi \kappa}{2}}}
\end{equation}
which are a class of polynomials that arise naturally in the continuous basis and are analogous to the Hermite polynomials for the number operator. These are orthogonal with respect to the  weight function
\begin{equation}
w(\kappa) = \left(\frac{2}{\kappa}\sinh \frac{\pi\kappa}{2}\right)^{-1}
\end{equation}
A generating functional for these polynomials is given by:
\begin{equation}
\sum_{\ninz_+} \frac{z^n}{n}y_n(\kappa) = \frac{1}{\kappa}(1-e^{-\kappa \arctan z}) = f_\kappa(z)
\end{equation}
and they satisfy the recurrence relation:
\begin{equation}
y_{n+1}(\kappa) + y_{n-1}(\kappa) = -\frac{\kappa}{n}y_n(\kappa),
\end{equation}
among many other relations listed in \cite{Erler:2003eq}. Setting $y_0(\kappa)=0, y_1(\kappa) = 1$, leads to the polynomials:
\begin{align}
y_1(\kappa) &= 1, \qquad & y_2(\kappa)&=-\kappa,\nonu
y_3(\kappa)& =\half \kappa^2-1,\qquad& y_4(\kappa)&=-\frac{1}{6}\kappa^3+\frac{4}{3}\kappa,\nonu
y_5(\kappa)& = \frac{1}{24}\kappa^4-\frac{5}{6}\kappa^2+1,\qquad&y_6(\kappa)&=-\frac{1}{120}\kappa^5 +\frac{1}{3}\kappa^3 -\frac{23}{15}\kappa,
\end{align}
and so on and so forth.

After this short exposition of these somewhat amusing polynomials, let us return to the evaluation of the matrix functions.
In terms of the new continuous basis, we can express functions of the $t$ matrices such as 
\begin{subequations}
	\begin{align}
	\left(F(t t^\top )\right)_{2n,2m}&= \int_{-\infty}^\infty \, d\kappa \,v_{2n}(\kappa) F\left(\tanh^2\left(\frac{\pi \kappa}{4}\right)\right)v_{2m}(\kappa),\\
	\left(t F(t^\top t)\right)_{2n,2m-1}&= \int_{-\infty}^\infty \, d\kappa\, v_{2n}(\kappa)\tanh\left(\frac{\pi \kappa}{4}\right) F\left(\tanh^2\left(\frac{\pi \kappa}{4}\right)\right)v_{2m-1}(\kappa), \mbox {~etc.}
	\end{align}
\end{subequations}
using which we have evaluated \eqref{eq:moyal q linear coeff} upto a WorkingPrecision of $16$ in \emph{Mathematica}. The resulting numbers for some matrix elements are listed in Table \ref{tab:moyalq}.
\begin{table}[h!]
	\centering	
	\renewcommand{\arraystretch}{1.124}
	\[\begin{array}{c||cccc}
	\hline
	\bR_{nm}^{{\rm Moy}|1} & 2 & 4 & 6 & 8 \\
	\hline\hline
	2 & 0.325154 & -0.0939333 & 0.0511147 & -0.0338815 \\
	4 & -0.187867 & 0.0542726 & -0.0295329 & 0.0195760 \\
	6 & 0.153344 & -0.0442994 & 0.0241059 & -0.0159787 \\
	8 & -0.135526 & 0.0391520 & -0.0213049 & 0.0141220 \\
	\hline
	\end{array}\]
	\caption{Numerical evaluation of the linear coefficient in a few even parity matrix elements $\bR_{nm}(q)$ using the continuous $\kappa$ basis for the Moyal $\ast$ .}
	\label{tab:moyalq}
\end{table}
Now, by using the oscillator based expansion in \eqref{eq:Rnmosc expr} and \eqref{eq:osc q expansion}, we obtain the linear coefficient to be in terms of the ghost Neumann matrices:
\begin{equation}
\bR_{2n,2m}^{(1)|osc} =-\left( X^{12}_{2n,1}X^{12}_{1,2m}+X^{21}_{2n,1}X^{21}_{1,2m}\right).
\end{equation}
These  rational numbers  are  tabulated in Table \ref{tab:oscqlin} and found to be the stabilizing value as the WorkingPrecision for the numerical integrations above is increased.
\begin{table}[h]
	\renewcommand{\arraystretch}{1.24}
	\begin{center}
		\[\begin{array}{c||cccc}
		\hline
		\bR_{nm}^{{\rm osc}|1} & 2 & 4 & 6 & 8 \\
		\hline\hline
		2 & \frac{6400}{19683} & -\frac{16640}{177147} & \frac{244480}{4782969} &
		-\frac{13126400}{387420489} \\
		4 & -\frac{33280}{177147} & \frac{86528}{1594323} & -\frac{1271296}{43046721} &
		\frac{68257280}{3486784401} \\
		6 & \frac{244480}{1594323} & -\frac{635648}{14348907} & \frac{9339136}{387420489} &
		-\frac{501428480}{31381059609} \\
		8 & -\frac{52505600}{387420489} & \frac{136514560}{3486784401} &
		-\frac{2005713920}{94143178827} & \frac{107688985600}{7625597484987} \\
		\hline
		\end{array}\]
	\end{center}
	\caption{Exact linear coefficients in a few even parity matrix elements $\bR_{nm}(q)$ obtained using the oscillator method in terms of Neumann coefficients. }
	\label{tab:oscqlin}
\end{table}
Indeed, we may also express the Neumann matrices $X^{(\pm)}$ in terms of the matrix
\begin{equation}
\hat{m}_0^\ast \coloneqq \left(\begin{array}{cc}
0&  -S\\ -T^{\top}
& 0
\end{array} \right)
\end{equation}
defined in \cite{Bars:2003gu}  to analytically prove that both expressions for the linear term coincide. This expansion can thus result in interesting  relations between the Neumann matrices and matrices arising from the Moyal structure which may be established by using the canonical way of expressing all the Neumann matrices in terms of the matrix $t$ and the frequency matrices $\ke$ and $\ko$ \cite{Bars:2003gu}.

Regarding studying the determinant factor in the integrand using a $q$ expansion(see also \ref{subsec:detremarks}), which is common for all matrix elements, we find that the lowest power of each matrix element \emph{do not decrease along a row or a column} which is required for a systematic expansion. Essentially, one cannot separate the degree zero piece as there is no nice way to express $\displaystyle \det (1+A^{-1}B)$ in terms of $\det A^{-1}B$. 

Although one can include the higher powers $((\bQ_0^\kappa)^{-1}\delta\bQ^\kappa)^s$ to obtain the exact coefficients for  a $q$ series, this would necessitate many more numerical integrations arising from collecting powers together and results in numerical uncertainties. The oscillator basis on the other hand furnishes the exact coefficients since the Neumann matrices are known exactly from CFT. We therefore simply contend ourselves with the zeroth order and the linear coefficient using the $\kappa$ basis and compare with the oscillator based expansion. This serves as a consistency check on the correctness of our expressions in the $t \to +\infty$ limit. Hence for the purpose of constructing a $q$-series, we  employ the oscillator based expressions in \eqref{eq:Rnmosc expr} expressed in terms of Neumann matrices in the following subsection. This can then be used to search for some hints of the non-analyticities expected from the underlying geometrical picture.

\vspace{-2mm}
\subsubsection{Another consistency check using factorization}\label{subsubsec:Factorization}
\vspace{-2mm}
Let us pause for a moment and do a quick check on the overall determinant factors near the  $t\to \infty$ or $q\to 0$ limit to make sure that the result is regulator independent. In this limit, we expect the integrand to factorize into the $3$-point function, with two legs on-shell with $p^2=1=-m^2$ for the ``lightest'' tachyon state and one off-shell tachyon state with $p=0$, and a tachyon propagator\footnote{This may be read off from open string partition function.} with $t\to \infty$. The off-shell $3$-tachyon amplitude has been known\cite{Cremmer:1986if,Samuel:1987uu,Gross:1986fk}
to be of the form:
\begin{equation}
g_{123}(k_i) = g_T K^3 \times K^{-(k_1^2+k_2^2+k_3^2)},
\end{equation}
where we recall that $g_T$ is the on-shell $3$-tachyon coupling (by definition) and $\displaystyle K = \frac{3\sqrt{3}}{4}$. Hence, we expect the leading asymptotics to be:
\begin{equation}
\frac{g_0}{3}\frac{(1+w^\top w)^{\frac{d+2}{2}}}{(2\pi)^{d(N+1/2)}}q^{-1}\frac{\det(2\cQ_\eta^{gh})}{\lvert\det(2\cQ^X_\eta)\rvert^{d/2}}(2\cQ_p)^{-d/2}\to g_T K^3 \times K^{- 2} \,\frac{q^{-1} (-2\log q)^{-d/2}}{(2\pi)^{d(N+1/2)}}
\end{equation}
as $q\to 0$, when $d=26$. The $2\pi$ factors arise from the $\eta^X$ and $p$ integrations and the manner in which the basis states $e^{i\xi^\top\eta}$ are normalized. We have also set $l_s = \sqrt{2}$ on the right hand side for consistency with our earlier conventions.

The factor $\cQ_p$ arising from the momentum integration is dominated by $-\log q$ and hence we require:
\begin{equation}
\lim_{t\to \infty}\frac{g_0}{3}(1+w^\top w)^{\frac{d+2}{2}}\frac{\det(2\cQ_\eta^{gh})}{\lvert\det(2\cQ^X_\eta)\rvert^{d/2}}= g_T \, K 
\end{equation}
The determinant factors involve the block matrices \eqref{eq:Q gh}, \eqref{eq:Qeta matter}:
\begin{small}
	\begin{align}
	\cQ_\eta^{gh} &=+ \aqua\left[\begin{array}{cc}
	\ke^{-1}f_3(\ke)+T\ko^{-1}f_4(\ko) T^\top & -\frac{i}{2}\left[f_1(\ke)-Tf_1(\ko)R\right]\\
	-\frac{i}{2}\left[f_1(\ke)-R^\top f_1(\ko) T^\top\right] & \frac{1}{4}\left[\ke f_4(\ke) + R^\top\ko f_3(\ko) R\right]
	\end{array} \right],\nonu
	\cQ_{\eta}^X&=+\frac{1}{2}\left[\begin{array}{cc}
	\ke^{-1}f_4+T\ko^{-1} f_3(\ko)T^\top& -\frac{i}{4} (f_1(\ke)-Tf_1(\ko)R)  \\ 
	-\frac{i}{4} (f_1(\ke)-Tf_1(\ko)R)^\top  & \frac{1}{16	}\left(\ke f_3 +R^{\top}\ko f_4R\right)
	\end{array} \right]
	\end{align}
\end{small}
In the $q\to 0$ limit, we have $f_1\to 1, f_3\to 1$ and $f_4\to 3$ and hence the ratio of the determinant factors above reduces to give:
\begin{small}
	\begin{align}
	\det(2)_{2N}^{-1}\frac{\det(1+3tt^\top) \det\left(\frac{3+r^\top r}{4}\right)}{\det(3+tt^\top)^{d/2}\det\left(\frac{1+3r^\top r}{16}\right)^{d/2}}&=2^{2N(d-2)}\frac{\det(1+3tt^\top)\det(3+r^\top r)}{\det(3+tt^\top)^{d/2}\det\left(1+3r^\top r\right)^{d/2}}\nonu
	&=2^{2N(d-2)}\det(1+3tt^\top)^2\det(3+tt^\top)^{-d}\det(tt^\top)^{\frac{d-2}{2}},
	\end{align}
\end{small}
where $r\coloneqq t^{-1} = \ko^{-1}t^{\top}\ke$ and we have substituted $r^\top r = (tt^\top)^{-1}$. Multiplying with the remaining factors and using $\det(tt^\top) = (1+w^\top w)^{-1/2}$, we have:
\begin{align}
\frac{g_0}{3}(1+w^\top w)^{\frac{d+2}{2}}\times 2^{2N(d-2)}\frac{\det(1+3tt^\top)^2}{\det(3+tt^\top)^{d}}\det(tt^\top)^{\frac{d-2}{2}}&=2^{2N(d-2)}\frac{g_0}{3}(1+w^\top w)^{-\frac{d}{8}+\frac{3}{4}}\,\frac{\det(1+3tt^\top)^2}{\det(3+tt^\top)^{d}},\nonu
& = - \mu_3\frac{g_0}{3},
\end{align}
where $\mu_3$ is the normalization factor that relates the interaction term in the Moyal and the oscillator formalisms(\S\ref{subsec:notations}), and which vanishes as $N\to \infty$. In terms of $\mu_3$, the couplings are related as $g_T =-\mu_3\times 2g_0 K^{-3}$ and hence the $N$ dependence is removed. The LHS now becomes:
$g_T/6 \, K^3$
which is off from the expected result of $g_T K$ by a factor of $\displaystyle \frac{1}{6} K^2 = 9/32=0.28125$.
The $6$ is because of the  symmetry factor $3!$ for the $3$-point function but  no such factors would arise for the tadpole case. We hope to return to this slight discrepancy when occasion offers itself.

Moving on, we can obtain the higher order terms more efficiently and exactly using the oscillator based expression, to which we turn next.

\subsection{A convergent expansion in $q$ using the oscillator expression}\label{subsec:osc expansions}
In this work, we have been mainly interested in the behaviour of the finite matrix elements as $t \to 0^+$. This corresponds to looking at the $q\to1^-$ limit, and hence may also be indirectly inferred from a series expansion near $q=0$  (the $t = +\infty$ limit) due to the expected non-analyticities. Physically, one would expect that the $t$ evolved string field becomes ill-defined when $\Re (t)<0$; the propagator would result in divergent sums while acting on a string field for $t<0$.
Thus,  intuitively we would expect the matrix elements to be uniformly convergent for $|q|<1$ and to have non-analytic behaviour everywhere on the unit circle $|q| = 1$ which obstructs an analytic continuation beyond the unit disc in the $q$ plane.

We therefore proceed to directly use the oscillator expression given in \cite{Ellwood:2003xc} to probe the $q\to 1^-$ limit. The matrix elements $\bR_{nm}(q)$, can be given a systematic expansion in powers of $q$ as follows.  The matrix whose powers are taken in the geometric series expansion has a minimal degree $1$. Therefore, the matrix powers start contributing only from higher and higher powers onwards as the infinite sums in the matrix products would not alter the order of the zeroes. This allows us to obtain the exact coefficients by adding up the contribution from a \emph{finite} number of matrix powers. 
\vspace{-2mm}
\subsubsection*{A q-series expansion}\label{subsubsec:qseriesosc}
\vspace{-2mm}
By a theorem of  Sierpi\'nski (see \cite[\S4.2]{remmert2012theory}), there can exist power series which converges at a single point on the boundary (say $z=1$) but diverges at every other point. In our particular case, we would have a series with radius of convergence $1$, that converges at $q=1$ to either $+1$ or $-1$ but exhibits discontinuous behaviour on the disc boundary.

The oscillator based expressions given in \cite[\S4]{Ellwood:2003xc} is naturally suited for systematically finding a $q$ series expansion for $\bR_{nm}(q)$ since the propagator is simple in this basis. Again, the ghost sector is relatively simpler as compared to the matter sector due to the absence of the momentum zero mode.

As the hatted matrices \eqref{eq:hatted X} appearing in \eqref{eq:Rnmosc expr} for $\bR(q)$ do not seem to satisfy any nice identities unlike the $\mathcal{M}_{0,\pm}$ matrices, we resort to a  geometric series for studying the matrix inverse $(\mathbb{1}-S\widetilde{X})^{-1}$. Inserting this formal  expansion into \eqref{eq:Rnmosc expr}, we have:
\begin{equation}
\bR(t) = X^{11} + \left[\begin{array}{cc}
\hat{X}^{12}(0,t)& \hat{X}^{21}(0,t)
\end{array} \right]\sum_{s=0}^{\infty}(S \widetilde{X})^sS\left[\begin{array}{c}\hat{X}^{21}(t,0)
\\ \hat{X}^{12}(t,0)
\end{array} \right]
\end{equation}
and let us introduce the infinite matrices $\bR^{(s)}_{nm}$ by rewriting:
\begin{equation}
\bR_{nm}(q)= \sum_{s=0}^\infty\bR_{nm}^{(s)}(q)
\end{equation}
in terms of the variable $q$.

\medskip
At the risk of further over-complicating the notation, let us also introduce a constant matrix $\mathscr{X}$ as follows:
\begin{equation}
\mathscr{X}\coloneqq \left[\begin{array}{cc}
X^{21}&  X^{11}\\ 
X^{11}& X^{12}
\end{array} \right],
\end{equation}
which is essentially the $S\widetilde{X}(t)$ matrix stripped off the $t$ dependent propagator pieces and the $C$ matrices. The $C$ matrices and the $q^{n/2}$ factors from  the propagator effectively make the contribution from the $s$th power term into:
\begin{align}\label{eq:osc q expansion}
\bR^{(s)}_{nm}(q)= \delta_{s,0} ~X^{11}_{nm}+ \sum_{p=s+1}^{\infty}(-1)^p q^p\sum_{|\vec{\mu}| = p} \Big[ X^{12}_{n,\mu_1}&\blp\underbrace{\sX \ldots\mathscr{X}}_{\mbox{s terms}}\brp_{11|\mu_1 \mu_{s+1}}X^{12}_{\mu_{s+1}m}\nonu
+X^{12}_{n,\mu_1}&\blp\underbrace{\sX \ldots\mathscr{X}}_{\mbox{s terms}}\brp_{12|\mu_1 \mu_{s+1}}X^{21}_{\mu_{s+1}m}\nonu
+X^{21}_{n,\mu_1}&\blp\underbrace{\sX \ldots\mathscr{X}}_{\mbox{s terms}}\brp_{21|\mu_1 \mu_{s+1}}X^{12}_{\mu_{s+1}m}\nonu
+ X^{21}_{n,\mu_1}&\blp\underbrace{\sX \ldots\mathscr{X}}_{\mbox{s terms}}\brp_{22|\mu_1 \mu_{s+1}}X^{21}_{\mu_{s+1}m}\Big],
\end{align}
where we are only summing over the set of integer partitions of the power $p$ into $s+1$ terms:
\[|\vec{\mu}|=\mu_1+\ldots+ \mu_{s+1}=p,\]
and its permutations. For performing these block matrix computations we have again used the \textit{NCAlgebra} package\footnote{I would like to thank the UC San Diego Mathematics department for making available this package using which parts of the computations in this work were performed.} \cite{NCAlgebra} which among its many powerful features handles block matrices in a somewhat more reliable and easier manner as compared to \emph{Mathematica}'s built-in functions. For instance, the block matrix powers which grow exponentially with the degree can be quickly evaluated as formal expressions using the ``NCDot''/``MM'' (MatrixMultiply) command. These can then be fed into a ``module'' for inserting the $X^{0,\pm}$ exact values. Essentially, the output of the \textit{NCAlgebra} commands are used to construct \emph{lists} and we apply the transformation rules on them to convert them to the  coefficients.

\medskip
For low values of $s$, one can use the ``Permutations'' and ``IntegerPartitions'' commands in \emph{Mathematica} to insert the appropriate indices and perform the (constrained) summations\footnote{One can also employ ``If'' conditionals to do these summations by brute-force for low enough $s$. The routine needs to check $\displaystyle{\sum\limits_{p=1}^r \sum\limits_{s=1}^p(p-s+1)^s}$ \emph{If} conditionals and also perform multiplication and addition for the size of the Permutations of Integer Partitions to obtain the first $r+1$ coefficients exactly. This number grows very quickly.}.  Again, this becomes computationally challenging since the number of terms in each block grows exponentially with $s$ as $2^{s-1}$ and we had to contend ourselves with $s\leq17$ truncation due to time and energy constraints.

To obtain till the $q^{18}$ coefficient exactly, one needs to include the $s = 0,\ldots,17$ contributions (the $s=18$ terms start only at $q^{19}$). Once we have an expansion in terms of exact coefficients, we can find the corresponding diagonal or near diagonal Pad\'e approximant ($n\approx m$) and look at its pole-zero structure in the complex $q$ plane as we do in App \ref{sec:Pade}. This is a useful exercice in general, when the available data is limited due to a multitude of reasons. 

\bigskip
We have obtained  the coefficients  till the $q^{18}$ term for a \emph{general matrix element} $\bR_{nm}$ symbolically.  For particular values of $n,m$, the expansions can then be readily obtained. We provide a few elements below for illustration:
\begin{align}
\bR_{11}(q)&=-\frac{11}{3^3}-\frac{ 2^7}{3^6}q-\frac{ 2^7\cdot 23}{3^8}q^2+\frac{ 2^9\cdot 7\cdot
	13}{3^{12}}q^3-\frac{ 2^7\cdot 13693}{3^{15}}q^4+\frac{ 2^8\cdot
	54503}{3^{17}}q^5\simbreak
+ \cdots 
+\frac{ 2^8\cdot 53\cdot 3469\cdot
	105251\cdot 28802532911}{3^{53}}q^{17}+\frac{ 2^7\cdot 20826099209\cdot
	1406808088061}{3^{56}}q^{18}+\cO(q^{19})\nonu
&\approx 
-0.407407-0.175583 q-0.448712 q^2+0.0876711 q^3-0.122149 q^4+0.108044 q^5\simbreak
-0.0360726
q^6-0.0321163 q^7+0.0250613 q^8+0.0228212 q^9-0.0179066 q^{10}\simbreak
-0.0218985
q^{11}+0.00985881 q^{12}+0.0211021 q^{13}-0.000638823 q^{14}-0.0163765 q^{15}\simbreak
-0.00652212
q^{16}+0.00736123 q^{17}+0.00716576 q^{18}
+\cO(q^{19}),\\
\bR_{22}(q)&=\frac{19}{3^5}+\frac{ 2^8\cdot 5^2}{3^9}q+\frac{ 2^8\cdot 269}{3^{11}}q^2-\frac{
	2^{10}\cdot 569}{3^{14}}q^3+\frac{ 2^8\cdot 107\cdot 2131}{3^{17}}q^4-\frac{ 2^9\cdot
	7\cdot 224617}{3^{20}}q^5\simbreak
+\cdots
+\frac{ 2^8\cdot
	204248123\cdot 1153179431133481}{3^{58}}q^{18}+\cO(q^{19})\nonu
&\approx 0.0781893\, +0.325154 q+0.388739 q^2-0.121819 q^3+0.452008 q^4-0.23088 q^5\simbreak+0.139741
q^6
+0.0208859 q^7-0.0978634 q^8+0.0156951 q^9+0.0705072 q^{10}\simbreak-0.0118808 q^{11}-0.0591811
q^{12}
-0.00238481 q^{13}+0.0496639 q^{14}+0.0171492 q^{15}\simbreak -0.0333904 q^{16}-0.0243469
q^{17}+0.0128015 q^{18} +\cO(q^{19}),
\end{align}
and for two non-diagonal elements, we have:
\begin{align}
\bR_{24}(q)&=-\frac{2^5\cdot 5^2}{3^9}-\frac{ 2^8\cdot 5\cdot 13}{3^{11}}q+\frac{ 2^8\cdot 5\cdot
	109}{3^{14}}q^2+\frac{ 2^{10}\cdot 5\cdot 67\cdot 199}{3^{18}}q^3+\frac{ 2^8\cdot
	5^2\cdot 137\cdot 181}{3^{20}}q^4\simbreak
-
\cdots +\frac{ 2^8\cdot 5\cdot 7^4\cdot 181\cdot
	846389\cdot 14954516415841}{3^{62}}q^{18}+\cO(q^{19})\nonu
&\approx -0.0406442-0.0939333 q+0.0291702 q^2+0.176204 q^3+0.0455149 q^4-0.051438 q^5\simbreak
-0.00973223
q^6-0.0752605 q^7-0.0205626 q^8+0.0851266 q^9-0.023345 q^{10}\simbreak
-0.0660465 q^{11}+0.032174
q^{12}+0.0440361 q^{13}-0.0130819 q^{14}-0.032953 q^{15}\simbreak
-0.00765715 q^{16}+0.019743
q^{17}+0.0184546 q^{18}+\cO(q^{19}),\mbox{~and for}\\
\bR_{26}(q)&=\frac{2^5\cdot 5\cdot 29}{3^{11}}+\frac{ 2^8\cdot 5\cdot 191}{3^{14}}q-\frac{ 2^8\cdot
	5\cdot 199}{3^{15}}q^2-\frac{ 2^{10}\cdot 5\cdot 7\cdot 37\cdot 53}{3^{19}}q^3-\frac{
	2^8\cdot 5\cdot 11\cdot 7489}{3^{23}}q^4\simbreak
-\cdots-\frac{ 2^8\cdot 5\cdot 7\cdot 31\cdot 18664747\cdot
	823481250069563}{3^{64}}q^{18}+\cO(q^{19})\nonu
&\approx 0.0785788\, +0.153344 q-0.0532556 q^2-0.181411 q^3-0.00336015 q^4-0.0665607 q^5\simbreak
-0.157655
q^6+0.0480861 q^7+0.179151 q^8+0.0564545 q^9-0.119223 q^{10}\simbreak
+0.0299126 q^{11}+0.132947
q^{12}-0.0913012 q^{13}-0.0771423 q^{14}+0.0626403 q^{15}\simbreak
+0.028327 q^{16}+0.0115182
q^{17}-0.00372998 q^{18}+\cO(q^{19}).
\end{align}
It is interesting to note that the coefficients are all nice rational numbers given that  the Neumann matrices are only  algebraic valued. We observe that there is a  (rather slow) non-monotonic fall-off of the coefficients.
However, we can see from Table \ref{tab:revenseries} below that for low $n, m$, they still approximate the function near $q=1$.
\begin{table}[h!]
	\renewcommand{\arraystretch}{1.12}
	\begin{center}	\[
		\begin{array}{c||c c c c}
		\hline \bR_{2n,2m}^{series}(1) & 1 & 2 &3 & 4\\
		\hline \hline
		1&	0.988788 & 0.0157688 & 0.00910705 & -0.00146615 \\
		2&	0.0315376 & 1.00062 & -0.0204256 & 0.0336059 \\
		3&	0.0273211 & -0.0306384 & 1.07677 & 0.0259935 \\
		4&	-0.0058646 & 0.0672117 & 0.034658 & 1.01014 \\
		\hline
		\end{array}\]
	\end{center}
	%
	%
	%
	\caption{A few of the purely even parity matrix elements $\bR_{2n,2m}$ evaluated at $q = 1$ or $t=0$ using the oscillator based expansion till $q^{18}$. The diagonal elements are all consistent with being $+1$ with the off-diagonal ones vanishing, since the twist matrix $C_{nm} = (-)^n \delta_{nm}$ reduces to $+\delta_{nm}$ in the even sector. In the odd sector, we have checked that there is consistency with $-\delta_{nm}$ as well.}
	\label{tab:revenseries}
\end{table}
We expect these Taylor series expansions to correspond to certain special combinations of elliptic functions. As it is difficult to identify the form of the function from the series---and it varies for each matrix element---we tried to look up the numbers in the OEIS\footnote{It offers a feature to check rational sequences by searching for the numerator sequence and denominator sequence separately.}. Although we haven't found any match so far, it may be possible that one can express these in terms of rational expressions\footnote{This is expected the case as the Schottky double is a torus and elliptic functions are the natural doubly periodic functions should appear in any physical quantity\cite{Bluhm:1989ws,Samuel:1989fea,Ellwood:2003xc}.} of elliptic functions and their derivatives, line integrals, etc.
Once one obtains the expression in terms of elliptic functions, one can convert them to Jacobi $\Theta$ functions and then apply the Jacobi imaginary transform to obtain the closed string contributions explicitly, similar to \cite{Bluhm:1989ws}.

\begin{figure}[h!]
	\begin{subfigure}[b]{0.48\textwidth}
		\centering
		\includegraphics[width=\textwidth]{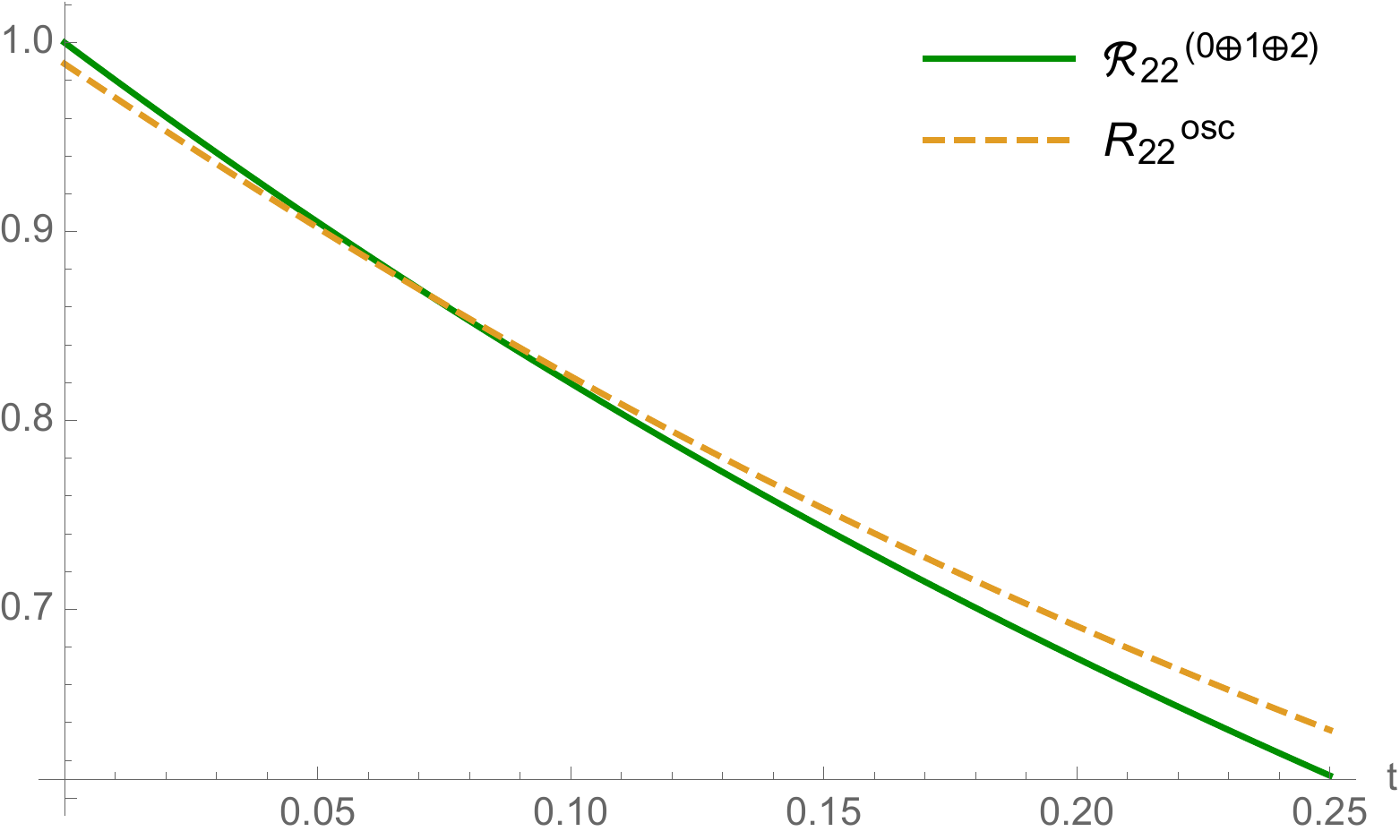}
		\caption{}
		\label{fig:r22oscs012}
	\end{subfigure}
	\begin{subfigure}[b]{0.48\textwidth}
		\centering
		\includegraphics[width=\textwidth]{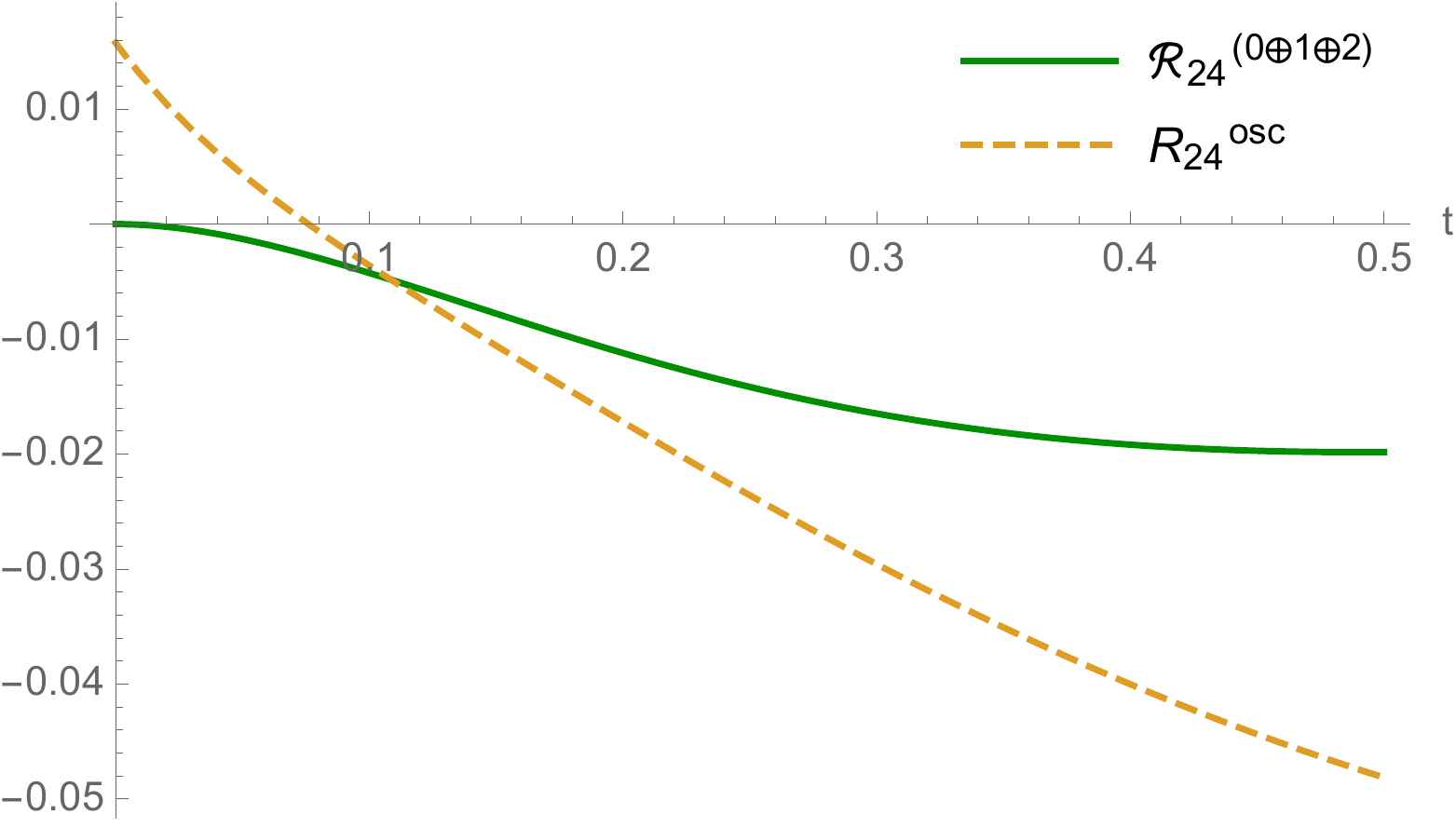}
		\caption{}
		\label{fig:r24oscs012}
	\end{subfigure}
	\label{fig:MoyalOscComparison}
	\caption{A comparison of the behaviour of the matrix element $\bR_{2n,2m}$ near $t=0$ obtained using the first three terms ($s=0, 1, 2$) in the Moyal basis (green) and using the first $19$ terms (till $q^{18} = e^{-18t}$) in the oscillator basis (orange, dashed) plotted for (\subref{fig:r22oscs012})$\bR_{22}(t)$ and (\subref{fig:r24oscs012})$\bR_{24}(t)$. The two furnish very similar values for the diagonal case but differ for the non-diagonal case, which was expected given the vanishing behaviour near $t=0$.}
\end{figure}

Furthermore, it is interesting to compare this expansion to the one we obtained in \S\ref{subsec:evenparity} for the general even parity matrix elements directly in the $t$ variable. We find that they do  all follow each other sufficiently closely near the $t\to0^+$ region (which maps to the $q\to1^{-}$ region) as can be seen from some sample matrix elements plotted in Fig.\ref{fig:r22oscs012}; in the non-diagonal case there is a numerical difference since the true function is expected to vanish as $q\to 1^-$, but notice that the scales differ. 

\section{Comments on the string propagator in the ghost sector}\label{sec:prop}

In this section, we write down the  ghost sector expressions for the corrections to open string propagator at the one-loop level ($N = 2, g = 1$). This corresponds to the self-energy diagram in QFTs and in case of the bosonic theory, the diagrams are similar\footnote{The extended nature of the world-sheet however also allows for ``twisting'' the internal propagators \cite{Gross:1970eg}.} to the $\phi^3$ theory.  We will begin by reviewing the covering of the bosonic moduli space using the four relevant string diagrams \cite{Bluhm:1989ws,Freedman:1987fr} after some preliminary remarks concerning ghost charges. Notice that there are two bosonic moduli $t_1$ and $t_2$ each ranging from $0$ to $\infty$ and the analytic structure becomes much more intricate (and interesting) consequently.

\subsubsection*{On ghost number assignments}

Recall that for one-loop\footnote{To be precise,  what we call ``one-loop'' here would correspond to the lowest order $\cO(\hbar^{1/2})$ correction if the relation between the open string and the closed string coupling are taken into account and hence would actually be ``half-loop'' ($\sqrt{\hbar}$) level, as per standard Polchinski conventions.} diagrams,  the perturbative quantization procedure dictates that states of all ghost number, $G_i\in \mathbb{Z}$, must propagate in the loop subject to the ghost number saturation condition for the corresponding genus by the Riemann-Roch theorem. These are the so called spacetime ``ghost strings'' which are different from the ordinary reparametrization $bc$ ghosts \cite{Giddings:1986bp} on the worldsheet.

Applying this rule for the one-loop $2$-point function, we require that the vertex operators for the two states $\ket{\Phi_1}$ and $\ket{\Phi_2}$ corresponding to the two propagators of ``length'' $t_1 = -\ln q_1$ and $t_2=-\ln q_2$ carry the ghost charges:

\begin{equation}
G_{2} = 3-1- G_1=2-G_1,
\end{equation}
when both external lines are connected to the loop but only $G_1 = 1 = G_{2}$ when only a single line is connected to the loop as in . This results from the requirement of total ghost number $+3$ for the Witten type vertex. For the first case, the condition requires that both states be of either even or odd ghost number which is true also for the Schnabl gauge analysis\cite{Kiermaier:2008jy}.

While considering the two diagrams of the first type, we will account for only the contribution from the ghost number $+1$ quantum states in this work. However, while constructing the quantum effective action it is essential that we remove this restriction. Hence, our analysis would necessarily be limited in its physical validity. The remaining two are not one particle irreducible and have the tadpole as a subgraph. Hence they share some of the structures. At the end, all the four diagrams should be added with equal weight ($=+1$) in order to match  with the first quantized results on-shell \cite{Gross:1970eg}.

\subsubsection*{Covering of moduli space}\label{subsubsec:moduli}
\begin{figure}[ht!]
	\begin{subfigure}[b]{0.48\textwidth}
		\centering	\includegraphics[scale=0.42]{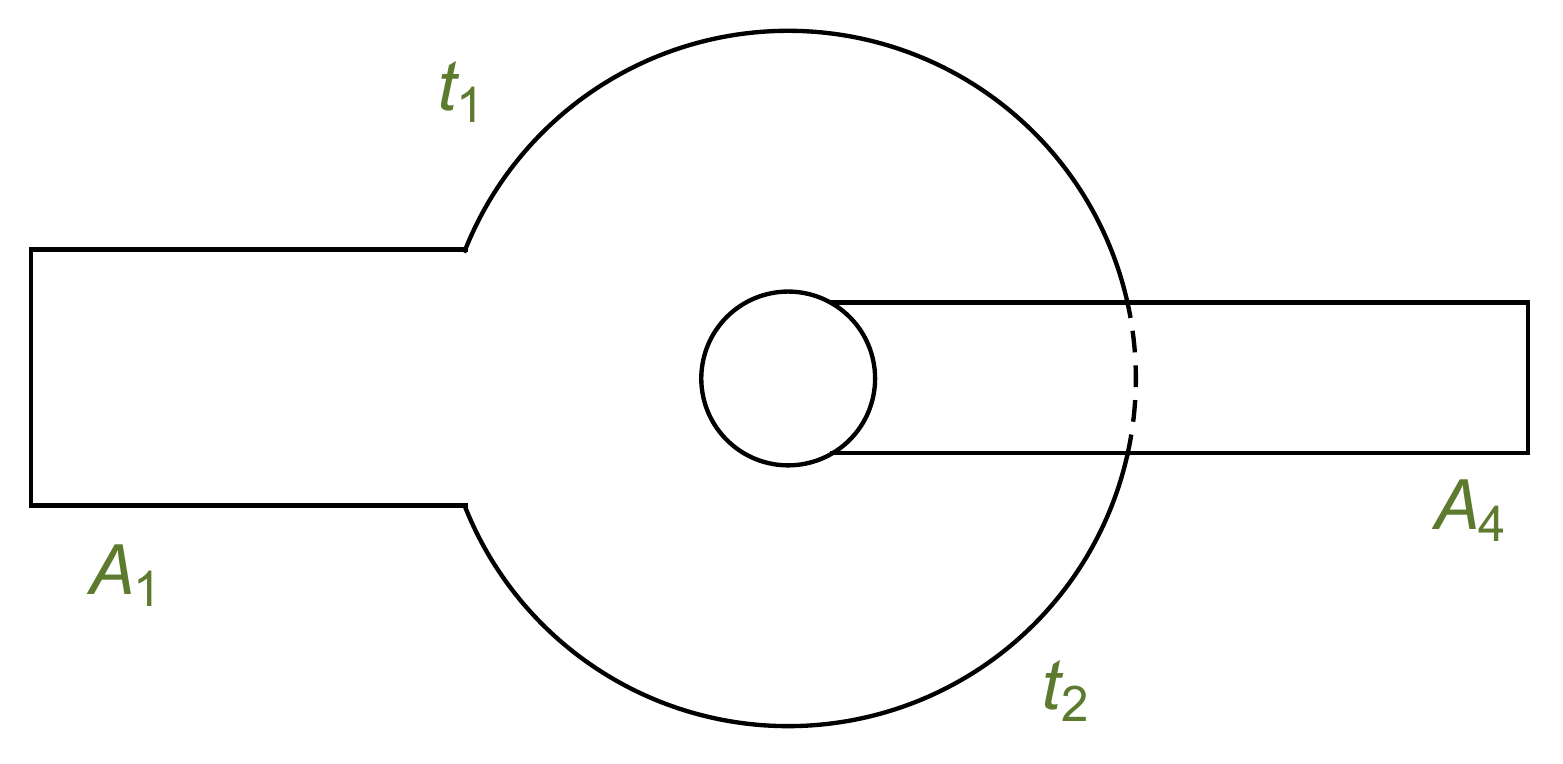}
		\caption{}
		\label{fig:nonplanar}
	\end{subfigure}
	\begin{subfigure}[b]{0.48\textwidth}
		\centering	\includegraphics[scale=0.48]{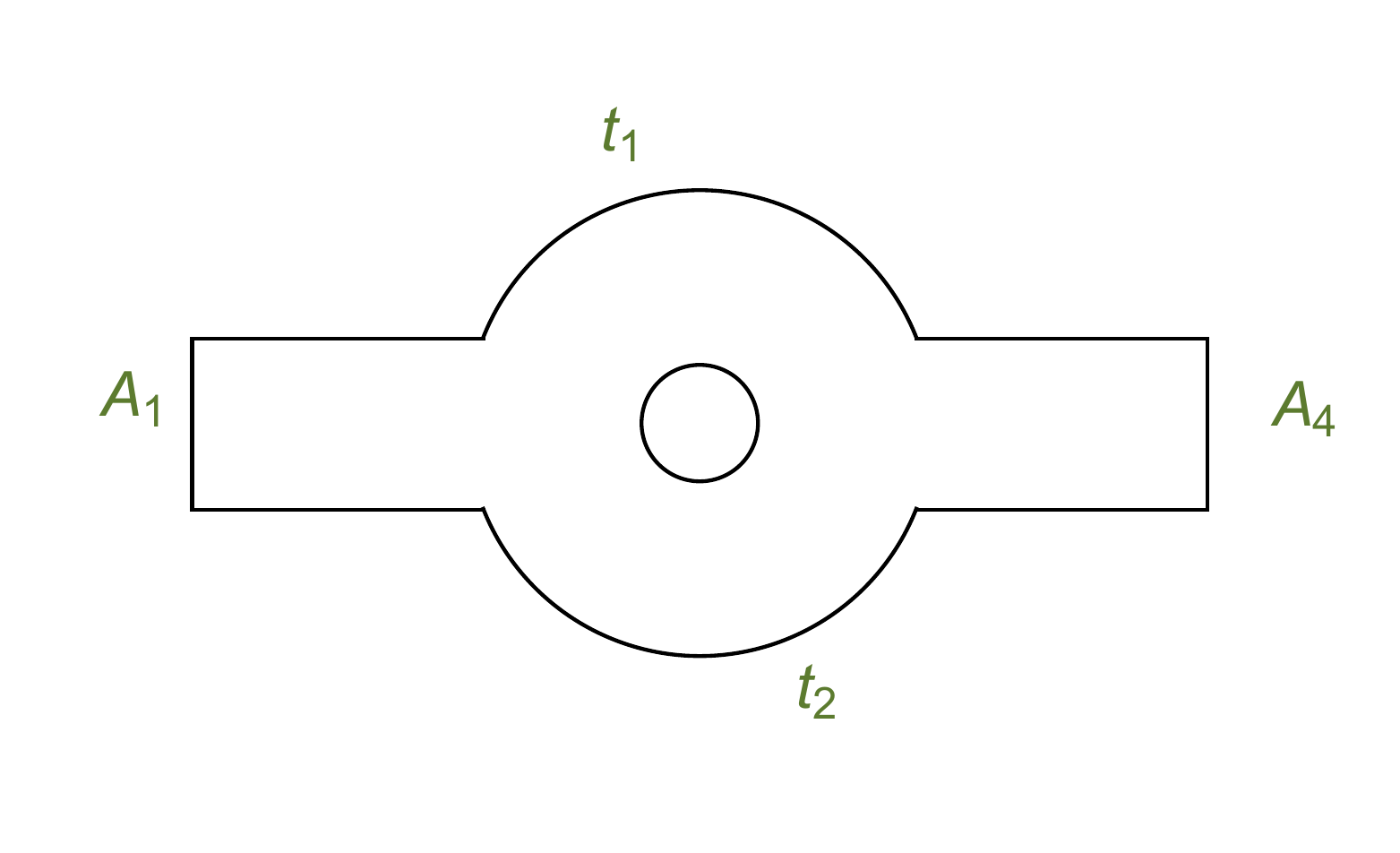}
		\caption{}
		\label{fig:planar0}
	\end{subfigure}
	\begin{subfigure}[b]{0.48\textwidth}
		\centering	\includegraphics[scale=0.44]{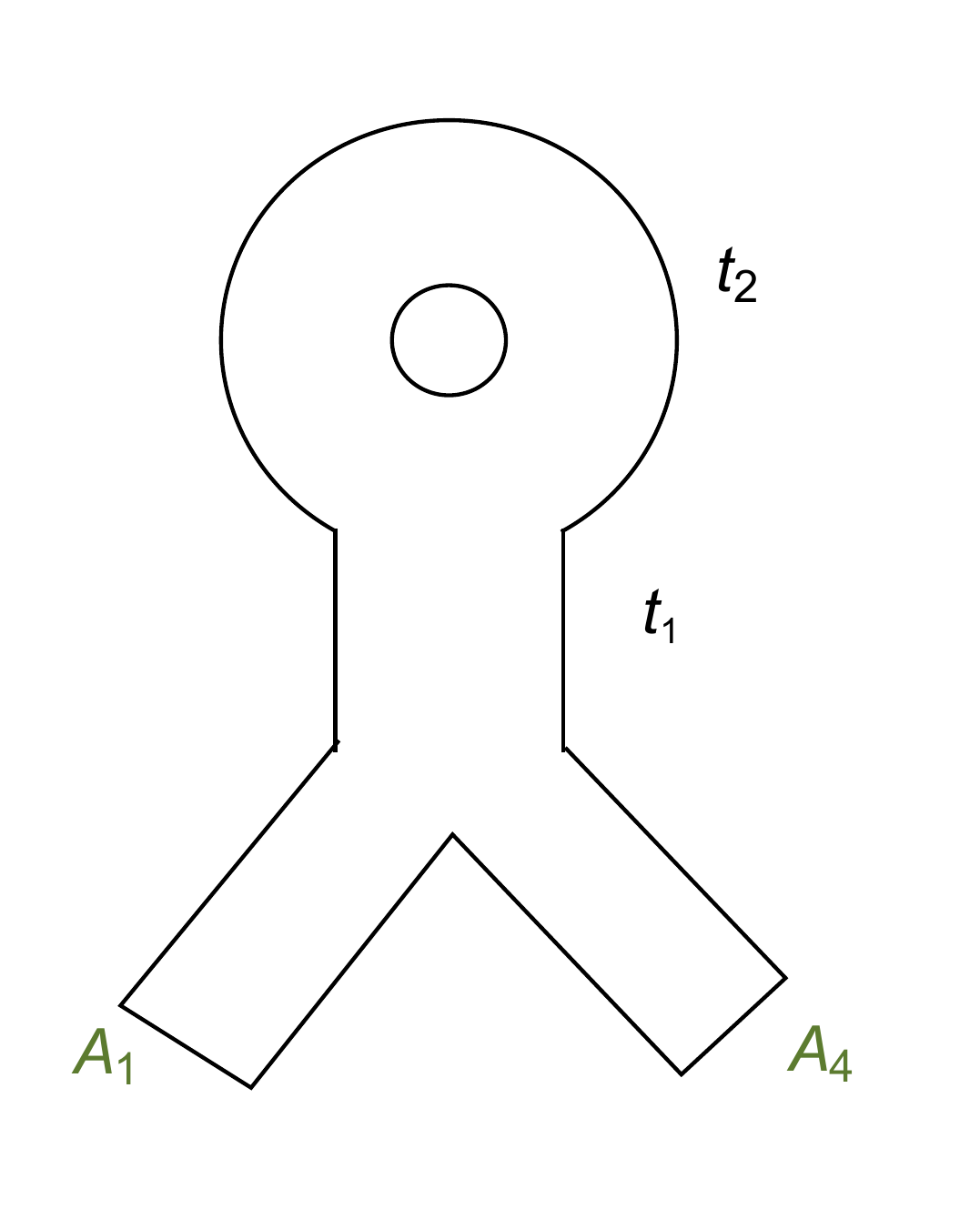}
		\caption{}
		\label{fig:planar1}
	\end{subfigure}
	\begin{subfigure}[b]{0.48\textwidth}
		\centering	\includegraphics[scale=0.42]{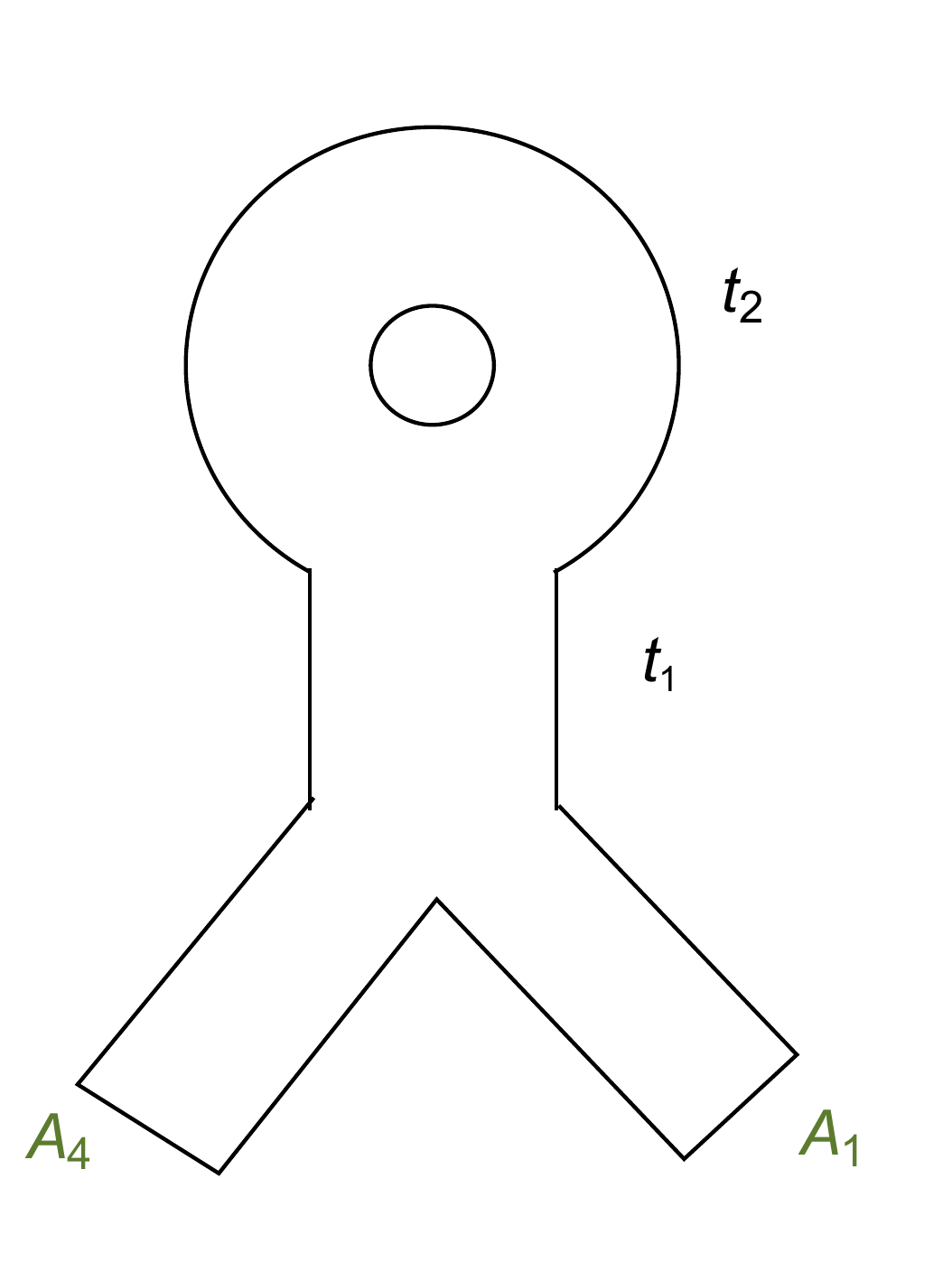}
		\caption{}
		\label{fig:planar2}
	\end{subfigure}
	\caption{The four diagrams contributing to one-loop $2$-point function. Diagrams (\subref{fig:nonplanar}) and (\subref{fig:planar0}) may be considered to arise from the $s$ channel and the last two: (\subref{fig:planar1}) and (\subref{fig:planar2}), from the $t$ channel.}
	\label{fig:stringprops}
\end{figure}

As expounded in \cite[\S5]{Bluhm:1989ws} by Samuel et al., the moduli space is covered by four string diagrams as depicted in Fig.\ref{fig:stringprops}, of which one is \emph{non-planar} and the rest three are planar.\footnote{Planar is used in the sense of Feynman graphs;  the string diagrams are still non-planar due to the unique structure of the Witten type vertex.} Of these three planar cases, two have the one loop tadpole as a subdiagram and hence has zero momentum transfer. With the appropriate change of variables, these are guaranteed to have the same form of the integrand. These diagrams smoothly cross-over as the modular parameters are varied in order to provide a single covering of the moduli space (Ref.\cite{Bluhm:1989ws} clearly demonstrates this). Additionally, the ghost factors are no longer trivial as in the tree level cases. 

We see that the last two diagrams differ by the way two legs of the \emph{off-shell} four-point function are glued together to form a loop. The Witten vertex is cyclic but not permutation symmetric and hence we find these inequivalent diagrams for obtaining a single covering of moduli space as required by consistency with the Polyakov amplitudes on-shell. In the following, we consider only pure ghost external states for convenience.
It was shown that in addition to the physical poles corresponding to an intermediate particle going on-shell, there are also \emph{unphysical} poles in the off-shell amplitude\cite{Bluhm:1989ws,Freedman:1987fr}. Hence these diagrams contribute to very interesting off shell structure.

\subsection{The non-planar integrand $\mathcal{I}^{(s)}_{12|43}$}\label{subsec:non-planar}
The non-planar contribution to the open string propagator is given by:
\begin{equation}
\mathcal{A}^{(s)}_{12|43}=\int_0^\infty dt_1 \int_0^\infty dt_2 \, \mathcal{I}^{(s)}_{12|43}(t_1,t_2)
=\int_0^1 \frac{dq_1}{q_1} \int_0^1 \frac{dq_2}{q_2} \, \mathcal{I}^{(s)}_{12|43}(q_1,q_2),
\end{equation}
where the legs $2$ and $3$ are identified, and the labels $2$ and $3$ are therefore redundant. The integrand (in the ghost sector) can be expressed as below:
\begin{equation}\label{eq:non-planar integrand}
\mathcal{I}_{12|43}^{(s)}(q_1,q_2)= \int  d \egh \,\Tr \left[A_{12}(q_1,\egh,\xigh,\lgh_1)\ast A_4(\xigh,\lgh_4) \ast A_3(q_2,\egh, \xigh)\right],
\end{equation}
and we have explicitly indicated the arguments for clarity.
Here we have defined the monoids:
\begin{align}
A_{12}(q_1,\egh,\xigh,\lgh_1)&= q_1^{L_0^{gh}}\left[A_1(\xi,\lgh_1)\ast e^{+\bxigh \egh}\right]\nonu
A_3(q_2,\egh,\xigh)&= q_2^{L_0^{gh}}\bls e^{-\bxigh \egh}\brs,
\end{align}
in terms of the simpler elements:
\begin{align}
A_1(\xigh,\lgh_1)&= \cN_0 e^{-\bxigh M_0^{gh}\xigh-\bxigh \lgh_1},\nonu
A_{2}(\egh,\xigh)&= e^{+\bxigh \egh},\nonu
A_3(\egh,\xigh)&= e^{-\bxigh \egh},\nonu
A_{4}(\xigh,\lgh_4)&= \cN_0 e^{-\bxigh M_0^{gh}\xigh -\bxigh \lgh_4}.
\end{align} 
where as before, $\lgh_1,\lgh_4$ are the sources which may be used to insert the specific asymptotic string states at the end of the calculations. Here again, we choose to remove the first $\ast$ product that appears between $A_{12}$ and $A_4$ while evaluating the trace under the assumption of associativity. We apply the sub-algebra rules for the monoid and the propagator rules to write down the parameters for the resulting string fields below:
\begin{itemize}
	\item[$A_{12}^{gh}(q_1):$]\qquad
	\begin{align}
	M_{12}(q_1)&=M_0^{gh},\qquad \lgh_{12}(t_1)=q_1^{\ktgh}(-(1-m_0^{gh})\egh+\lgh_1),\nonu
	\cN_{12}(q_1)&=\cC_\eta^{(12)} \exp\left[- \begh\cQ_{\eta}^{(12)} \egh + \lambda_1^{gh\top}\cL_\eta^{(12)\top}\egh\right],
	\end{align}
\end{itemize}
where the coefficient matrices appearing in the normalization factor $\cN_{12}$c are given by:
\begin{subequations}
	\begin{align}
	\cQ_\eta^{(12)}&= -\aqua\sigma m_0^{gh} -\frac{1}{8} (1-m_0^{gh\top})M_0^{gh-1}f_3(q_1)(1-m_0^{gh})\\
	\cL_\eta^{(12)\top}&= -\half \left[\sigma+\frac{1}{2}M_0^{gh-1}f_3(q_1)\right]\\
	\mathcal{C}_\eta ^{(12)}&=\cN_0\exp\left[\frac{1}{8}\lambda^{gh\top}_1 M_0^{-1}f_3(q_1)\lgh_1\right],
	\end{align}
\end{subequations}
and for the monoid
\begin{itemize}
	\item [$A_3^{gh}(q_2)$:]\qquad
	\begin{align}  M_3(q_2)&= \frac{f_3(q_2)}{f_2(q_2)} M_0^{gh},\qquad \lgh_3(q_2) = \frac{2q_2^{\ktgh}}{f_2(q_2)} \egh\nonu
	\cN_3(q_2)&= \det\left[\half f_2(q_2)\right]\exp\left[+\aqua \begh M_0^{-1}\frac{f_3(q_2)}{f_2(q_2)} \egh \right].
	\end{align}
\end{itemize}
Now, let us proceed to evaluate the string field resulting from taking  $A_4\ast A_3(q_2)\eqqcolon A_{43}(q_2)$. The parameters for the resulting expression are the following:
\begin{subequations}
	\begin{align}
	m_{43}(q_2)&= \left[m_0^{gh}+\frac{f_3(q_2)}{f_2(q_2)}(\mgh)^2\right]\left[1+\frac{f_3(q_2)}{f_2(q_2)}(\mgh)^2\right]^{-1}\simbreak
	+
	\left[\frac{f_3(q_2)}{f_2(q_2)}m_0^{gh}-m_0^{gh}\frac{f_3(q_2)}{f_2(q_2)}m_0^{gh}\right]\left[1+\mgh\frac{f_3(q_2)}{f_2(q_2)\mgh}\right]^{-1},\\
	\lambda_{43}(q_2)&= 2[1-\mgh]\left[1+\frac{f_3(q_2)}{f_2(q_2)}(\mgh)^2\right]^{-1}\frac{q_2^{\ktgh}}{f_2(q_2)}\egh + \left[1+\frac{f_3(q_2)}{f_2(q_2)}\mgh\right]\left[1+\mgh \frac{f_3(q_2)}{f_2(q_2)}\mgh\right]^{-1}\lambda_4^{gh},\\
	\cN_{43}&= \cN_0\cN_3(q_2)\det\left[1+\frac{f_3(q_2)}{f_2(q_2)}(\mgh)^2\right]\exp\left[+\aqua \lambda_\alpha^\top\sigma K_{\alpha \beta}\lambda_\beta\right],
	\end{align}	
\end{subequations}
where in the last expression, we must substitute:
\begin{align}
K_{44} & = \left(\mgh+(\mgh)^{-1}\frac{f_2(q_2)}{f_3(q_2)}\right)^{-1},\qquad
&K_{43}  &= \left(1+\frac{f_3(q_2)}{f_2(q_2)}(\mgh)^2\right)^{-1},\nonu
K_{34} & = - \left(1+\mgh\frac{f_3(q_2)}{f_2(q_2)}\mgh\right)^{-1},\mbox{~and~}
& K_{33}&= \left(\frac{f_3(q_2)}{f_2(q_2)}\mgh+(\mgh)^{-1}\right)^{-1}.
\end{align}
Combining the two string fields by taking the ordinary product of functions and taking the $\xigh$ trace, we are left with
\begin{align}
\cC ^{(12|43)}_\eta	\int (d\eta)\, \exp\left[-\eta^\top\cQ^{(12|43)}_\eta\eta+\cL^{(12|43)\top}_\eta\eta\right]
\end{align}
where the $\lambda_{1,4}$ dependences are implicit. Thus, the final contribution from the ghost sector becomes the following:

\begin{equation}\label{eq:non-planar ghost contribution}
\det\blp2\cQ^{(12|43)}_\eta\brp\exp\left[+\aqua \cL_\eta^{(12|43)\top}\cQ_{\eta}^{(12|43)-1}\cL_\eta^{(12|43)}\right].
\end{equation}
Here, the argument of the exponential mixes the components of the ``vector'' 
\[\vec{\lambda}=\left(\begin{array}{c}
\lambda_1^{gh}\\ \lambda_{4}^{gh}
\end{array}\right) \]
and for a general one-loop $n$-point function, we obtain an $n$ component vector. This is similar in spirit to working in Fourier space but here we only work with $2N\times 2N$ matrices. In certain cases, we can use these formal expressions for numerical calculations; the advantage of this representation is the straightforward application of the transformation rules, although they involve several inverses of infinite matrices and intermediate matrix multiplications.

\subsection{The planar graphs}\label{subsec:planar}
One may observe that the planar amplitude with both external states on the same boundary of the annulus, $\mathcal{A}^{(s)}_{12|34}$,
comes with a relative positive sign with respect to the amplitude above. Hence, the combined integrand can be written as
\begin{equation}
\mathcal{I}^{(s)}_{12|43}+\mathcal{I}^{(s)}_{12|34} = 
\int  d \egh \,\,\Tr \left[A_{12}(q_1,\egh,\lgh_1) \left\{A_4(\lgh_4) , A_3(q_2,\egh)\right\}_\ast\right].
\end{equation}
This is special for the $2$-point function since for general diagrams, the permutation \emph{non-invariance} of the Witten vertex requires that we treat such diagrams, with lines on different boundary components, as contributing to separate amplitudes in general (colour ordering). The anti-commutator structure in  the amplitude allows for taking advantage of the partial twist symmetry of these monoid elements. Here, we simply remark that we can write:
\begin{equation}
\hat{A}_3\ast \hat{A}_4 = (-)^2\Omega\left(\Omega(\hat{A}_4)\ast \Omega(\hat{A_3})\right),
\end{equation}	
where we have included the ghost zero modes $\xi_0^i$ in the form of $\hat{A}_i = - \xi_0^{(i)}A_{i}$ for clarity. This leads to some partial simplifications and we hope to report in this direction in the future.

As mentioned earlier, the two remaining planar graphs have the one-loop tadpole as a subdiagram and are related by interchange of the external states labelled $1$ and $4$.  We consider the integrand
\begin{equation}
\cI^{(t)}_{41|23}(q_1,q_2) = \int  d \egh \,\,\Tr \left[A_{41}(q_1,\lgh_1,\lgh_4)\ast A_2(\egh) \ast A_3(q_2,\egh)\right].
\end{equation} 
which one can think of as being obtained by identifying the $2$ and $3$ legs of a $t$ channel diagram (see Fig. \ref{fig:stringprops}). One can again write down  formal expressions for the parameters in the integrand in terms of lightcone like variables, although we are unable to simplify them for further analysis at this point.

%

\section{Concluding remarks}\label{sec:disc apology}
In this work, we have primarily focussed on the finite contributions from the squeezed state matrix elements $\bR_{nm}(t)$ characterizing the tadpole state in the ghost sector of OSFT, and looked for hints of non-analyticity as a function of the modular parameter $t$. Using the Moyal representation of the star product, we were able to write down formal expressions for the generating functionals for correlators. Since all integrals in this formalism are of the Gaussian kind, we obtained these in terms of determinants and inverses of infinite matrices, which is one of the main difficulties with these methods.  

Due to the partial analytic control we have over the infinite matrices, we were able to study the behaviour of $\bR_{nm}(t)$ near the two boundaries of moduli space by employing expansions in $t$ and in $q = e^{-t}$ for the matrix inverse---although conformal techniques become awkward in this basis. In particular, we were able to demonstrate the utility of the formalism by correctly capturing the linear order behaviour (\S\ref{subsec:linear order}) near $t=0$ which matches with BCFT prediction \eqref{eq:R linear behaviour}. However, we are now able to see this purely from the OSFT perspective. In the oscillator representation, this expansion  becomes ill-defined and produces results that differed by a factor of $2$.
In the process of identifying the zeroth and linear order coefficients, we have thus uncovered a subtle difference between the Moyal and the oscillator methods, owing to the Fourier transform \eqref{eq:Fourier} and the resulting somewhat peculiar form of the propagator \eqref{eq:peculiar propagator}.

Ideally, one would like to see the signatures of the closed string states by generating an expansion involving the closed string variable $\hat{q}\coloneqq e^{2\pi^2/\ln q}$, starting from a closed form expression in the $q$ variables and doing the Jacobi imaginary transform. This way one could recover the off-shell physics associated with the closed string spectra. Unfortunately, the algebraic approach we employ in this work is not tailored for this endeavour and hence we have studied the effects of closed string physics only indirectly. 

Nonetheless, we have performed consistency checks of our analytic expressions by examining various limiting regimes of interest and found general agreement with the oscillator and BCFT results. Beyond the linear order, we are able to successively approximate the matrix elements of $\bR_{nm}(t)$. However, the algebra becomes quite unwieldy as may be expected from the fact that the aforementioned infinite matrices are constructed out of non-commuting blocks. We have also employed the oscillator expression \eqref{eq:Rnmosc expr} to generate a series in $q$ till the $18th$ degree (for general $n, m$) and used it to analyse hints of non-analyticity. We however, refrain from making any claims pertaining to the margin of errors or the efficiency yet, since these are much less clear.

To summarize, the present work makes a modest attempt at answering perturbative questions in OSFT using the Moyal formalism and complements the CFT and oscillator investigations. Due to the strong divergences from the closed string tachyon, the full amplitude is unphysical but still serves as a useful probe of the structure of this very special string field theory. Recently, more physical superstring field theories have been fully constructed which can describe the Ramond sector \mcite{ossft,*Kunitomo:2015usa,*Erler:2016rxg,*Erler:2016ybs,konopka2016construction,Konopka:2015tta,Erler:2017pgf}. The work in \cite{Kunitomo:2016bhc} has correctly reproduced the $4$-point amplitude involving spacetime fermions. It would be of utmost interest to study quantization of this theory from which the tachyon is projected out.

One promising avenue would be extending the recent progress made in the direction of \textit{partial gauge fixing} \mcite{matsunagakun,*Matsunaga:2015kra,*Matsunaga:2016zsu,*Matsunaga:2018hlh,*Goto:2015hpa}. This still remains somewhat mysterious and a better understanding of the gauge algebra at the quantum level may also shed more light on how closed string degrees of freedom are encoded in open superstring field theories.

\section*{Acknowledgements}
I wish to thank Itzhak Bars, Loriano Bonora, Ted Erler, Yuji Okawa, and Martin Schnabl for helpful conversations and discussions. I am grateful to the participants and organizers of \href{http://www.hri.res.in/~strings/mschedule.html}{\textit{SFT 2018, HRI, Allahabad}}, for providing a kind and inspiring environment during which part of this work was done. I also thank the UCSD Mathematics Department for making available the \href{http://math.ucsd.edu/~ncalg/}{\textit{NCAlgebra}} package using which parts of the calculations in this paper were performed.

\newpage
\appendix

\section{Determinant factors}\label{sec:determinant}
In order to compute the divergent part of the integrand, we must include the matter contribution as well. Here we present this computation; we will use this combined result in the following to look at the convergence properties of the finite $N$ regularization for the simplest loop amplitude. 
\vspace{-4mm}
\subsection*{Matter sector Gaussian integrals}
\vspace{-2mm}
Similar to the ghost sector, we evaluate the matter sector integrand by performing the state sum over matter degrees of freedom by choosing a Fourier basis: $e^{-i\bxi \eta}e^{ip\bar{x}}$. Here, however, the presence of the matter zero mode results in additional terms which were absent in the ghost sector by virtue of the Feynman-Siegel gauge. 

\noindent
The matter contribution to the integrand is given by:
\begin{equation}
\cI_e^X(q) = \int d^d \bar{x}\, \int \frac{d^dp}{(2\pi)^d}\,\frac{(d\eta)}{(2\pi)^{2dN}} \Tr \left[A_e\ast e^{-i\bxi \eta-ip\bar{x}}\ast (q^{\hat{L}_0^X-1}e^{i\bxi \eta+ip\bar{x}})\right]
\end{equation}
Once again, we choose the monoid elements appropriate for excitations on the perturbative vacuum:
\begin{align}\label{eq:matter monoids}
A_1(\xi, \lambda)&= \cN_0^X e^{-\bxi M_0^X \xi - \bxi \lambda},\nonu
A_2(\xi,\eta,p)&= e^{-i\bxi \eta}e^{-ip\bar{x}},\nonu
A_3(\xi, \eta,p)&= e^{+i\bxi \eta}e^{+ip\bar{x}},
\end{align}
where we have assumed Lorentz symmetry over all the matter indices in $\xi^\mu$. Note the extra factors of $i$ as compared to the ghost sector and the loop momentum $p^\mu$. The monoid $A_1(\xi)$, which serves as a generating functional, is chosen to have zero momentum since this is the one-point function for the $D25$ brane case. We recall for convenience that the matrix $M_0^X$ and the normalization factor  used in defining the matter vacuum are given by
\begin{equation}
M_0^X = \left[\begin{array}{cc}\aqua \ke
&  \mzero\\ \mzero
& 4T\ko^{-1}T^\top
\end{array} \right],\qquad \cN_0^X = \det(4\sigma M_0^X)^{d/4}=2^{Nd} (1+w^\top w)^{d/8}.
\end{equation}
As before, we sequentially apply the monoid algebra rules for doing the string products. The rules in the matter sector are identical to the ghost sector with the choice of basis we are using, including the signs in the exponentials from Gaussian integrations. In \cite{Bars:2002nu} to obtain  the parameters for the monoids $A_{12}\coloneqq A_1\ast A_2$:
\begin{itemize}
	\item [$A_{12}(\xi,p)$:] \hspace{2 cm}
	\begin{align}
	M_{12} &=  M_0^X,  &\lambda_{12}&=(1-m_0^X)(i \eta)+\lambda, \nonu \mathcal{N}_{12} &= \mathcal{N}_0 \exp \left(-\frac{1}{4}\eta^\top\sigma m_0 \eta +\frac{i}{2}\lambda^\top\sigma\eta\right), &p_{12} &= -p.
	\end{align}
\end{itemize}
For the propagator rules\cite{Bars:2002qt}, there are extra terms from the momentum $p$: 
\begin{subequations}
	\begin{align}
	M(t) &= \left[\shkt +(\shkt +M_0^X M^{-1}\ckt)^{-1}\right](\ckt)^{-1}M_0^X,\\
	\lambda(t) &= \left[(\ckt+MM_0^{X-1}\shkt)^{-1}(\lambda+i w p)\right] - i w p,\\
	\cN(t) &= \frac{\cN e^{-p^2 t}\exp\left[\aqua (\lambda+i pw)^\top (M+\coth t\kt M_0^X)^{-1}(\lambda+i w p)\right]}{\det\left(\half(1+MM_0^{X-1})+\half(1-MM_0^{X-1})e^{-2t\kt}\right)^{d/2}}.
	\end{align}
\end{subequations}
Applying this to $A_3$ in \eqref{eq:matter monoids} and rewriting the hyperbolic functions in terms of the functions $f_i(\kt; q)$, we have the parameters for
\begin{itemize}
	\item [$A_3(\xi,p,q)$:] \hspace{2cm}
	\begin{align}
	M_3(q)&=  \frac{f_3(q)}{f_2(q)} M_0, \qquad \lambda_3(q)= \frac{2q^{\kt}}{f_2(q)}(-i\eta+i w p)-i wp, \nonu
	\mathcal{N}_3(q) &= \frac{2^{dN}q^{p^2}}{\det(f_2(q))^{d/2}}\exp \left[-\frac{1}{4}(\eta-w p)^\top M_0^{-1}\frac{f_3(q)}{f_2(q)} (\eta-w p)\right], \qquad p_3 = +p.
	\end{align}
\end{itemize}
As before one may now remove the remaining $\ast$ product in the trace and simply set $A_{12}A_3(q)\eqqcolon A_{123}(q)$ with parameters:
\begin{align}
M_{123}(q) &= \frac{2}{f_2(q)} M_0, \nonu
\lambda_{123}(q) &= i \left(\frac{f_1(q)}{f_2(q)}-m_0\right) \eta-i \frac{f_1(q)}{f_2(q)}w p +\lambda, \nonu
\mathcal{N}_{123}(q) &=  \mathcal{N}_{12} \mathcal{N}_3(q),\qquad p_{123}=0. 
\end{align}
The trace operation is simply a functional Gaussian integral\footnote{with the appropriate factors of $2\pi$ and $i$ s inserted in the measure.} and produces
\begin{align}\label{eq:matter trace}
\Tr [A_{123}(q)] &= \frac{\mathcal{N}_{123}(q)}{\det(2M_{123}(q)\sigma)^{d/2}}\exp \left(\frac{1}{4}\lambda^\top_{123} M_{123}^{-1}\lambda_{123} \right)\nonu
&\coloneqq \cC_\eta \exp \left[-\eta^\top\cQ_\eta \eta + \cL^\top_\eta \eta\right],
\end{align} 
where we suppress the $q$ dependence for typographical simplicity. Collecting the $\eta$ dependence from the various factors, the coefficient matrices which appear in the quadratic exponential above are the following
\begin{align}
\cQ_\eta &= \cQ_{\eta|1}+\cQ_{\eta|2}+\cQ_{\eta|3}, \mbox{~~with}\nonu
\cQ_{\eta|1}&=\aqua \sigma m_0, \qquad \cQ_{\eta|2} = \aqua M_0^{-1}\frac{f_3(q)}{f_2(q)},\qquad
\cQ_{\eta|3}= \frac{1}{8}\left(\frac{f_1}{f_2}-m_0\right)^\top M_0^{-1}f_2\left(\frac{f_1}{f_2}-m_0\right),\nonu
\cL_\eta &= \cL_{\eta|1}+\cL_{\eta|2}+\cL_{\eta|3}, \mbox{~~ with}\nonu
\cL_{\eta|1}^\top & = \frac{i}{2}\lambda^\top \sigma,\qquad \cL_{\eta|2}^\top= \frac{p}{2}w^\top M_0^{-1}\frac{f_3}{f_2},\qquad
\cL_{\eta|3}^\top = \frac{i}{4}\left(\lambda - i\frac{f_1(q)}{f_2(q)}w p\right)^\top M_0^{-1}f_2(q)\left(\frac{f_1(q)}{f_2(q)}-m_0\right), \mbox{~ and}\nonu
\cC_\eta & = \cC_{\eta|1}\cdot \cC_{\eta|2}\cdot \cC_{\eta|3},\mbox{~~with}\nonu
\cC_{\eta|1}& = \cN_0,\qquad \cC_{\eta|2}= \frac{2^{dN}q^{p^2}}{\det(f_2(q))^{d/2}}\exp\left[-\frac{p^2}{4}w^\top M_0^{-1}\frac{f_3}{f_2}w\right],\nonu
\cC_{\eta|3}& = \det\left(\aqua M_0^{-1}f_2(q)\right)^{d/2}\exp\left[\frac{1}{8}\left(\lambda-i\frac{f_1}{f_2}wp\right)^\top M_0^{-1} f_2 \left( \lambda-i \frac{f_1}{f_2}wp\right)\right].
\end{align}
We can rewrite the above expressions for the \emph{symmetric}  matrix $\cQ_\eta$ after some matrix algebra as:
\begin{equation}
\cQ_\eta^X = \frac{1}{8}\left[M_0^{-1}f_4(q)+\sigma  f_3(q) M_0 \sigma +\sigma f_1(q)-f_1(q)^\top \sigma\right]
\end{equation}
where we remind the reader that $\displaystyle f_4(q)\coloneqq\frac{f_1^2+2f_3}{f_2}$.

\medskip
\noindent
Combining the four terms, we have the block matrix form:
\begin{equation}\label{eq:Qeta matter}
\cQ_\eta^X(q)
=\frac{1}{2}\left[\begin{array}{cc}
\ke^{-1}f_4+T\ko^{-1} f_3(q;\ko)T^\top& -\frac{i}{4} (f_1(\ke)-Tf_1(\ko)R)  \\ 
-\frac{i}{4} (f_1(\ke)-Tf_1(\ko)R)^\top  & \frac{1}{16	}\left(\ke f_3 +R^{\top}\ko f_4R\right)
\end{array} \right]\nonu
\end{equation}
%
%
Performing the Gaussian integration of \eqref{eq:matter trace} over $\eta$ gives
\begin{equation}
\frac{1}{(2\pi)^{dN}} \cC_\eta\left(\det 2Q_\eta\right)^{-d/2}\exp\left[\frac{1}{4}\cL^{\top}_\eta \cQ_\eta^{-1}\cL_\eta\right]
\end{equation}
Now rewriting
\[
\cL^{\top}_\eta =\frac{i}{2} \lambda^\top\alpha^\top+\frac{p}{2}w^\top \beta^\top,
\]
for compactness using a little algebra in terms of
\begin{subequations}
	\begin{align}
	\alpha^\top(q) &\coloneqq \half\left(M_0^{-1}f_1+f_3^\top \sigma\right)
	\\
	\beta^\top(q) &\coloneqq \half\left(M_0^{-1}f_4-f_1^\top \sigma\right)
	\end{align}
\end{subequations}
the argument of the exponential factor involving $\cQ^{-1}_\eta$ becomes:
\[
\frac{1}{16}\left\{-\lambda^\top\alpha^\top\cQ_\eta^{-1}\alpha
\lambda + p^2 w^\top\beta^\top\cQ_\eta^{-1}\beta w + 2 i pw^\top\beta^\top \cQ_\eta^{-1}\alpha \lambda \right\}
\]
where the Lorentz contraction with $\lambda$ is understood. Then identifying the quadratic and linear pieces in the centre of mass momentum $p$ (conjugate to the \emph{matter} zero mode) as follows:
\begin{subequations}
	\begin{align}
	\cQ_p&=-\ln q +\frac{1}{8} w^\top M_0^{-1}f_4 w - \frac{1}{16}w^\top \beta^\top \cQ_\eta^{-1}\beta w \\
	\cL_p&=-\frac{i}{4}w^\top M_0^{-1}f_1\lambda +\frac{i}{8}w^\top \beta^\top \cQ_\eta^{-1}\alpha \lambda
	\end{align}
\end{subequations}
we can finally perform  the integration over $p$, to yield the matter contribution to the generating functional:
\begin{align}
\mathcal{W}^X(t,\lambda)&= \frac{(1+w^\top w)^{d/8}}{(4\pi)^{d(N+1/2)}}\frac{1}{|\det (\cQ_\eta) \cQ_p|^{d/2}}\exp\left[-\lambda^\top\cQ_\lambda^X \lambda \right] \mbox{\qquad where,}\\
\cQ_\lambda^X &= \aqua \left\{\alpha^\top \cQ_\eta^{-1}\alpha - \frac{1}{2}M_0^{X-1}f_2(q)+\frac{1}{16}\frac{\alpha^{\top}\cQ_\eta^{-1}\beta ww^{\top}\beta^{\top}\cQ_\eta^{-1}\alpha}{\cQ_p}\right\},
\end{align}
where $d=26$ is required for having $c=0$ for the BCFT.

\subsection*{Combined integration over $\eta, p$}
\vspace{-2mm}
Another equivalent  form that would be more suitable for numerical calculations of the determinant factor is obtained by performing the integration over $\eta$ and $p$ after combining them into a single $(2N+1)\times 1$ vector $\psi$:
\begin{equation}
\psi \coloneqq \left(\begin{array}{c}\eta
\\ p
\end{array} \right)
\end{equation}
Now, we can trade using the inverse $\cQ_\eta^{-1}$, which is numerically extensive, in favour of working with a larger size matrix while evaluating determinants.
In terms of $\psi$, we can write the expression obtained by taking the trace over $\xi$ \eqref{eq:matter trace} as
\begin{equation}
\Tr (A_{123}(q))= \Gamma_\psi \exp\left[-\psi^\top \cQ_\psi\psi +\cL_\psi^\top \psi\right],
\end{equation}
where the matrix $\cQ_\psi$ would now be $(2N+1)\times (2N+1)$ dimensional and can be written as:
\begin{align}
\cQ_\psi &=\left[\begin{array}{cc}
\bA&  \bB\\ \bB^\top
& \bC
\end{array} \right], \mbox{~where}\nonu
\bA&= \cQ_\eta^X, \mbox{~above in \eqref{eq:Qeta matter}},\nonu
\bB&= \frac{1}{8}(M_0^{-1}f_4 +\sigma f_1)w, \mbox{~ a $2N\times 1$ column vector},\nonu
\bC&=-\ln q+\frac{1}{8}w^\top M_0^{-1}f_4w, \mbox{a scalar}
\end{align}
and the $(2N+1)\times 1$ column vector $\cL_\psi$ would simply be of the form:
\begin{align}
\cL_\psi &=\frac{i}{4}\left[\begin{array}{c}
\left(M_0^{-1}f_1-\sigma f_3\right)\lambda\\
-w^\top M_0^{-1}f_1\lambda
\end{array} \right]\nonu
&=\frac{i}{4}\left[\begin{array}{cc}
\left(M_0^{-1}f_1-\sigma f_3\right)&\mathbb{0}  \\ -w^\top M_0^{-1}f_1
& 0
\end{array} \right] \left[\begin{array}{c}\lambda
\\ 0
\end{array} \right]\eqqcolon \hat{\cL}_\psi \left[\begin{array}{c}\lambda
\\ 
0
\end{array} \right].
\end{align}
The remaining factor $\Gamma_\psi$ is then given by:
\begin{equation}
\cC_\psi = (1+w^\top w)^{d/8}\exp\left[\aqua \lambda^\top M_{123}^{-1}\lambda\right]
\end{equation}
Now, performing the Gaussian integration over $\psi$ as 
\begin{align}
\int \frac{(d\psi)}{(2\pi)^{(2N+1)d}}\, \cC_\psi \exp\left[-\psi^\top \cQ_\psi \psi +\cL_\eta^\top \psi\right]&=\frac{(1+w^\top w)^{d/8}}{(4\pi)^{d(N+1/2)}}\det(\cQ_\psi)^{-d/2}\nonu
&\!\!\!\!\!\!\!\!\!\!
\!\!\!\!\!\!\!\!\!\!\!\!\times\exp\left\{\aqua \lambda^\top \left[\left(\hat{\cL}_\psi \cQ_\psi^{-1}\hat{\cL}_\psi\right)_{2N\times 2N}+\half M_0^{-1}f_2(q)\right]\lambda\right\}.
\end{align}
Here, the first term in the exponential $\displaystyle \left(\hat{\cL}_\psi \cQ_\psi^{-1}\hat{\cL}_\psi\right)_{2N\times 2N}$ denotes the first $2N\times 2N$ square block in the $(2N+1)\times (2N+1)$ matrix $\hat{\cL}_\psi \cQ_\psi^{-1}\hat{\cL}_\psi$ which turns out to be the non-zero entry in that matrix. We use only the $\lambda$ independent factors while numerically computing the matter contribution to the integrand next.
\vspace{-2mm}
\subsection*{Some Numerical Results}
\vspace{-2mm}
In analogy with the analysis done in the oscillator formalism \cite[\S4]{Ellwood:2003xc} to study the determinant factor (the scalar part) to the one-loop tadpole, we plot $\log S_0(t)$ vs $1/t$ for various values of the finite cut-off $N$. This is an interesting exercice as both the oscillator and the Moyal representations have their advantages and disadvantages. In the oscillator case, we have the exact Neumann matrices and the analytical expression for the matter-ghost determinant involves lesser number of inverses and matrix multiplications (which makes a numerical analysis more reliable). However, the level truncated Neumann matrices do not satisfy the Gross-Jevicki non-linear identities and so wouldn't be fully internally consistent. For the Moyal representation, the finite $N$ deformation is in a sense consistent since the matrices satisfy identical algebraic relations as the open-string ($N\to \infty$) limit whenever they do not lead to associativity anomalies. But associativity anomalies are essential to obtain the correct closed string physics in OSFT and hence we examine the convergence rate in the Moyal formalism as well.

We use the finite $N$ versions of the matrices and vectors given in \cite{Bars:2002nu}  for our analysis:
\begin{align}\label{eq:finiteNmatrices}
T_{eo} &=\frac{w_e v_o \ko^2}{\ke^2-\ko^2},  \qquad R_{oe} = \frac{w_e v_o \ke^2}{\ke^2-\ko^2},\nonu
w_e &= i^{2-e}\frac{\prod_{\op}\lvert\ke^2/\kappa_{\op}^2-1\rvert^{\half}}{\prod_{\ep\neq e}\lvert \ke^2/\kappa_{\ep}^2-1\rvert^{\half}}, \qquad v_o= i^{o-1}\frac{\prod_{\ep}\lvert 1-\ko^2/\kappa_{\ep}^2\rvert^{\half}}{\prod_{\op\neq o}\lvert 1-\ko^2/\kappa_{\op}^2\rvert^{\half}}.
\end{align}
Now, in addition to the matter-ghost contribution we have, we need to insert the extra factors which relate the Witten vertex and the Moyal star. Hence, while doing the numerical analysis we have multiplied by the additional factor $-\mu_3^{-1} K^3$, where $K = \frac{3\sqrt{3}}{4}$ and $\mu_3$ is given in  \eqref{eq:vertexnormalization}.
\begin{figure}[h!]
	\centering
	\includegraphics[scale=0.44]{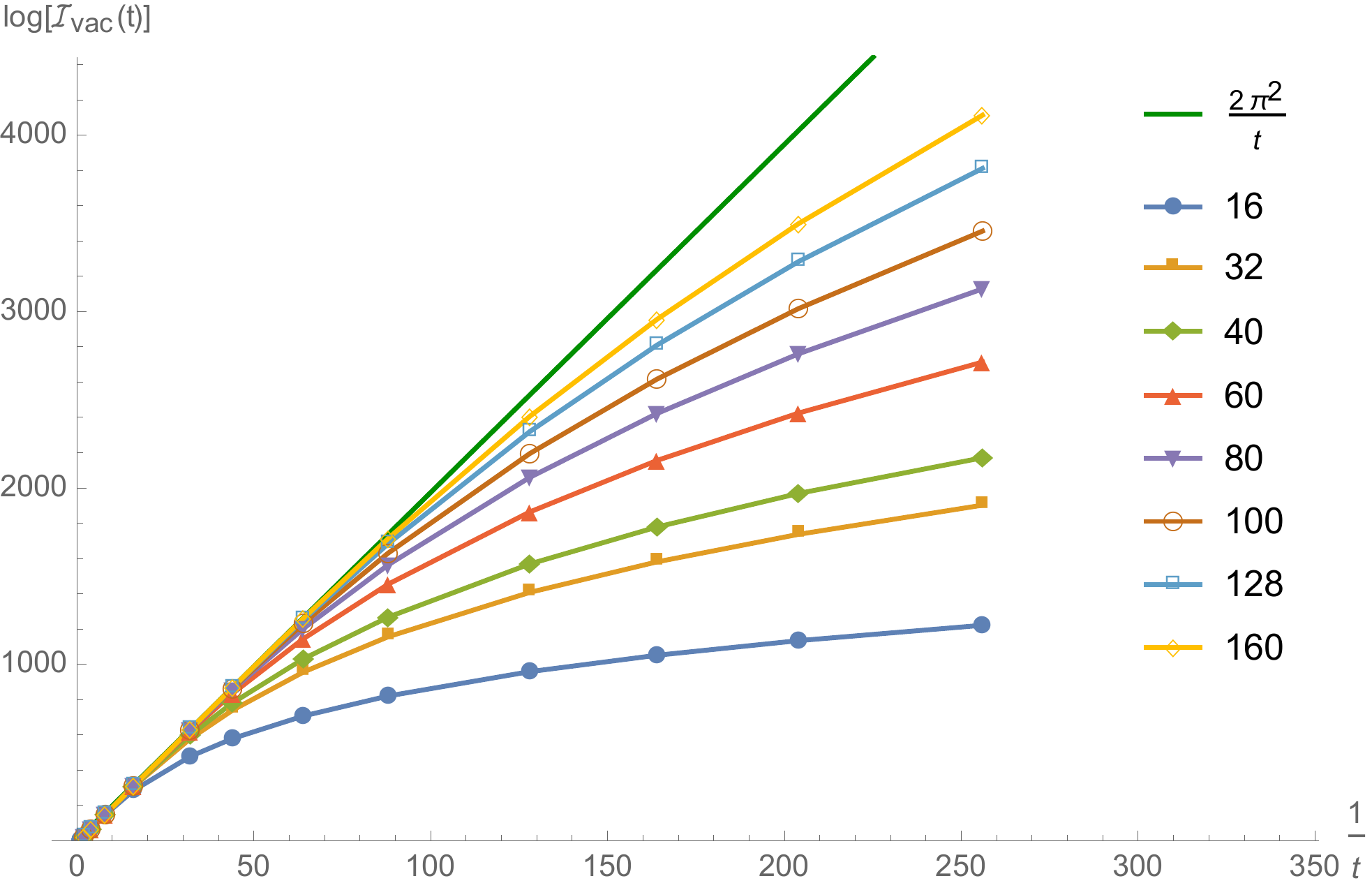}
	\caption{The log of the overlap amplitude with the  perturbative vacuum plotted against $1/t$ for various values of the matrix size $N$. The green line is the expected infinite $N$ behaviour with slope $2\pi^2$. We see that the result steadily approaches this line as $N$ increases.}
	\label{fig:tadpoledet}
\end{figure}
We find that the results are comparable although the convergence rate is not as good. This was to be expected as because of the substructure of the Neumann matrices, more number of inverses and matrix multiplications are required which also affects the convergence rate. However, the relation between the size of the matrices $L$ and $2N$ in the two regularizations is not direct as in the latter, the matrix identities are satisfied even when $N<\infty$. Therefore, there is no analogue of the $UV$ cut-off seen in the level truncation approach as the determinant is still singular for finite $N$ as $t \to \infty$.


\vspace{-2mm}
\section{Pad\'e approximants for $\bR_{nm}(q)$}\label{sec:Pade}
In this appendix, we shall try to infer the analytic properties of the functions represented by the expansions from \S\ref{subsec:osc expansions} in the $q$ plane by considerations of their \emph{Pad\'e approximants}. These are meromorphic functions of the expansion parameter that have identical Taylor series coefficients till the finite data  generated for an unknown function (by using various computational techniques).

Specifically, the $r/s$ Pad\'e approximant till order $N$ is a rational function,  constructed as the quotient of two polynomials of degree $r, s$ respectively such that $r + s=N$ and:
\begin{equation}
P^r_s(z) \coloneqq \frac{A_r(z)}{B_s(z)} =\frac{a_0+a_1 z +\cdots+a_r z^r}{1+b_1 z+\cdots + b_s z^s}= p_0 + p_1 z +\cdots +p_N z^N + R_N(z),
\end{equation}
where the expansion coefficients $p_i$, ($i=0,\cdots, N$) coincide with the series expansion at hand. Generally, the diagonal/symmetric case $r = s \approx \lfloor N/2\rfloor$ captures the zeros and poles of the unknown function more accurately and provides the fastest convergence to the true function as $N$ increases. One can estimate if the poles so obtained are spurious or not by roughly checking how much they overlap with the zeros in the complex $z$ plane as the value of $N$ increases. An accumulation of non-spurious poles could signal an essential singularity or a branch cut \mcite{pade,*yamada2014numerical,*baker1996pade,*van2006pade}.

\begin{figure}[h!]
	\centering
	\begin{subfigure}[b]{0.48\linewidth}
		\centering
		\includegraphics[scale=0.44]{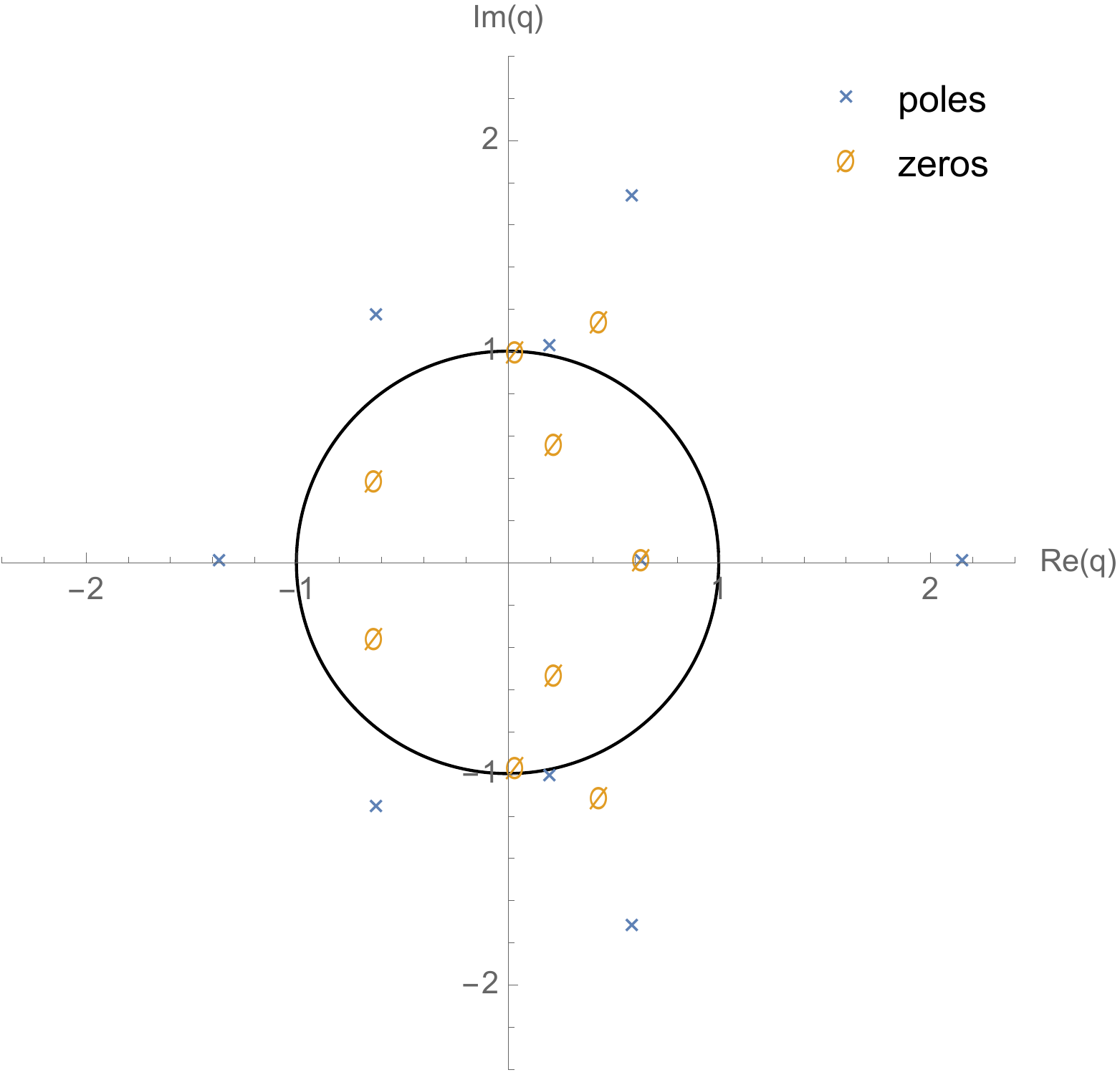}
		\caption{\label{fig:polezeroR33}}
	\end{subfigure}
	\begin{subfigure}[b]{0.48\linewidth}
		\centering\includegraphics[scale=0.44]{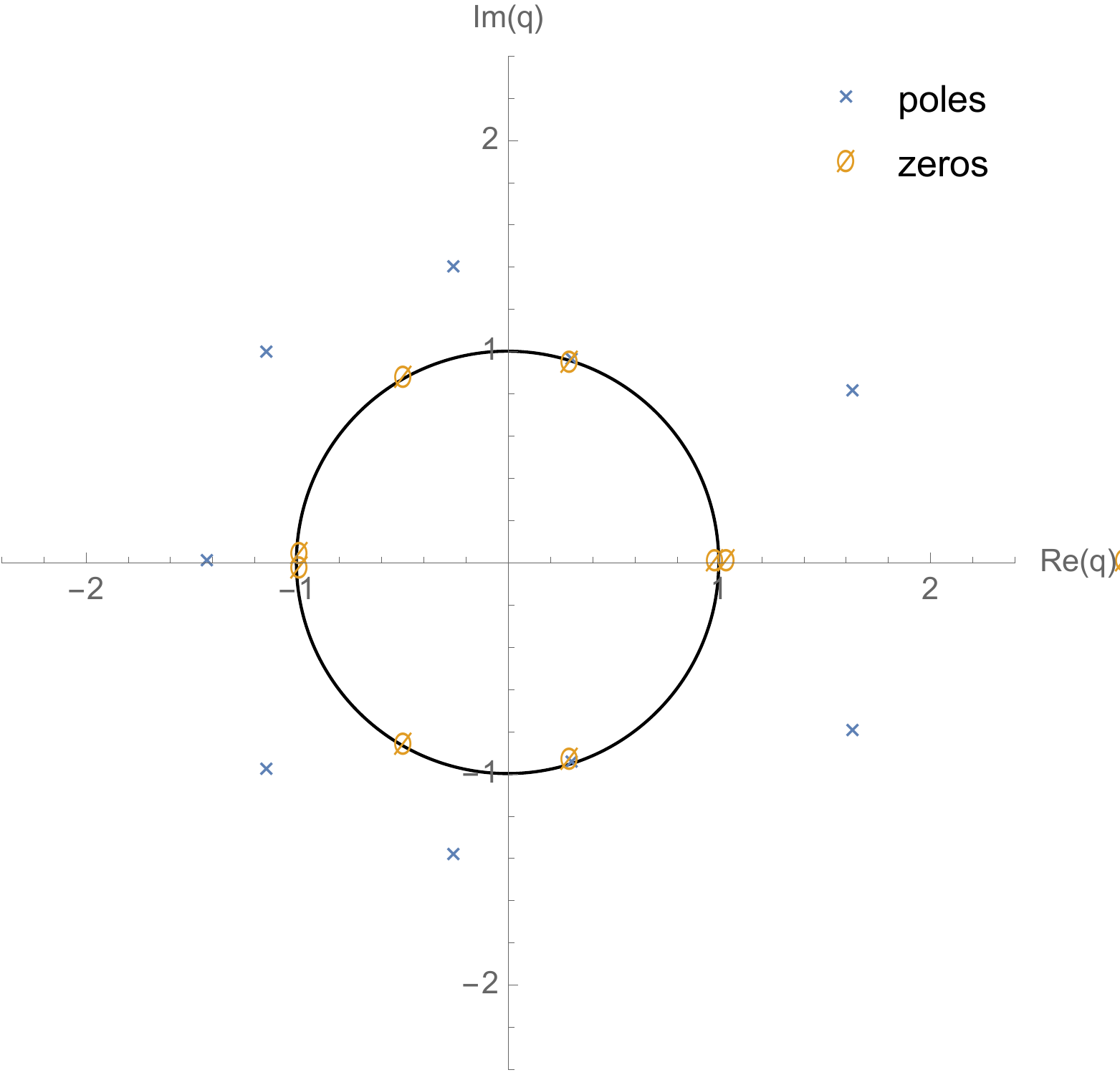}
		\caption{\label{fig:polezeroR13}}
	\end{subfigure}
	\caption{The zeros and poles in the complex $q$ plane of the $9/9$ Pad\'e approximant to the matrix elements (\subref{fig:polezeroR33}) $\bR_{33}(q)$ and (\subref{fig:polezeroR13}) $\bR_{13}(q)$ obtained using the oscillator  expansions based on the exact Neumann matrices.}
	\label{fig:polezero}
\end{figure}

We have constructed the Pad\'e approximants for a few matrix elements in Table \ref{tab:padeeval} to demonstrate their utility and to show that these provide a better approximation compared to relying on  the Taylor series as can be seen by comparing to Table \ref{tab:revenseries} above.
\begin{table}[h!]
	\begin{center}\[	\begin{array}{c||c c c c}
		k&\bR_{11}^P&\bR^P_{22}&\bR^P_{13}&\bR^P_{24}\\
		\hline\hline
		6 & -0.999853 & 1.00166 & -0.00221739 & 0.0244735 \\
		7 & -1.00018 & 1.00347 & -0.0218953 & -0.0581234 \\
		8 & -0.999948 & 1.00009 & 0.000125679 & 0.00170096 \\
		9 & -1.00003 & 1.00014 & -0.000261842 & -0.00459688 \\
		\end{array}\]
	\end{center}
	\caption{The Pad\'e approximant $P^k_k$ evaluated at $q=1$ for various matrix elements. The values are consistent with what one expects for the diagonal and non-diagonal elements, namely $(-)^n$ and $0$ respectively, although the convergence as $k$ increases is not uniform.}
	\label{tab:padeeval}
\end{table}
In order to look for hints of non-analyticity, we can study the poles and zeros of these rational functions as we do in Fig. \ref{fig:polezero} for two purely odd parity matrix elements. As can be observed from the plots, the poles do not appear to accumulate near the unit circle (or near $q=+1$ for that matter) at this order, and a few of them even seem to be somewhat spurious since they overlap a nearby zero. But notice that the poles are still consistent with being outside the unit disc.
Let us therefore consider the absolute values of the corresponding residues at these poles to ascertain the relative strength of the poles.  We divide out by the constant terms in these expansions---which is always $X^{11}_{nm}$ as can be seen immediately from \eqref{eq:Rnmosc expr}---in order to provide the numbers more intuitive. Furthermore, it is useful to consider the absolute values of the location of the poles to see if they are indeed approaching the boundary of the unit disc. We have performed these checks for several matrix elements and have presented the data in Table \ref{tab:R13residues} corresponding to the  $\bR_{13}$ case above. Because the off-diagonal functions vanish at the point $t=0$, we may expect to see stronger signals for these. Once again, we remark that with the current limited data there is not a robust behaviour that may be claimed to hold  and also that the residues may not be representative of the (non)analytic structure due to possible rapid oscillations. 
\begin{table}[h!]
	\centering
	\[	\begin{array}{c c c}
	\hline
	q_i& \lvert q_i\rvert &\mbox{Rescaled residue at\quad} q_i\\
	\hline\hline
	-1.42723 & 1.42723 & 1.73178 \\
	-1.14524-0.985928 i & 1.51117 & 4.92979 \\
	-1.14524+0.985928 i & 1.51117 & 4.92979 \\
	-0.259587-1.39047 i & 1.4145 & 4.20125 \\
	-0.259587+1.39047 i & 1.4145 & 4.20125 \\
	0.300928\, -0.953144 i & 0.99952 & 0.0630495 \\
	0.300928\, +0.953144 i & 0.99952 & 0.0630495 \\
	1.63103\, -0.80282 i & 1.81791 & 1.96839 \\
	1.63103\, +0.80282 i & 1.81791 & 1.96839 \\
	\hline
	\end{array}
	\]
	\caption{The location of the poles of the $9/9$ Pad\'e approximant to $\bR_{13}$, their absolute values and the corresponding residues. For being more useful, we have rescaled all the residues with the constant term $X^{11}_{13} = \frac{80}{729}\approx 0.10974$.}
	\label{tab:R13residues}
\end{table}
To get a better idea of the strength of these poles, we can also consider a plot\footnote{The code for generating this plot was taken from a \href{https://mathematica.stackexchange.com/questions/3458/plotting-complex-quantity-functions}{\emph{Mathematica} Stack Exchange page}.} of the absolute value (rescaled by $X^{11}_{nm}$) of these approximants in the complex $q$ plane as in Fig. \ref{fig:absoluteplot} below.
\begin{figure}[h!]
	\centering
	\begin{subfigure}[b]{0.48\textwidth}
		\includegraphics[width=\textwidth]{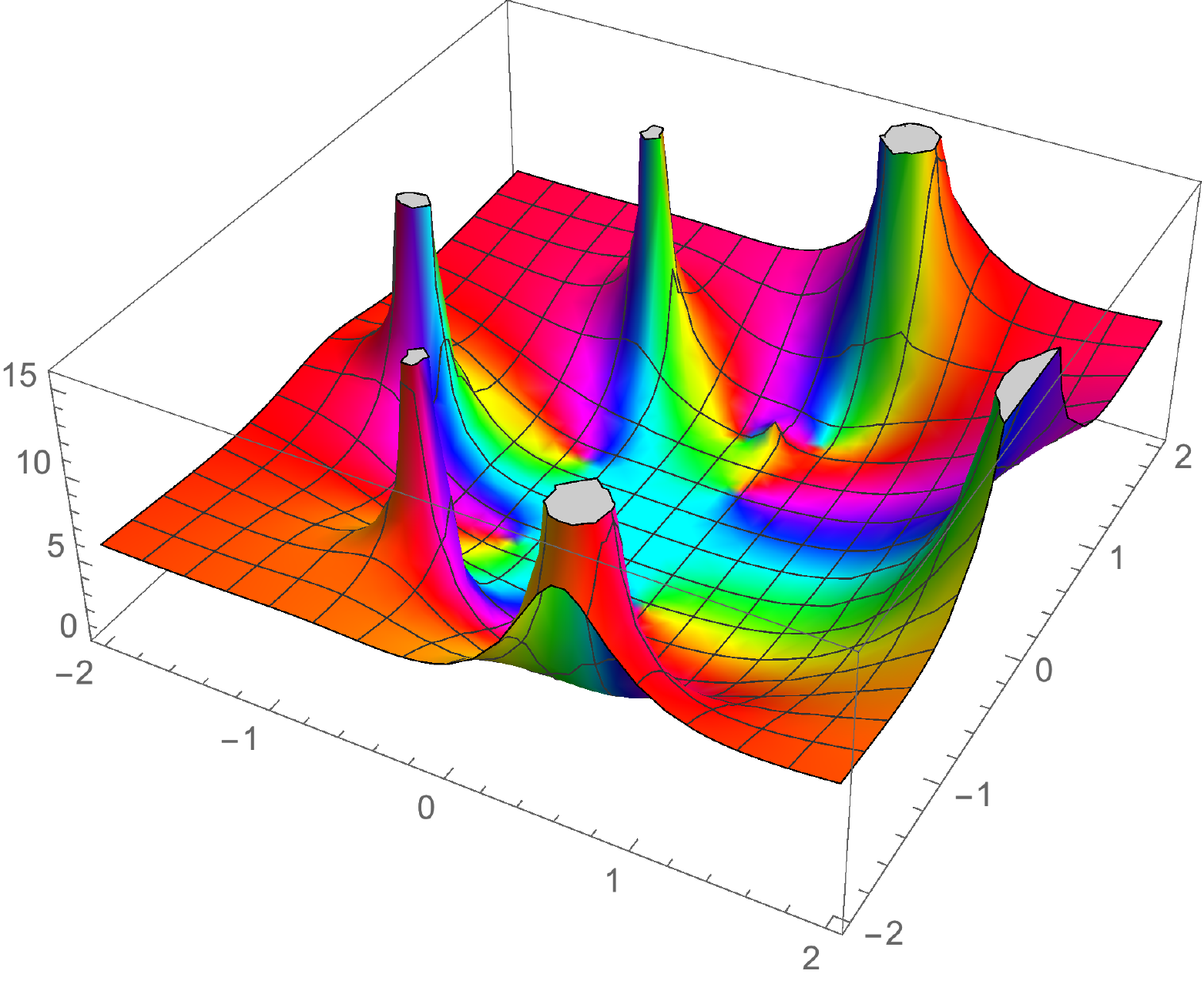}
		\caption{	\label{fig:r33complex}}
	\end{subfigure}
	\hfill
	\begin{subfigure}[b]{0.48\textwidth}
		\includegraphics[width=\textwidth]{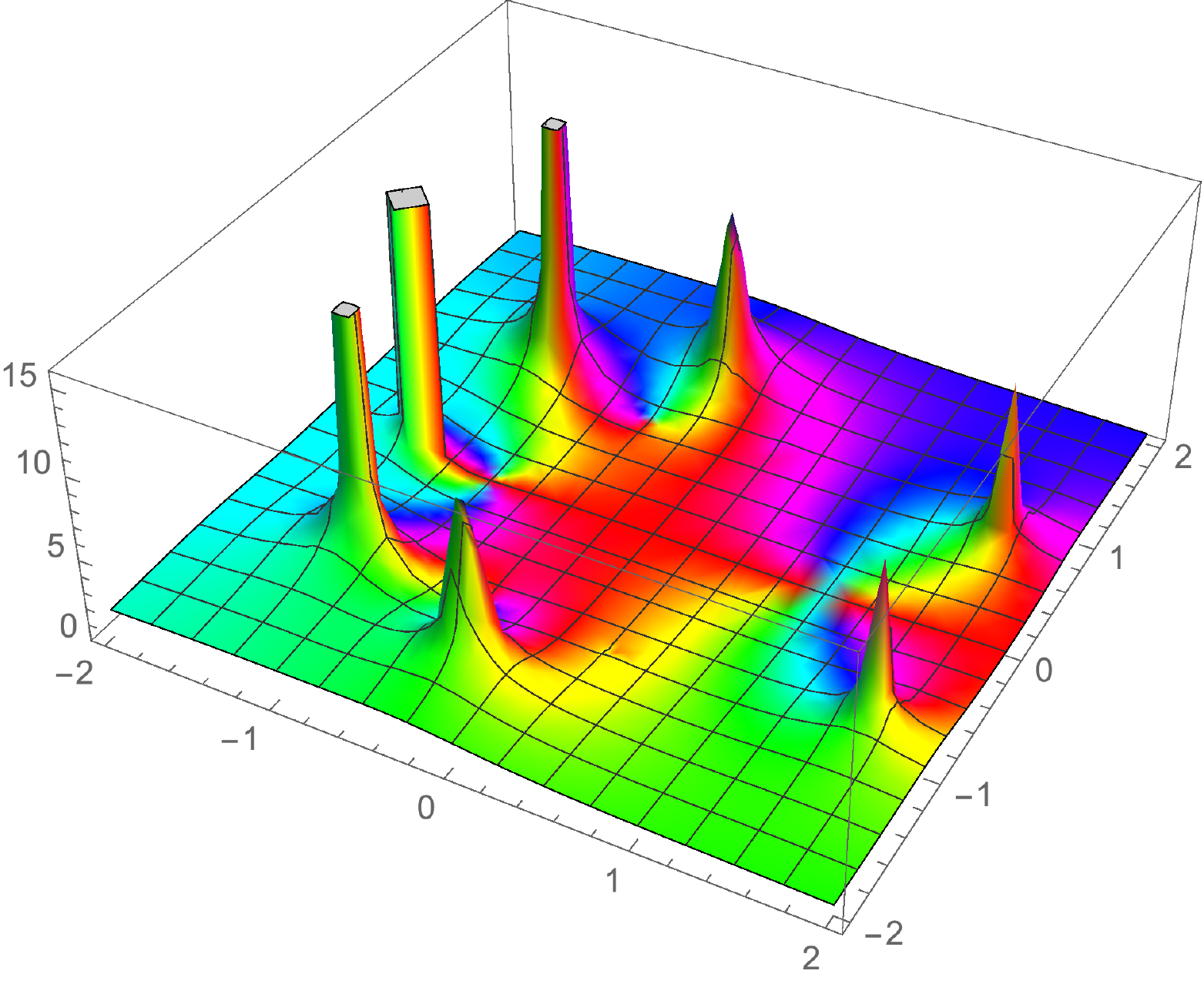}
		\caption{\label{fig:r13complex}}
	\end{subfigure}
	\caption{A plot displaying the absolute value and phase for the $9/9$ Pad\'e approximants to (\subref{fig:r33complex}) $\bR_{33}$ and (\subref{fig:r13complex}) $\bR_{13}$ in the complex $q$ plane. The ``spikes'' correspond to the location of the (simple)poles and the strength of the residues can be visually estimated by noticing how fast these are diverging. The phases are indicated using  colours  such that positive real numbers are assigned red, negative real numbers are assigned cyan, and the hue varies linearly. All the numbers for absolute values are  rescaled by the constant piece $X^{11}_{nm}$.}
	\label{fig:absoluteplot}
\end{figure}
Next, we have at our disposal another approximation scheme which is known to work better for low values of $N$: the Borel-Pad\'e approximation. In this method, one combines Pad\'e approximants with the Borel transform by first taking the Borel transform of the truncated series, then finding its Pad\'e approximant and finally doing the inverse Borel transform.\footnote{See the discussion in \cite[\S3.1]{Erler:2009uj} whose notation we shall try to follow.} The Borel transform of the truncated power series in $q$ is obtained by replacing each coefficient $p_k$ by $p_k/k!$, i.e:
\begin{equation}
\sum_{k=0}^N p_k q^k \to \sum_{k = 0}^N
\frac{p_k}{k!}q^k
\end{equation}
whose Pad\'e approximant $P^r_s(q)_{\rm{Borel}}$ can be obtained in a similar manner as above.

The final step is to perform the inverse Borel transformation that involves an integration along the positive real axis:
\begin{equation}
\tilde{P}^r_s(q) = \int_0^\infty dt \, e^{-t}\, P^r_s(t q)_{\rm{Borel}}.
\end{equation}
The interesting case is when the integrand has poles on the positive real axis which can correspond to ambiguities from subleading terms, not ordinarily seen in a power series expansion. Hence, we have analysed the pole structure of $P^r_s(q)_{\rm{Borel}}$ but have found that although there appears to be poles at certain positive values of $q$, these are not stable as the order $r/s$ is varied. For $\bR_{33}$, for instance, we have 
in Table \ref{tab:PadeBorelPositivePoles}
\begin{table}[h!]
	\centering
	\[\begin{array}{c c}
	\{r, s\} & \mbox{Pole of~} (P^r_s)_{\rm Borel} \mbox{~on}~ \mathbb{R}_+
	\vspace{0.16cm}\\
	\hline
	\{8, 7\} & {\rm none}\\
	\{8, 8\} & 18.526252\\
	\{9, 8\} & 17.710812\\
	\{9, 9\} & 18.450791\\
	\hline
	\end{array}\]
	\caption{The location of the poles on the positive real axis for the Pad\'e approximant to the Borel transform, $(P^r_s)_{\rm Borel}$, as $r, s$ is varied for $\bR_{33}(q)$.}
	\label{tab:PadeBorelPositivePoles}
\end{table}
but for many other matrix elements we have checked, this behaviour is much less clear as the imaginary parts are not stable.

However, we have evaluated the above integral for $q = 1$ and have found the expected result of $C_{nm} = (-)^n\delta_{nm}$ to good enough accuracy. We have also examined the (scaled) residues of $P^r_s(q)_{\rm{Borel}}$ towards this line of analysis. In short, the essential singularity expected for $q=+1$ and branch cuts due to  $\log(-\log q)$  do not show up conclusively at this order, indicating the need for much higher order coefficients or some other underlying features of the functions. 

%
%

\section{The bc system and $\hat{\beta}$ oscillators}\label{sec:bcbetaosc}
\vspace{-1mm}
In the BRST formulation, the worldsheet ghosts are introduced  as part of the gauge-fixing procedure analogous to the Faddeev-Popov ghosts in gauge field theories. In the first quantized theory, the worldsheet ghost and the anti-ghost  are denoted by $c(z)$ and $b(z)$ respectively\footnote{We follow Polchinski conventions \cite{Polchinski:1998rq} for the $bc$ ghost CFT.}. These are anti-commuting  fields with conformal weights $-1$ and $+2$.

After setting the time coordinate $\tau$ of the underlying worldsheet theory to $0$, we can have the mode expansion for these fields as follows:
\begin{equation}
b_{\pm\pm}(\sigma) = \sum_{n\in\mathbb{Z}}\hat{b}_n e^{\pm i n \sigma} = \pi_c(\sigma)\mp i b(\sigma),\qquad c^{\pm}(\sigma) = \sum_{n\in\mathbb{Z}}\hat{c}_n e^{\pm i n \sigma}=c(\sigma) \pm i \pi_b (\sigma).
\end{equation}
\subsection*{Moyal coordinates}
\vspace{-2mm}
Analogous to the matter sector, we can have ``positions'' and ``momenta'' linear combinations \cite[\S2.2]{Bars:2003gu} that we denote by $\hat{x}_n,\hat{p}_n, \hat{y}_n$ and $\hat{q}_n$ as follows:
\begin{align}\label{eq:xpyq modes}
\hat{x}_n = \frac{i}{\sqrt{2}}(\hat{b}_n - \hat{b}_{-n}),\quad \hat{p}_n = \frac{i}{\sqrt{2}}(\hat{c}_n-\hat{c}_{-n}), \quad\hat{y}_n
= \ortwo (\hat{c}_n+\hat{c}_{-n}),\quad \hat{q}_n = \frac{1}{\sqrt{2}}(\hat{b}_n+\hat{b}_n)
\end{align}
so that we may write:
\begin{align}
b(\sigma)& = i\sqrt{2}\sum_{\ninz_+}\hat{x}_n \sin n\sigma,\qquad
&\pi_b(\sigma)&= -i \sqrt{2}\sum_{n\in \mathbb{Z}_+}\hat{p}_n\sin n\sigma,\nonu
c(\sigma)& = \hat{c}_0 + \sqrt{2}\sum_{\ninz_+}\hat{y}_n\cos n \sigma,\qquad & \pi_c(\sigma)& = \hat{b}_0 +\sqrt{2}\sum_{\ninz_+}q_n\cos n \sigma.
\end{align}
Schematically, we may represent \cite{Erler:2003eq} this as:
\begin{align}
b & \to x, ~~ c \to y \oplus c_0\nonu
\pi_b & \to p, ~~ \pi_c \to q \oplus b_0
\end{align}
After choosing the Siegel gauge, we take the physical string field to be dependent only on the $c_0$ mode. Here we understand that the $b_0$ factor has been explicitly ``factored'' out. By virtue of the canonical (anti-)commutation relations \[\{\hat{c}_n,\hat{b}_m\}=\delta_{n+m,0},\qquad n, m \in \mathbb{Z},\]
we have the corresponding  structure:
\begin{equation}
\{\hat{x}_n,\hat{p}_m\}=\delta_{nm},\qquad \{\hat{y}_n,\hat{q}_m\}= \delta_{nm}, \qquad \mbox{~but now~} n, m \in \mathbb{Z}_+.
\end{equation}
At this point, it is essential to introduce the $SL(2,\mathbb{R})$/conformal vacuum and the associated ghost vacua constructed out of it. The conformal vacuum $\ket{\Omega}$ is the vacuum invariant under the global conformal group generated by the $L_{0,\pm}$ Virasoro generators. Because of the two ghost zero modes $\hat{c}_0$ and $\hat{b}_0$, we can have the two fold degenerate vacua $\ket\pm$ on top of this:
\begin{equation}
\ket{-}=\hat{c}_1\ket{\Omega},\qquad \ket{+}=\hat{c}_0\hat{c}_1\ket{\Omega},
\end{equation}
at ghost numbers $+1$ and $+2$ respectively. One has the freedom to work with either of these two vacua and henceforth we define \emph{states} by using the $\ket{-}$ vacuum, conventionally denoted as $\ket{\hat{\Omega}}$ or sometimes $\ket{\hat{0}}$.

From the underlying BCFT based on worldsheet path integrals,  we require   three ghost insertions to account for the conformal killing vectors (CKVs) for the disc (tree level)  amplitudes. In the Fock space language, this translates to 
the additional normalization condition \footnote{We thus set the total spacetime volume to $1$ through this normalization, which may be accomplished by a toroidal compactification of all $26$ bosonic coordinates, including the timelike direction. In general, for Dp branes, the tangential/longitudinal directions may be compactified.} on the vacua: 
\begin{equation}
\bra{+}-\rangle = 1\qquad \Longleftrightarrow \qquad \langle \hat{c}_{-1} \hat{c}_0 \hat{c}_1\rangle = 1.
\end{equation}
Thus, in every non-vanishing inner product, it is assumed that the ghost number requirement is saturated to $+3$ in this form\footnote{ The ghost number assignments are understood to be for the vertex operators as per modern conventions.}.

\subsection*{$\hat{\beta}$ oscillators}
\vspace{-2mm}
The relation between the matrix elements $\cF_{nm}$ corresponding to the half-phase space degrees of freedom $\xi^{gh}$ and the usual Fock space matrix elements $\bR_{nm}(t)$ can be obtained by using the form of the $\hat{c}_n, \hat{b}_n$ oscillators in the diagonal basis. To this end, we must employ the action of the linear maps between the two bases on these operators. The Moyal images of the Fock space states can be obtained by acting on the vacuum monoid with the so-called $\hat{\beta}$ oscillators:
\begin{equation}
\hat{c}_n \mapsto \hat{\beta}^c_n,\qquad\hat{b}_n\mapsto \hat{\beta}^b_n,  \qquad\mbox{where~~}
\hat{\beta}_{\mathcal{O}}A(\xi)\coloneqq \langle\xi| \hat{\mathcal{O}}|\psi\rangle.
\end{equation}
These are thus simply the counterparts for $\hat{c}_n, \hat{b}_n$ and the usual $\hat{\alpha}$ oscillators (in the matter sector) used in bosonic string theory and may be expressed either as \emph{differential operators} or phase space \emph{fields} with left and right $\ast$ action on the string field in the $\xi$ basis. We choose the differential operator representation in our discussion that follows.

In \cite{Bars:2003gu}, the oscillators are given for the \emph{odd} parity degrees of freedom $x^o, p^o, y^o$ and $q^o$, that can also be used to represent the $bc$ ghost system in the Moyal language. We have applied the canonical transformation that takes the odd basis to the even basis (\S\ref{subsec:moyal structures}) to rewrite them as follows:

\begin{align}\label{eq:betamap}
\hat{\beta}^c_e&\coloneqq \ortwo \left[-\frac{2i}{\theta^\prime}\sign (e)\ke^{-1} p^b_{|e|}+\frac{\theta^\prime}{2}\frac{\partial}{\partial p^c_{|e|}}\right],\qquad
&\hat{\beta}^c_o&\coloneqq \ortwo \left[R_{|o|e}x^c_e -i \sign(o)S^\top_{|o|e}\ke^{-1}\frac{\partial}{\partial x^b_e}\right],\nonu
\hat{\beta}^b_e&\coloneqq\ortwo \left[\frac{2}{\theta^\prime}p^c_{|e|} -i \sign(e)\frac{\theta^\prime}{2}\ke \frac{\p}{\partial p^b_{|e|}}\right], \qquad
&\hat{\beta}^b_o &\coloneqq \ortwo \left[-i\sign(o)S^\top_{|o|e}\ke x^b_e+ T^\top_{|o|e}\frac{\p}{\p x^c_e}\right],
\end{align}
where we have restored the non-commutativity parameter $\theta^\prime$ for the ghost sector and the summations over repeated indices are restricted to only the positively modded variables.\footnote{We only give the ghost parts without the zero-mode contribution since once we have chosen the Siegel gauge, only this form would be relevant to the discussion that follows.} We remind the reader that the matrix $S_{eo}$ arises naturally while defining the Moyal product in the ghost sector and simply equals
$\displaystyle
S_{eo} = \ke T_{eo}\ko^{-1}
$. It also satisfies 
$\displaystyle
SS^\top = \mathbb{1}_e,  S^\top S = \mathbb{1}_o
$, which can be proven from the properties of the $T$ and $R$ matrices.

\section{The twisted ghost butterfly case}\label{sec:twisted}
It is an interesting exercice to consider the overlap with the twisted ghost butterfly state instead of the perturbative vacuum.
This is one of the simplest star algebra projectors and is defined by the following state in Fock space:
\begin{equation}
|\psi_B\rangle = \exp\left[-\half L_{-2}^\prime\right]|\Omega^\prime\rangle,
\end{equation}
where the prime refers to the twisted ghost conformal field theory studied by Gaiotto, Rastelli, Sen, and Zwiebach (GRSZ) \cite{Gaiotto:2001ji}(see also \mcite{butterfly,*Okawa:2003cm,*Schnabl:2002ff,Bars:2003gu}). Because the (total) stress tensor on the canonical strip coordinate $w$ is twisted as:
\begin{equation}
T^\prime(w) = T(w)-\partial j_g(w),
\end{equation} 
where $j_g = c\, b$ is th ghost number current in the original CFT, (and similarly for the anti-holomorphic component), the new Virasoro operator above is given by
\begin{equation}
L_{-2}^\prime = L_{-2}-2j_{-2}.
\end{equation}
In Moyal space, this state is represented by the twist even and $SU(1,1)$ symmetric string field
\begin{equation}
\hat{A}_B^\prime=\xi_0\,\cN_B e^{-\bxigh M_B \xigh},
\end{equation}
where
\begin{equation}
\cN_B = 2^{-2N},\qquad M_B = -\half\left[\begin{array}{cc}
\ke & \mzero \\ 
\mzero& 4\ke^{-1}
\end{array} \right].
\end{equation}
We remark that this string field satisfies
\begin{equation}
\beta^b_e\ast \hat{A}^\prime_B = \beta^c_e\ast \hat{A}^\prime_B = \hat{A}_B^\prime \ast \beta^b_{-e} = \hat{A}_B^\prime \ast \beta^c_{-e} = 0, \forall e>0,
\end{equation}
where now  the $\beta^{b,c}_e$ are \emph{fields} in Moyal space instead of differential operators:
\begin{equation}
\beta^b_e = p^c_{|e|} -\frac{i}{2}\sign(e)\ke x^b_{|e|}, \qquad \beta^c_{e} = \half x^c_{|e|}-i\sign(e) \ke^{-1}p^b_{|e|}.
\end{equation}
Moving on, let us write 
$
M_B \eqqcolon \chi \Mgh,
$
where we introduce the matrix 
\begin{equation}
\chi =\left[\begin{array}{cc}
\Gamma^\top&  \mzero\\ \mzero
& \mone
\end{array} \right], \mbox{~with~} \Gamma\coloneqq T\ko^{-1}T^{\top}\ke.
\end{equation}
We mention that the matrix elements of $\Gamma^\top$ can be evaluated exactly in the infinite $N$ limit and are given by:
\begin{equation}
\Gamma^\top_{2n,2m} = (-)^{n+m+1}\frac{2n}{\pi^2(n^2-m^2)}\left[\psi\left(\half+n\right)+\psi\left(\half-n\right)-\psi\left(\half+m\right)-\psi\left(\half-m\right)\right].
\end{equation}
Then, we have $M_B (\Mgh)^{-1}=\chi$. 
Now, let us consider the monoid defined by 
\[\hat{A}_B^\prime\ast e^{\bxigh\egh}\cdot( q^{L_0}e^{-\bxigh\egh}).\]
This has the parameters:
\begin{align}
M_{123} &= M_B + \frac{f_3(q)}{f_2(q)}\Mgh = \left[\chi +\frac{f_3(q)}{f_2(q)}\right]\Mgh,\nonu
\lambda_{123}&= \left[m_B - \frac{f_1(q)}{f_2(q)}\right]\eta = \left[\chi\mgh-\frac{f_1(q)}{f_2(q)}\right]\eta, \mbox{~and}\nonu
\cN_{123}& = 4^{-2N}\det(f_2(q))\exp\left[\aqua\begh\left((\Mgh)^{-1}\frac{f_3(q)}{f_2(q)}+\sigma\chi \Mgh \sigma\right)\egh\right].
\end{align}
As for the perturbative vacuum state, we can next take the trace over $\xigh$ and then perform the Gaussian integral over $\egh$. This time, we set $\lgh =0$ and have the non-vanishing coefficient matrices as follows:
\begin{subequations}
	\begin{align}
	\cQ_\eta^B & = \aqua \left[(\Mgh)^{-1}\frac{f_1(q)}{f_2(q)}+\sigma\chi\right]\left[\chi + \frac{f_3(q)}{f_2(q)}\right]^{-1}\left[\frac{f_1(q)}{f_2(q)}-\chi \Mgh\sigma\right]\simbreak
	+\aqua \sigma \chi \Mgh\sigma+\aqua (\Mgh)^{-1}\frac{f_3(q)}{f_2(q)},\\
	\cC_\eta^B & = 2^{-2N}(1+w^\top w)^{1/2}\det\left(f_2(q)\chi + f_3(q)\right).
	\end{align}
\end{subequations}
Finally, the Gaussian integration results in  $\cC_\eta^B \det(2\cQ_\eta^B)$ which may be looked at numerically. Here, we can also consider excitations on top of this state by having a general $\lgh$ but keeping in mind that the state already contains excitations created by the odd oscillator modes $\hat{c}_{-o}, \hat{b}_{-o}$. Since it is annihilated by all the even oscillators $\hat{c}_e, \hat{b}_e$, we notice that for excitations created by the corresponding creation operators, we recover the same even parity matrix elements $\bR_{2n,2m}$ as in the earlier analysis.

\printbibliography

\end{document}